\documentclass[10pt]{emulateapj}
\usepackage{apjfonts}
\usepackage{natbib}
\usepackage{amsfonts}
\usepackage{amsmath}
\usepackage[usenames]{color}
\def\mmsun{\mathrm{M}_\odot}

\def\Dwa{$\,$\uppercase\expandafter{\romannumeral5}$\,$}

\def\sless{\lower2pt\hbox{$\buildrel {\scriptstyle <}
   \over {\scriptstyle\sim}$}}

\def\sgreat{\lower2pt\hbox{$\buildrel {\scriptstyle >}
   \over {\scriptstyle\sim}$}}
\def\sharpnull#1{}

\newcommand{\sn}{$\mathrm{S}_n$}

\begin{document}
\slugcomment{Submitted to ApJ. March 31, 2008, accepted June 24, 2008.}

\title{2D Multi-Angle, Multi-Group Neutrino Radiation-Hydrodynamic 
Simulations of Postbounce Supernova Cores}

\author{Christian D. Ott\altaffilmark{1},
Adam Burrows\altaffilmark{2,1}, Luc Dessart\altaffilmark{2,1},
and Eli Livne\altaffilmark{3}}
\altaffiltext{1}{Department of Astronomy and Steward Observatory, 
The University of Arizona, 933 N. Cherry Ave., 
Tucson, AZ 85721; cott@as.arizona.edu, 
luc@as.arizona.edu}
\altaffiltext{2}{Department of Astrophysical Sciences,
Princeton University, Peyton Hall, Ivy Lane,
Princeton, NJ 08544, burrows@astro.princeton.edu} 
\altaffiltext{3}{Racah Institute of Physics, The Hebrew University,
Jerusalem, Israel; eli@phys.huji.ac.il}

\begin{abstract}
We perform axisymmetric (2D) multi-angle, multi-group neutrino
radiation-hydrodynamic calculations of the postbounce phase of
core-collapse supernovae using a genuinely 2D discrete-ordinate (\sn)
method. We follow the long-term postbounce evolution of the cores of
one nonrotating and one rapidly-rotating 20-M$_\odot$ stellar model
for $\sim$400 milliseconds from 160~ms to $\sim$550~ms after bounce.
We present a multi-D analysis of the multi-angle neutrino radiation
fields and compare in detail with counterpart simulations carried out
in the 2D multi-group flux-limited diffusion (MGFLD) approximation to
neutrino transport. We find that 2D multi-angle transport is superior
in capturing the global and local radiation-field variations
associated with rotation-induced and SASI-induced aspherical
hydrodynamic configurations. In the rotating model, multi-angle
transport predicts much larger asymptotic neutrino flux asymmetries
with pole to equator ratios of up to $\sim$2.5, while MGFLD tends to
sphericize the radiation fields already in the optically
semi-transparent postshock regions. Along the poles, the multi-angle
calculation predicts a dramatic enhancement of the neutrino heating by
up to a factor of 3, which alters the postbounce evolution and results
in greater polar shock radii and an earlier onset of the initially
rotationally weakened SASI. In the nonrotating model, differences
between multi-angle and MGFLD calculations remain small at early times
when the postshock region does not depart significantly from spherical
symmetry. At later times, however, the growing SASI leads to
large-scale asymmetries and the multi-angle calculation predicts up to
30\% higher average integral neutrino energy deposition rates than
MGFLD.
\end{abstract}


\section{Introduction}
\label{section:intro}

Four decades after the first pioneering neutrino
radiation-hydrodynamic calculations of stellar collapse
(\citealt{colgate:66,arnett:66, leblanc:70,wilson:71}), the details of
the core-collapse supernova explosion mechanism remain
obscure. However, certain essentials are clear. The collapse of the
evolved stellar core to a protoneutron star (PNS) and its evolution to
a compact cold neutron star provides a gigantic reservoir of
gravitational energy, $\sim$3$\times$10$^{53}$~erg, a mass-energy
equivalent of $\sim$0.17~$\mmsun$. Any core-collapse supernova
mechanism must tap this energy and convert the fraction needed to
match Type-II supernova observations ($\sim$10$^{51}$~erg $\equiv$ 1
Bethe [B]) into kinetic and internal energy of the exploding stellar
envelope.

There is general agreement that the prompt hydrodynamic explosion
mechanism does not work and that the bounce shock always stalls,
falling short of blowing up the star (e.g.,
\citealt{bethe:90,janka:07}), and must be re-energized to lead to a
supernova. However, there is no agreement on the detailed mechanism
that revives and endows the shock with sufficient energy to make a
canonical $\sim$1-Bethe supernova. For decades, the
``neutrino-driven'' mechanism, first proposed in its direct form by
\cite{colgate:66}, and in its delayed form by \cite{wilson:85} and
\cite{bethewilson:85}, seemed compelling.  It relies on a subtle
imbalance of neutrino heating and cooling that leads to a net energy
deposition behind the stalled shock, sufficient to revive it and drive
the explosion on a timescale of hundreds of milliseconds. While
appealing, it has been shown to fail for regular massive stars in
spherical symmetry (1D) when the best neutrino physics and transport
are used \citep{ramppjanka:00,liebendoerfer:01a,
liebendoerfer:05,thompson:03}. Yet, weak explosions may be obtained in
1D for the lowest mass progenitors, O-Ne-Mg cores
(\citealt{kitaura:06,burrows:07c}).

It is now almost certain that the canonical explosion mechanism must
be multi-dimensional (2D/3D) in nature.  The multi-D dynamics
associated with convective overturn in the postshock region (e.g.,
\citealt{herant:94, bhf:95,jankamueller:96,buras:06b}) and the
recently identified standing accretion shock instability (SASI,
e.g., \citealt{foglizzo:00,foglizzo:07,
scheck:08,blondin:03,burrows:07a,iwakami:08}) lead to a
dwell time of accreting outer core material in the postshock region
that is larger on average than in the 1D case. This results in a
greater neutrino energy deposition efficiency behind the shock and,
thus, creates more favorable conditions for explosion
\citep{burrows:93, janka:01,thompson:05,marek:07}.

The first generation of multi-dimensional supernova calculations,
still employing gray flux-limited diffusion (or yet simpler schemes)
for neutrino transport, indeed found that neutrino-driven convective
overturn in the region between the stalled shock and the PNS
sufficiently increased the neutrino energy deposition rate to lead to
a delayed explosion (\citealt{herant:94,bhf:95, jankamueller:96,fh:00,
fryerwarren:02, fryerwarren:04}). The more sophisticated studies that
followed changed this picture. Recent long-term axisymmetric (2D)
supernova calculations with multi-group, multi-species neutrino
physics and transport find it difficult to explode garden-variety
massive stars via the neutrino mechanism. \cite{buras:06b} report
explosion only for the low-mass (11.2~$\mmsun$) progenitor of
\cite{wwh:02}, while \cite{marek:07} report the onset of explosion in
a 15-$\mmsun$ model of \cite{ww:95}, given moderately fast rotation
and the use of the Lattimer-Swesty equation of state (EOS;
\citealt{lseos:91}) with a nuclear compressibility modulus $K_0$ of
180~MeV, which is significantly softer than the current best
experimental values ($K_0 = 240 \pm 20$~MeV; \citealt{shlomo:06}).  On
the other hand, \cite{bruenn:06} obtain explosions for 11-$\mmsun$ and
15-$\mmsun$ progenitors from \cite{ww:95} only when they take silicon
and oxygen burning into account and due to a synergy between nuclear
burning, the SASI, and neutrino heating.

\cite{burrows:06,burrows:07a} do not obtain neutrino-driven explosions
(except in the case of O-Ne-Mg cores and accretion-induced collapse;
\citealt{dessart:06b}), but observe the excitation of PNS core
$g$-modes. In their calculations, the PNS core oscillations reach
non-linear amplitudes and damp via the emission of strong sound waves
that propagate through the postshock region and efficiently deposit
energy into the shock, eventually leading to late explosions at
$\sim$1~second after bounce. This \emph{acoustic} mechanism appears to
be robust enough to blow up even the most massive and extended
progenitors \citep{burrows:07a,ott:06b}, but remains controversial and
needs to be confirmed by other groups (see, e.g., 
\citealt{yoshida:07,weinberg:08}).

In the context of rapid progenitor rotation, \cite{burrows:07b},
\cite{dessart:08a}, and \cite{dessart:07} (the latter for the
accretion-induced collapse scenario) have shown that energetic
MHD-driven explosions may be obtained if field-amplification by the
magneto-rotational instability \citep{bh:91} is as efficient in the
core-collapse context as suggested \citep{akiyama:03}. Whether 
rotation alone and without strong magnetic fields favors or disfavors
a neutrino-driven explosion remains to be seen \citep{walder:05,
dessart:06b,ott:06spin}, but rapid rotation has been shown to damp
convection \citep{fh:00} and weaken the SASI \citep{burrows:07b}.

\subsection{Core-Collapse Supernova Theory and Neutrino
Radiation Transport}

Neutrinos, their creation, propagation, and interactions with
supernova matter, are of paramount importance to the core-collapse
supernova problem. They carry away $\sim$99\% of the final neutron
star's gravitational binding energy and $\sim$1\% of this energy would
be sufficient to blow up the star. Depending on progenitor
characteristics that set the postbounce rate of mass accretion onto
the PNS, a successful supernova explosion should occur within
$\sim$1--1.5~s after bounce to match observational and theoretical
neutron star upper mass limits around $\sim$2--2.5~$\mmsun$
(\citealt{lattimer:07}, and references therein).  Consequently, the
explosion mechanism must deliver canonical 1-B explosions on this
timescale and, if the explosion is neutrino-driven, the neutrino
heating efficiency\footnote{We define the heating efficiency as the
ratio of the energy deposition rate and the summed electron-neutrino
and anti-electron neutrino luminosities.
The $\mu$ and $\tau$ neutrinos and their
anti-particles do not contribute much to the heating.} must be on the
order of 10\% to yield an explosion that achieves an energy of 1~B
within $\sim$1~s.

The neutrinos travelling through the postshock region in a postbounce
supernova core are not in thermal equilibrium with the baryonic
matter. They should ideally be treated with full kinetic theory,
describing the neutrino distributions and their temporal distribution
with the Boltzmann equation (\citealt{mihalas:99}). Boltzmann
transport is in its most general form a 7-dimensional problem. The 6D
neutrino phase space (usually split up into 3D spatial coordinates,
neutrino energy, and 2 angular degrees of freedom) 
and time. In addition, there are up to 6 neutrino
types (3 particle species, and their anti-particles) to deal
with. Spherically-symmetric Boltzmann transport schemes have been
devised and implemented in the core-collapse context
\citep{mezzacappa:93b,messer:98,burrows:00,
yamada:99,mezzacappa:99,ramppjanka:02,liebendoerfer:04,hb:07}, but
general Boltzmann transport in multiple spatial dimensions is
computationally challenging and will remain so in the intermediate
term.  Hence, approximations must be made in devising computationally
tractable neutrino transport schemes for multi-D simulations.

A highly sophisticated approximation that arguably comes close to full
Boltzmann transport in the case of quasi-spherical configurations in
2D is that presented in \cite{buras:06a}, and based on earlier work by
\cite{ramppjanka:02}. These authors solve equations for the zeroth and
first angular moments of spherically-symmetric radiation fields along
multiple radial rays (ray-by-ray approach; \citealt{bhf:95}) and
perform a variable Eddington factor closure \citep{mihalas:99} via a
single spherically symmetric Boltzmann solution on an averaged 1D
profile of the 2D hydrodynamics data.  Neighboring rays are coupled to
provide for limited treatment of latitudinal transport. Their
multi-group (multi-energy and multi-neutrino species) scheme includes
inelastic neutrino-electron scattering, aberration, gravitational
redshift, and frame effects to ${O}(\mathrm{v}/c)$.

\cite{livne:04} implemented a genuinely 2D direct solution of a
reduced Boltzmann equation via the method of discrete ordinates
(S$_n$: see, e.g.,
\citealt{yueh:77,mezzacappa:93b,adamslarsen:02,castor:04}, and
references therein) in the code VULCAN/2D, neglecting energy
redistribution and fluid-velocity dependence.

A common, more approximate way to handle neutrino transport that has a
long pedigree in 1D core-collapse studies is multi-group
(energy/neutrino species) non-equilibrium flux-limited diffusion
(MGFLD; \citealt{mihalas:99}, \citealt{arnett:66},
\citealt{bowerswilson:82}, \citealt{bruenn:85},
\citealt{myra:87,myra:90,baron:89,cooperstein:92}). FLD schemes solve a
diffusion equation for the mean radiation intensity, the zeroth
angular moment of the specific radiation intensity. Hence, they drop
all local angular dependence of the radiation field, while, in the
MGFLD case, retaining the spectral neutrino distribution. MGFLD
accurately describes the radiation field at high optical depth where
the diffusion approximation is exact. In the free-streaming limit, the
flux must be limited to maintain causality and an interpolation must
be performed between diffusion and free-streaming regimes by an
ad-hoc prescription (using a flux limiter).

2D FLD schemes were pioneered in the core-collapse context by
\cite{leblanc:70} and modern MGFLD implementations can be found in
\cite{swesty:06} and in \cite{burrows:07a}.  It is not a priori clear
whether MGFLD is an accurate enough prescription to yield postbounce
supernova dynamics in qualitative and quantitative agreement with a
more accurate multi-angle treatment.
Since net energy deposition by neutrinos is favored only in the
semi-transparent gain layer, the quality of a MGFLD scheme may
sensitively depend on the flux limiter chosen \citep{burrows:00}.
The fact that 2D gray FLD schemes have in the past led to
neutrino-driven explosions
\citep{herant:94,bhf:95,fh:00,fryerwarren:02,fryerwarren:04}, while
MGFLD schemes appear not to \citep{walder:05,burrows:06,burrows:07a},
emphasizes the importance of a spectral treatment of neutrino
transport.

In 1D, MGFLD and Boltzmann neutrino transport were compared on static
hydrodynamic postbounce backgrounds by \cite{janka:92},
\cite{yamada:99}, \cite{messer:98}, and \cite{burrows:00}.  Also in
1D, \cite{mezzacappa:93a} compared Boltzmann transport and MGFLD
evolutions in the collapse phase, while \cite{liebendoerfer:04}
performed the only comparison to date of 1D long-term Boltzmann and MGFLD
supernova evolutions.  The static studies all agree that
Boltzmann transport yields larger instantaneous neutrino heating rates
in the gain region, mostly because of a more slowly decreasing inverse
flux factor ($c$ over the ratio of flux to neutrino energy density),
a quantity that can be related to the rate of energy absorption. On the other
hand, \cite{liebendoerfer:04} find no significant dynamical
differences between MGFLD and Boltzmann transport evolutions in their
long-term comparison study with the 13-$\mmsun$ progenitor
model of~\cite{nomoto:88}.

In this paper, we present 2D \emph{multi-angle}, multi-group neutrino
transport supernova calculations using the Newtonian axisymmetric
VULCAN/2D code \citep{livne:93,livne:04,burrows:07a}.  Comparing
multi-D Boltzmann and MGFLD treatments, we perform postbounce
simulations with VULCAN/2D and compare 2D steady-state snapshots, as
well as fully-coupled dynamical 2D radiation-hydrodynamics evolutions,
for non- and rapidly-rotating 20-$\mmsun$ models whose precollapse
profiles are taken from \cite{wwh:02}.  We analyze our angle-dependent
neutrino radiation fields and provide for the first time local 2D map
projections of the specific intensity $I_\nu$.

In \S\ref{section:methods}, we describe our hydrodynamic and
radiation-transport schemes and the microphysics that we use in this
postbounce core-collapse supernova study. In \S
\ref{section:initialmodels}, we introduce the presupernova models and
the postbounce configurations, the setup, and the methodology of our
Boltzmann-transport--MGFLD comparisons. In \S\ref{section:snapshots},
we present results of snapshot Boltzmann transport calculations and
compare them with their MGFLD counterparts. In
\S\ref{section:evolution}, we then discuss time-dependent calculations,
the dynamical differences between Boltzmann transport and MGFLD
runs, and the consequences for postbounce supernova model
evolution. We wrap up in
\S\ref{section:summary} with a summary and critical discussion of the
work presented in this paper.

\section{Methods}
\label{section:methods}

\subsection{Hydrodynamics}

We employ the arbitrary Lagrangian-Eulerian (ALE, with second-order
total-variation-diminishing [TVD] remap) radiation-hydrodynamics code
VULCAN/2D. The hydrodynamics module was first described by
\citet{livne:93}\footnote{For details and an extension to
magneto-hydrodynamics not employed here, see \cite{livne:07}.}.
The 2D time-explicit hydrodynamics scheme is second-order accurate (in
smooth parts of the flow), unsplit, and implements a finite-difference
representation of the Newtonian Euler equations with artificial
viscosity on arbitrarily structured grids and in cylindrical
coordinates.  The computational grid employed here is set up to
resemble a spherical-polar grid at radii greater than 20~km and
gradually transitions to a Cartesian structure at smaller radii
\citep{ott:04}.  This (a) avoids hydrodynamic timestep restrictions
due to focussing of angular grid lines and (b) liberates the PNS core,
thus allowing mass motion along the axis of symmetry.

Self-gravity is implemented via direct grid-based solution
of the Newtonian Poisson equation, as described in
\citet{burrows:07a}, and we employ the finite-temperature
nuclear equation of state of \citet{shen:98a,shen:98b}.
The calculations are run with 230 logarithmically-spaced 
radial and 120 angular zones (including the
inner, quasi-Cartesian region). The grid encompasses
a radial extent of 4000~km and the full 180$^\circ$
of the axisymmetric domain.

\subsection{Neutrino Transport and Microphysics}

VULCAN/2D contains two multi-group, multi-species neutrino
radiation-transport options.  As we discuss below
(\S\ref{section:initialmodels}), both modules are used in this study.
The module implementing 2D transport in the MGFLD approximation, 
evolving the zeroth moment of the radiation
field, is discussed in \cite{burrows:07a}.  The angle-dependent
transport module that evolves the specific neutrino radiation
intensity,
$I(\mathbf{r},\mathbf{\Omega},\varepsilon_\nu,\mathrm{species},t)$, via
the method of discrete ordinates (S$_n$), was first discussed by
\citet{livne:04} (see also \citealt{morel:96,adamslarsen:02,castor:04}).

\begin{figure}
\centering
\includegraphics[width=7cm]{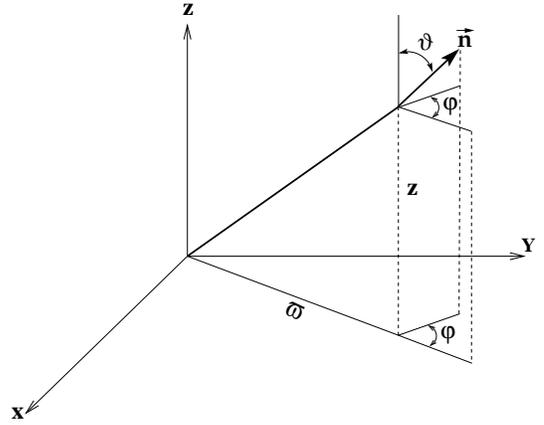}
\caption{Coordinates used in the axisymmetric
\sn\ transport scheme implemented in VULCAN/2D.  The radiation
direction vector $\vec{n}$ is defined in terms of $\vartheta$ and
$\varphi$. $\vartheta$ is the angle with respect to the
coordinate-grid $z$-axis at all spatial positions ($z$,$\varpi$).  At each
($z$,$\varpi$), the local momentum-space unit sphere is covered by $n$
zones in $\vartheta$ and at each $\vartheta$ location by a number
$m(\vartheta)$ of $\varphi$-zones, so that each zone in
$(\vartheta,\varphi)$ covers roughly the same solid angle.
\label{fig:coord}} 
\end{figure}

For convenience and future reference,
we define the zeroth, first, and second moments of the
radiation field,
\begin{eqnarray}
J_{\nu} &\equiv& \frac{1}{4\pi}
\oint_{4\pi} d\Omega\,\, I_{\nu}\,,\label{eq:J}\\ \vec{H}_{\nu}
&\equiv& \frac{1}{4\pi} \oint_{4\pi} d\Omega\,\, \vec{n}\,
I_{\nu}\,,\label{eq:H}\\ {\textsf
K}_{\nu} &\equiv& \frac{1}{4\pi} 
\oint_{4\pi} d\Omega\,\, \vec{n}\,\vec{n}\,I_{\nu}\,\label{eq:eddy}.
\end{eqnarray}
Note the vector and tensor natures of $\vec{H}_\nu$ and
${\textsf K}_\nu$, respectively. $\vec{n}$ is the
radiation field unit vector whose coordinate-dependent
components are given in \cite{hb:07} for various common
coordinate systems. Here we employ cylindrical coordinates
(see Fig.~\ref{fig:coord}). The radiation-pressure tensor
${\textsf K_\nu}$ obeys the trace condition $J_\nu =
\mathrm{Tr}[K_\nu]$~\citep{mihalas:99}. The spectral
neutrino flux is defined as $\vec{F}_\nu = 4\pi \vec{H}_\nu$.

As explained in \citet{livne:04}, the
time-implicit S$_n$ solver in VULCAN/2D updates the
specific intensity in the laboratory
frame via the Boltzmann transport equation
(\citealt{castor:72}) without
fluid-velocity dependence,
\begin{equation}
\label{eq:trans_master}
\frac{1}{c}\frac{\partial I}{\partial t} + \vec{n}\cdot \vec{\nabla} I
+ \sigma I = S\,\,, 
\end{equation}
where we have dropped the neutrino group index $\nu$.
 $\sigma = \sigma^a + \sigma^s$, where
$\sigma^a(\mathbf{r},\varepsilon_\nu, \mathrm{species})$ 
is the inverse absorption mean-free path and 
$\sigma^s(\mathbf{r},\varepsilon_\nu,\mathrm{species})$
is the inverse scattering mean-free path (both equivalent to the
corresponding cross section multiplied by the number density).
 We assume scattering to be isotropic and
employ the transport cross section $\sigma^s = (1 - \langle \cos
\vartheta\rangle)\sigma^s_T$ instead of the total scattering cross
section $\sigma^s_T$. This approach has been shown to work well in
spherically-symmetric core-collapse supernova calculations
\citep{burrows:00,thompson:03}.  The right-hand side source term $S$
equals $S_\mathrm{em}(\mathbf{r},\varepsilon_\nu,\mathrm{species}) +
\sigma^s J$, where $S_\mathrm{em}$ is the emissivity. The transport
grid is identical to the hydrodynamics grid.  The specific intensity
and its moments are defined at cell centers, facilitating
spatially-consistent coupling with the scalar hydrodynamics variables,
as discussed in \citet{livne:04}.  Radiation stress at cell corners is
computed via linear interpolation employing cell-centered values of
the radiation flux.

As a consequence of the neglect of $O(\mathrm{v}/c)$ terms in our transport
formulation, neutrino advection, Doppler shifts and aberration effects
are not considered. This greatly limits the computational complexity
of the problem, but its impact on the transport solution depends on
the particular choice of reference frame and was examined in
\cite{hb:07}. It is clear that around core bounce and neutrino
breakout, during the non-linear phase of the SASI hundreds of
milliseconds after bounce, and in the case of rapid rotation,
including $O(\mathrm{v}/c)$ terms is advisable. We leave them out here in order
to make long-term multi-angle radiation-hydrodynamics simulations
feasible and allow direct comparison with the MGFLD variant of
VULCAN/2D. Full $O(\mathrm{v}/c)$ Boltzmann transport with energy
redistribution will be addressed using the code BETHE currently under
development by a subset of our group (\citealt{hb:07,murphy:08}).

We discretize the angular radiation distribution
evenly in $\cos{\vartheta}$ from -1 to 1 and in
$\varphi$ evenly from 0 to $\pi$ (treating only
one hemisphere because of axial symmetry).
We make the number of $\varphi$-bins a function of
$\cos{\vartheta}$ to tile the hemisphere more
or less uniformly in solid angle. 
In our time-dependent runs we employ
8 $\cos \vartheta$ bins, resulting in a total
of 40 angular zones. Steady-state radiation fields
are computed with 8 $\cos \vartheta$ bins, 
12 $\cos \vartheta$ bins (92 total angular zones) 
and 16 $\cos \vartheta$ bins (162 total angular zones)
at each spatial grid point.

The standard set of neutrino-matter interactions
listed in \cite{thompson:03} is included and all
computations are performed with 16 discrete neutrino 
energy bins, approximately 
logarithmically spaced from 2.5~MeV to 220~MeV.
Electron neutrinos ($\nu_e$) and electron-antineutrinos
($\bar{\nu}_e$) are treated independently while
we lump together the heavy-lepton $\mu$, $\bar{\mu}$, $\tau$, and 
$\bar{\tau}$ neutrinos into one 
group (``$\nu_{\mu}$'').
The code is very efficiently parallelized via MPI 
in energy groups and species.
As an additional simplification, we do not include 
energy redistribution by inelastic neutrino-electron
scattering. Such energy redistribution and
scattering are of modest ($\sim 10\%$) relevance
for the trapped electron fraction ($Y_e$) and
entropy of the inner core at core bounce, but otherwise
arguably quite subdominant \citep{thompson:03}.

\subsection{A Hybrid Approach -- Combining
{\rm S}$_n$ and MGFLD Neutrino Transport}

The time-implicit S$_n$ scheme in VULCAN/2D
is iterative and suffers convergence problems in
regions where the transport problem is 
scattering-dominated and the optical depth
is high $(\tau \sgreat 5)$. As a consequence, 
\citet{livne:04} limited the timestep
at postbounce times to $\sim$0.1--0.3~$\mu$s 
to ensure accuracy and stability. In the present
study, we take a different approach and
introduce a hybrid S$_n$--MGFLD transport scheme
that treats the quasi-isotropic transport problem 
in the optically-thick PNS interior 
in the diffusion approximation and transitions to full
multi-angle S$_n$ transport in a region
of moderate optical depth ($\tau\,\, \sgreat\, 2$), but
that is still significantly interior to the neutrinospheres
($\tau \sim 2/3$) where the neutrinos decouple from
matter and begin to stream. 

\begin{figure*}[t]
\centering
\includegraphics[width=8.8cm]{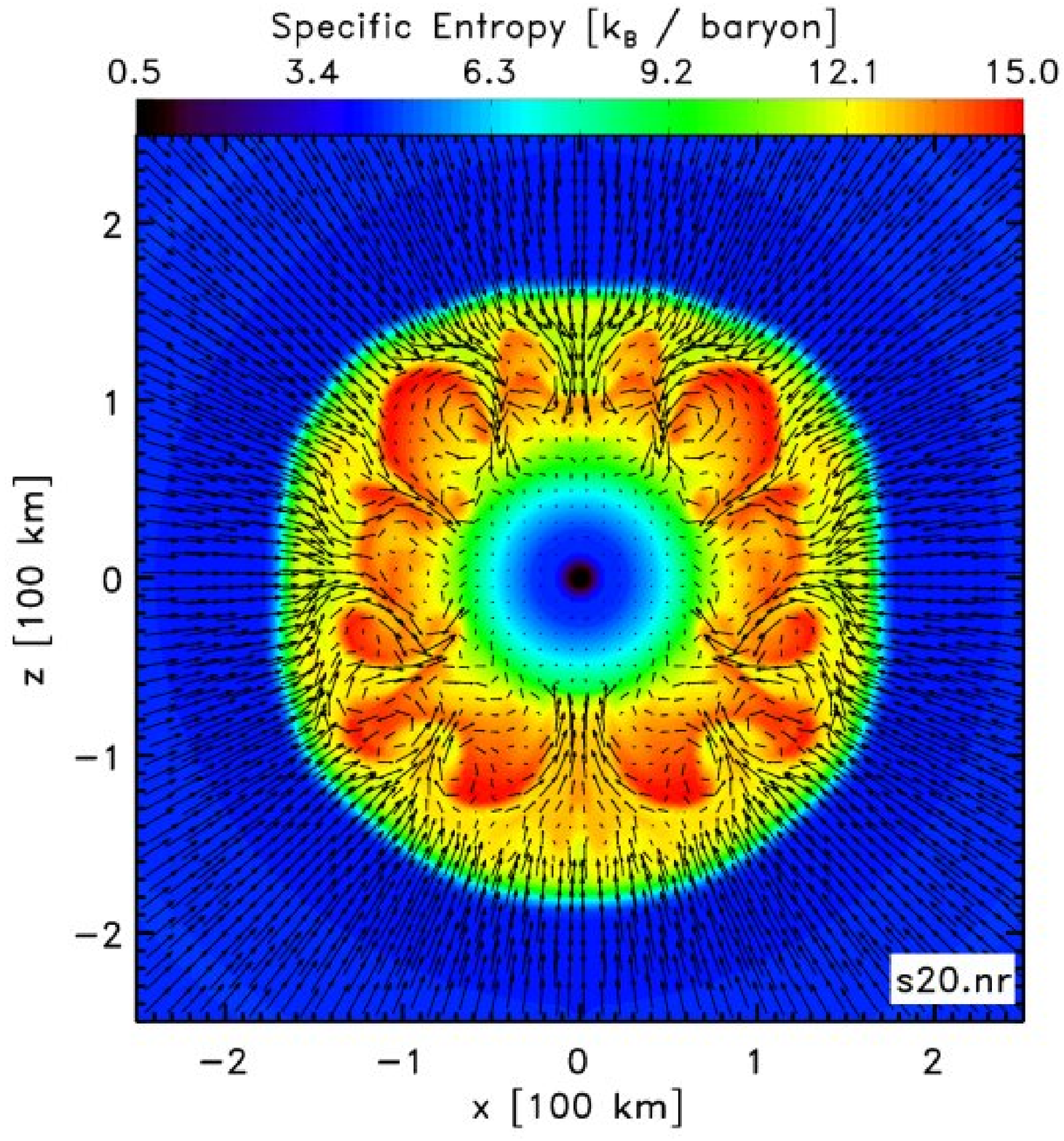}
\includegraphics[width=8.8cm]{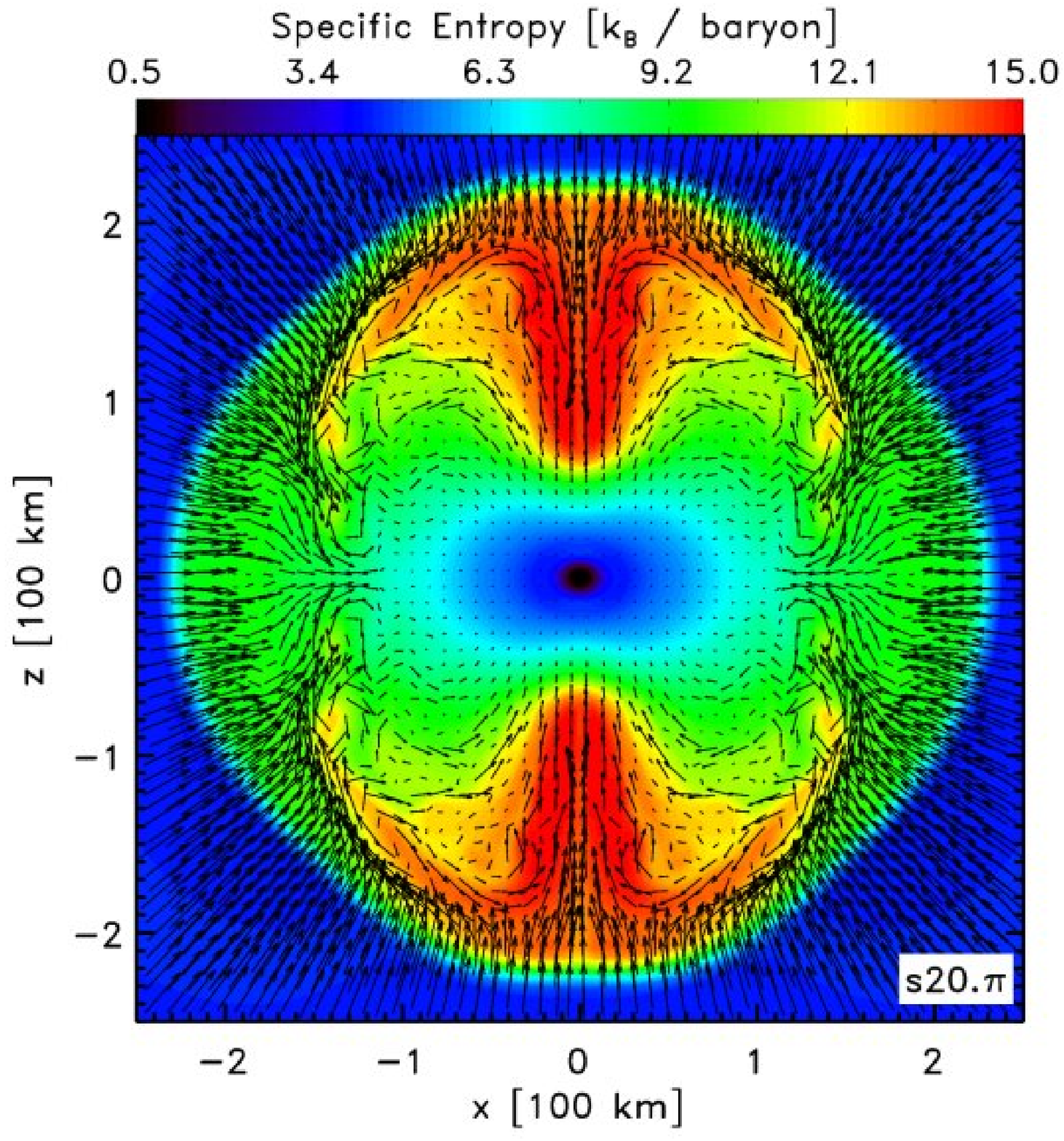}
\caption{Entropy colormaps of the nonrotating model s20.nr (left) and the
rotating model s20.$\pi$ (right) at 160~ms into their postbounce evolution
computed with MGFLD. Velocity vectors are superposed with vector
lengths saturated at 1.0$\times$$10^{9}$~cm~s$^{-1}$. Model s20.nr has
a practically spherical PNS and shows features of violent overturn in
the convectively unstable postshock region. The shock radius in this
model is $\sim$175~km at this point and the onset of the SASI is
apparent from the slightly deformed shock. Model s20.$\pi$, on the
other hand, has a strongly rotationally-flattened PNS and convective
overturn is confined to polar regions. These regions exhibit
the globally highest entropies and greatest entropy gradients, since
the polar velocity divergence at the shock is the highest. The shock
radius at this time in model s20.$\pi$ is $\sim$230~km and no SASI
features are visible.
\label{fig:intro2D}}
\end{figure*}

We chose a radius of 20~km in our calculations for the transition from
MGFLD to S$_n$. This is a sensible choice, (1) because the
neutrinosphere radii of all groups (energies/species) remain larger
than 20~km throughout the postbounce period our simulations cover and,
(2) because 20~km also marks the radius at which the transition from
the inner irregular quasi-Cartesian grid to the outer regular grid is
complete. This boundary is smooth and the S$_n$--MGFLD transition does
not suffer from Cartesian cornerstone effects.

The transition is implemented by setting up for each energy group and
species an approximate specific intensity $I_\nu$ at the centers of the
zones below the S$_n$--MGFLD interface using the information available
from MGFLD. This approximate $I_\nu$ is obtained via its angular expansion
to first order in $\vec{n}$ (the Eddington approximation): 
\begin{equation}
I_\nu = I_0 + 3 (\vec{n}\cdot\vec{H})\,\,.
\label{eq:approxI}
\end{equation}
Here, $I_0 = J_\mathrm{MGFLD}$ and 
$\vec{H} = \vec{F}_\mathrm{MGFLD}/4\pi$,
where $\vec{F}_\mathrm{MGFLD}$ is the flux, 
and $\vec{H}$ is the first moment of
$I_\nu$. In MGFLD, $\vec{F}_\mathrm{MGFLD}$ is computed via
\begin{equation}
\vec{F}_\mathrm{MGFLD} = - \mathrm{FL}[D] \vec{\nabla} J\,\,,
\end{equation} where \begin{equation} D = \frac{1}{3\sigma}\,\,, \end{equation}
with Bruenn's flux limiter\footnote{We use Bruenn's flux limiter in
VULCAN/2D, because \cite{burrows:00} found it to perform best in their
comparison of flux limiters with angle-dependent transport.}
\citep{bruenn:85}, 
\begin{equation} \mathrm{FL}\big[D\big] =
\frac{D}{1+D|\vec{\nabla} J|/J}\,\,.  
\end{equation} 
The first angular moment of eq.~(\ref{eq:approxI}), $\vec{F}_\mathrm{S_n} = \int
\vec{n}\, I d\Omega$, is then equal\footnote{Given the limited number
of angular zones of $I$ and the fact that we are not using
Gaussian-quadrature-type angular zoning, the integrals of $I$ are only
accurate to $\sim$5\% when 8 $\vartheta$-zones are used and accurate
to $\sim$1\% when 12 and more $\vartheta$-zones are employed. To
ensure conservation of energy in the S$_n$--MGFLD matching, we employ
purely geometrical and temporally constant correction factors to
enforce $\vec{F}_\mathrm{S_n} = \vec{F}_\mathrm{MGFLD}$ at the
interface.} to $\vec{F}_\mathrm{MGFLD}$ and the S$_n$--MGFLD matching
is consistent and provides a representation of the specific intensity
$I$ that is accurate to first order in $\vec{n}$.  Given the
essentially isotropic neutrino radiation field deep inside the PNS,
this approximation yields excellent results. We note that the scheme
makes the implicit assumption that the radial gradient of the mean
intensity at the transition radius is always negative or zero. This
condition is generally fulfilled in PNSs.

\section{Initial Models and Setup}
\label{section:initialmodels}

We employ the spherically-symmetric solar-metallicity
$\mathrm{20-}\mmsun$ (at ZAMS) model s20.0 from the stellar
evolutionary study of \cite{wwh:02}, who evolved it to the onset of
core collapse. At that moment, its iron core mass\footnote{Determined
by the discontinuity in the electron fraction, $Y_e$, at the outer
edge of the iron core where $Y_e$$\sim$0.5.} is $\sim$1.46~$\mmsun$
and its central density has reached
$\sim$8.4$\times$10$^{9}$~g~cm$^{-3}$. A graph of the progenitor's
precollapse density stratification as a function of enclosed mass can
be found in Fig.~1 of \cite{burrows:07a}. Note that in the study of
\cite{wwh:02}, iron core mass and extent vary non-monotonically in the
10--20~$\mmsun$ ZAMS mass range and that their solar-metallicity
20-$\mmsun$ model has, in fact, a more compact central configuration
than the corresponding 15-$\mmsun$ model.  Stellar evolution theory of
massive stars has yet to converge and studies by different groups do
not presently yield the same presupernova structures.

We set up two initial models in VULCAN/2D: s20.nr and s20.$\pi$. Both
models are mapped from 1D onto our 2D hydrodynamic grid under
the assumption of spherical symmetry. Model s20.nr is kept
nonrotating, while we impose an initial angular velocity profile
in model s20.$\pi$ according to the rotation law,
\begin{equation}
\Omega(\varpi) = \Omega_0 \frac{1}{1+(\varpi/A)^2}\,\,,
\label{eq:rotlaw}
\end{equation} where $\varpi$ is the distance from the rotation axis and $A$ is a
parameter governing precollapse differential rotation.  This rotation
law enforces constant angular velocity on cylindrical shells and, for
sensible choices of $A$, reproduces qualitatively
(\citealt{ott:06spin}) predictions from presupernova models that
include rotation in a 1D fashion~\citep{heger:00,heger:05}. Since the
computational complexity of this study inhibits us from performing a
sweep of the $\Omega_0$--$A$ parameter space, we chose $A = 1000$~km
and $\Omega_0=\pi$~rad~s$^{-1}$. Hence, the initial central period is
2~s -- this is an identical rotational setup to the fiducial model in
\cite{burrows:07b}.  As discussed in \cite{ott:06spin}, 2~s is rather
short, leads to a rapidly rotating postbounce configuration with a
millisecond-period PNS, and, unless significant postbounce spin-down
(e.g. via MHD torques) occurs, is inconsistent with average pulsar
birth spin estimates. We chose such rapid rotation simply because we
wish to study a postbounce supernova core with significant
rotationally-induced asymmetry. Key model parameters and
characteristics are summarized in Table~\ref{table:models}.

We collapse both models with the MGFLD variant of VULCAN/2D and evolve
them to $\sim$160~ms after core bounce.
Then, we transition to \sn\ Boltzmann transport and solve for
the stationary neutrino radiation field based on the artificially
frozen hydrodynamics data at this postbounce time. Once we have
obtained a converged angle-dependent radiation field, we activate
neutrino-matter coupling and hydrodynamics and evolve in time the coupled
radiation-hydrodynamics equations. For direct comparison, we also
continue the MGFLD simulations to later times. 
All steady-state snapshots are computed in three momentum-space 
angular resolutions, S$_{16}$, S$_{12}$, and S$_8$, while the
long-term evolution calculations could only be performed with
S$_8$, due to computational constraints.

In Fig.~\ref{fig:intro2D}, we show entropy colormaps of both models at
160~ms after bounce. Fluid velocity vectors are superposed, providing
a snapshot of the flow. By 160~ms after bounce, in the nonrotating
model s20.nr convection in the high-entropy ($O$(10)
$\mathrm{k}_\mathrm{B}$/baryon) gain layer has developed fully. The
shock sits at $\sim$175~km and is slightly deformed by the onset of
the SASI. Not visible on the scale of this figure is the
lepton-gradient-driven convective region deep inside the PNS,
which was extensively  discussed in \cite{dessart:06a}.

The PNS in the rapidly rotating model s20.$\pi$ is rotationally
flattened, with unshocked low-entropy inner-core
pole/equator asymmetry ratios below $\sim$0.5. The shock
is slightly prolate and has attained an average radius of 
$\sim$230~km.  The moment-of-inertia-weighted 
mean period of the unshocked (specific entropy
$s \le$~3~k$_\mathrm{B}$) inner core is $\sim$2.0~ms. Differential
rotation between $\sim$20 and 200~km is very large, with the
angular velocity  $\Omega$ dropping from $\sim$1600~rad~s$^{-1}$
to a mere $\sim$15~rad~s$^{-1}$ over this radial equatorial interval.
Yet, the specific angular momentum $j$ is still monotonically and rapidly 
increasing. It flattens, but does not decrease, only at radii greater
than $\sim$100~km.  This positive gradient in $j$ stabilizes the postbounce
core against convective instability at low latitudes \citep{fh:00}, 
confining overturn to the polar regions and large equatorial radii where
the $j$ gradient is less steep.

\begin{deluxetable}{llcllll}
\tabletypesize{\scriptsize}
\tablecaption{Model Summary\label{table:models}}
\tablehead{
\colhead{Model Name}&
\colhead{Progenitor}&
\colhead{$\Omega_0$}&
\colhead{$A$}&
\colhead{$t_b$}&
\colhead{$t_\mathrm{snap}$}&
\colhead{$t_f - t_b$}\\
\colhead{}&
\colhead{}&
\colhead{(rad s$^{-1}$)}&
\colhead{(km)}&
\colhead{(ms)}&
\colhead{(ms)}&
\colhead{(ms)}
}
\startdata
s20.nr&s20.0&0.0&---&179.2&160.0&500.0\\
s20.$\pi$&s20.0&$\pi$&1000&193.7&160.0&550.0
\enddata
\tablecomments{Summary of model parameters. The progenitors
are taken from \cite{wwh:02}. $\Omega_0$ is the initial
central angular velocity, $A$ is the differential rotation
parameter of the rotation law (eq. \ref{eq:rotlaw}). $t_b$ is
the time of core bounce, $t_\mathrm{snap}$ is the time after
$t_b$ at which the postbounce snapshots are taken, and $t_f - t_b$
is the point at which we stop our simulations.}
\end{deluxetable}

\begin{figure}[t]
\centering
\includegraphics[width=8.5cm]{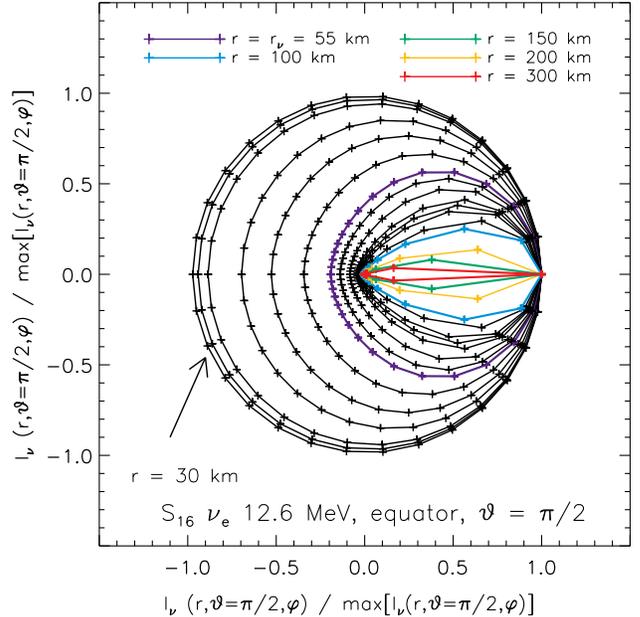}
\caption{Polar plot of the normalized specific intensity
$I_\nu(r,\vartheta,\varphi)/\mathrm{max}[I_\nu(r,\vartheta,\varphi)]$
in model s20.nr at 160~ms after core bounce, at selected equatorial
radii, and for $\nu_e$ neutrinos at $\varepsilon_\nu = 12.6$~MeV.
At each radius, we normalize the specific intensity by its local
maximum. Shown is the variation with $\varphi$ at fixed $\vartheta =
\pi/2$. The graphs are based on a S$_{16}$ calculation.  At $r
=$~30~km, the radiation field is practically isotropic, but is already
appreciably forward-peaked at the neutrinosphere ($r_\nu = 55$~km;
optical depth $\tau = 2/3$) and thereafter smoothly transitions over
$\sim$200--300~km to the free-streaming limit.
 \label{fig:s20nr_psi1}}
\end{figure}

\vspace*{.8cm}

\section{Results: Snapshots}
\label{section:snapshots}

In this section, we present our \sn\ multi-angle transport results for
steady-state model snapshots at 160~ms after core bounce.  We diagnose
the angle-dependent neutrino radiation fields and carry out a 
comparison between multi-angle and MGFLD transport results based on
local and global radiation-field variables.

\begin{figure*}[t!]
\centering
\includegraphics[width=17cm]{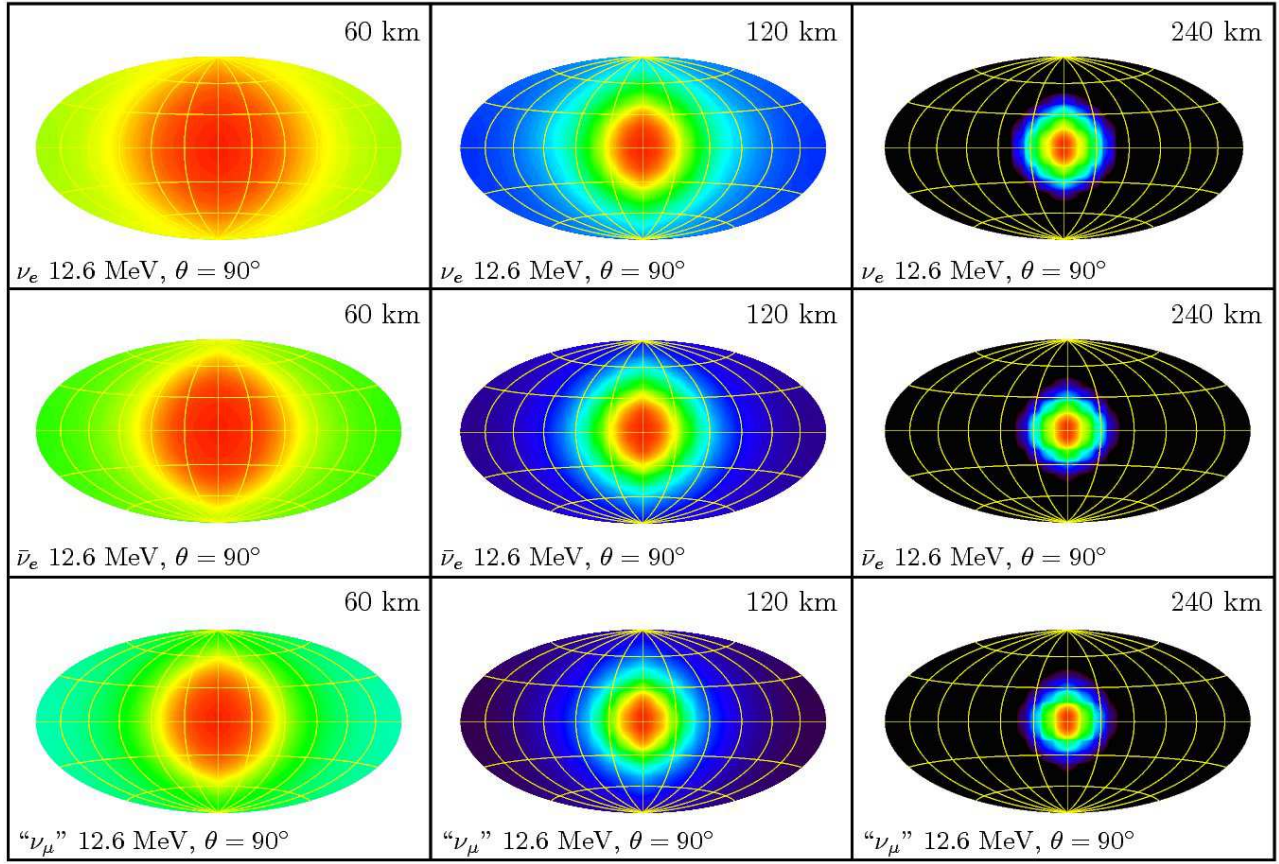}
\caption{ Hammer-type interpolated (smoothed) map projections of the
normalized specific intensity $I_\nu(\vartheta,\varphi) / J_\nu$ in
model s20.nr at 160 ms after bounce. The colormap is logarithmic and in
each individual projection is setup to range from
$\mathrm{max}(I_\nu(\vartheta,\varphi) / J_\nu)$ (red) to
10$^{-4}\mathrm{max}(I_\nu(\vartheta,\varphi) / J_\nu)$ (black).
Shown is the specific intensity of $\nu_e$, $\bar{\nu}_e$, and
``$\nu_\mu$'' neutrinos at $\varepsilon_\nu = 12.6$~MeV (rows) on the
equator ($\theta$=90$^\circ$, measured from the pole) and at radii of
60, 120, and 240 km (columns). The Hammer projection is set up in such
a way that $\vartheta$ varies in the vertical from $0^\circ$ (top) to
$180^\circ$ (bottom) and $\varphi$ varies horizontally from
$-180^\circ$ (left) to $+180^\circ$ (right). Grid lines are drawn in
$\vartheta$- and $\varphi$-intervals of 30$^\circ$. Note (a) the
increasing forward-peaking of $I_\nu$ with increasing radius (and
decreasing optical depth) and (b) that at any given radius $I_\nu$ of
``$\nu_\mu$'' is more forward-peaked than that of the $\bar{\nu}_e$
component, which, in turn, is always more forward-peaked than the
$\nu_e$ component. This fact is a consequence of a transport mean-free
path that varies with species (and energy; not shown here) and is
smallest for the electron neutrinos.
\label{fig:hammer1}}
\end{figure*}

\subsection{Angular Distributions}

The quintessential problem in treating neutrino radiation transport in
core-collapse supernova cores is the fact that the neutrino transport
mean-free path, the average distance a neutrino can travel without
experiencing scattering or absorption, changes by orders of magnitude
from inside to outside.  Moreover, the neutrino transport
mean-free-path $\lambda_\nu$ varies locally strongly with neutrino
energy ($\propto \varepsilon_\nu^2$) and matter
density.  As a consequence, gray transport schemes are problematic,
since neutrino-energy averages can be defined only locally and the
mean neutrino energy varies significantly throughout the supernova
core.

From a more geometric point of view, the radiation field in momentum
space goes from being completely isotropic (net flux $\sim$zero) to
being focussed into the radial direction (``forward-peaked'') in the
free-streaming regime. In the MGFLD approximation, the mean intensity
$J_\nu$ is evolved in time and the angular information, in particular
the information on the degree of forward-peaking, is captured only
by computing spatial gradients in $J_\nu$ and employing a flux
limiter to interpolate between diffusion and free streaming.

\begin{figure*}[t!]
\centering
\includegraphics[width=10cm]{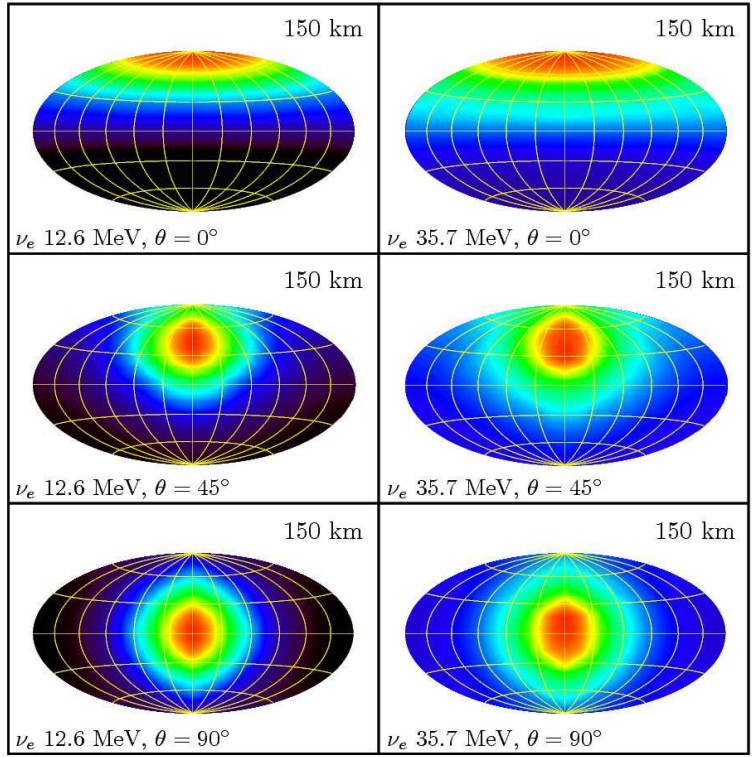}
\caption{ Hammer map projections of the interpolated (smoothed)
normalized specific intensity $I_\nu(\vartheta,\varphi) / J_\nu$ at
160~ms postbounce in model s20.nr. The projections are set up in
identical fashion to Fig.~\ref{fig:hammer1}. Shown here is the
variation of the angular distribution with energy group (columns) and
angular position (rows) for electron neutrinos. The radius is fixed to
150~km.  As expected in the coordinates used for the S$_n$ transport
in VULCAN/2D (see Fig.~\ref{fig:coord}), $I_\nu$ becomes
forward-peaked into $\vartheta=0^\circ$ and degenerate in $\varphi$
along the pole ($\theta=0^\circ$), forward-peaked into
$\vartheta=45^\circ,\varphi=0^\circ$ on the diagonal
($\theta=45^\circ$), and forward-peaked
$\vartheta=90^\circ,\varphi=0^\circ$ on the equator
($\theta=90^\circ$). The degree of the radiation anisotropy and its
variation from forward-peaked at $\varepsilon_\nu = 12.6$~MeV to less
forward-peaked at $\varepsilon_\nu = 37.5$~MeV is apparent.
\label{fig:hammer2}} 
\end{figure*}

The \sn\ Boltzmann solver in VULCAN/2D is able to 
self-consistently handle the transition from isotropic to
forward-peaked radiation.  Figure~\ref{fig:s20nr_psi1} depicts the
angular distribution in the azimuthal angle $\varphi$ (see
Fig.~\ref{fig:coord}) of the normalized specific spectral neutrino
intensity $I_\nu$ for electron neutrinos at 12.6~MeV. In
Fig.~\ref{fig:s20nr_psi1}, the polar angle $\vartheta$ is set equal to
$\pi/2$ and the $\varphi$-distribution is given at various radii in
the equatorial plane of model s20.nr.  At 30~km from the center, the
radiation field of $\nu_e$'s at $\varepsilon_\nu$~=~12.6~MeV is nearly
isotropic,which corresponds to a circle in
Fig.~\ref{fig:s20nr_psi1}. With increasing radius (and, of course,
decreasing matter density) the transport mean-free path at fixed
$\varepsilon_\nu$ increases and the radiation field gradually departs
from isotropy and becomes more and more forward-peaked. We define the
neutrinosphere as the surface at which the optical depth $\tau_\nu$,
given by \begin{equation} \tau_\nu = \int_\infty^R \frac{dr}{\lambda_\nu}\,\,, \end{equation}
is equal to 2/3. At around this $\tau_\nu$, the neutrinos decouple
from matter and begin to stream freely. At 160~ms after bounce in
model s20.nr, the 12.6~MeV $\nu_e$ neutrinosphere is located at
$r\sim$55~km. As is obvious from Fig.~\ref{fig:s20nr_psi1}, the
radiation field at the neutrinosphere is not yet dramatically
forward-peaked, but becomes so with increasing radius. However,
complete forward-peaking only obtains at radii $\sgreat$~250--300~km,
beyond which the angular resolution of our \sn\ scheme becomes
suboptimal, even with $n=16$. However, calculations with varying
number of $\vartheta$ (and, hence, $\varphi$) angles reveal that the
transition from isotropy to moderate and large anisotropy is
adequately reproduced at small and intermediate radii (out to
$\sim$200~km) even in the case of S$_8$.

For the purpose of displaying and studying the local neutrino
radiation field, we provide equal-area Hammer-type map projections
\citep{hammer:1892}.  Such map projections are new to the field of
neutrino radiation transport and beautifully reveal the multi-D
angular-dependence of the radiation field.  In Fig.~\ref{fig:hammer1},
we present such Hammer projections on the equator (spatial $\theta =
90^\circ$) of model s20.nr at radii of 60, 120, and 240~km for the
three neutrino species included in our simulations at $\varepsilon_\nu
= 12.6$~MeV . In each plot, we normalize the specific intensity to the
mean intensity to set a common scale. The colormap is logarithmic and
chosen to have regions on the sphere with high intensity appear red
and regions of low intensity appear black.

For neutrinos on the equator, the momentum-space forward direction is
$(\vartheta = 90^\circ, \varphi = 0)$. Electron neutrinos generally
have the shortest transport mean-free path of all species in the
core-collapse context and decouple from matter at the lowest
densities. The Hammer projection in the top-left corner of
Fig.~\ref{fig:hammer1} of the $\varepsilon_\nu$ = 12.6~MeV equatorial
radiation field at 60~km corresponds roughly to the blue line graph in
Fig.~\ref{fig:s20nr_psi1}, which portrays only its variation with
$\varphi$.  At fixed neutrino energy group $\varepsilon_\nu$, electron
anti-neutrinos and ``$\nu_\mu$'' neutrinos decouple at smaller
radii. Hence, as Fig.~\ref{fig:hammer1} shows, at 60~km, they already
manifest greater local anisotropy than the $\nu_e$s. This trend
continues at all considered radii in Fig.~\ref{fig:hammer1}.

In Fig.~\ref{fig:hammer2}, we again present Hammer projections of the
normalized specific intensity, but this time consider only $\nu_e$s,
keep the radius fixed at 150~km, and vary the neutrino energy and the
angular position on the grid. The bottom row of Fig.~\ref{fig:hammer2}
shows the normalized $I_\nu$ at the equator ($\theta=90^\circ$) and
for the 12.6-MeV and 35.7-MeV $\nu_e$ energy groups. The center and
top rows show the same groups at $\theta=45^\circ$ and at
$\theta=0^\circ$, respectively.  From the discussion of
Fig.~\ref{fig:hammer1}, we are already familiar with the overall
radiation field geometry.  The transport mean-free path scales roughly
inversely with $\varepsilon_\nu^2$. Hence, at any given position in
the postbounce supernova core, more energetic neutrinos should be
locally more isotropically distributed in momentum space than less
energetic ones. The less forward-peaked angular $I_\nu$ distribution
of the higher-energy neutrinos reflects this.

The degree of forward-peaking in $\vartheta$ and $\varphi$ of the
radiation field in the quasi-spherically symmetric nonrotating model
s20.nr is essentially independent of the angular position on the grid
and the radiation fields at any given radius can be transformed into
one another by simple rotation. Because of the aspherical and oblate
distribution of matter in the rotating model s20.$\pi$, the
forward-peaking is also a function of polar angle. Due to its PNS's
oblateness (see Fig.~\ref{fig:intro2D}), the neutrinos generally
decouple at significantly smaller radii near the pole than near the
equator, in turn leading to more strongly forward-peaked radiation
fields in the polar than in the equatorial regions
\citep{janka:89a,janka:89b,walder:05,dessart:06b,dessart:07}.

\begin{figure*}[t]
\centering
\includegraphics[width=5.95cm]{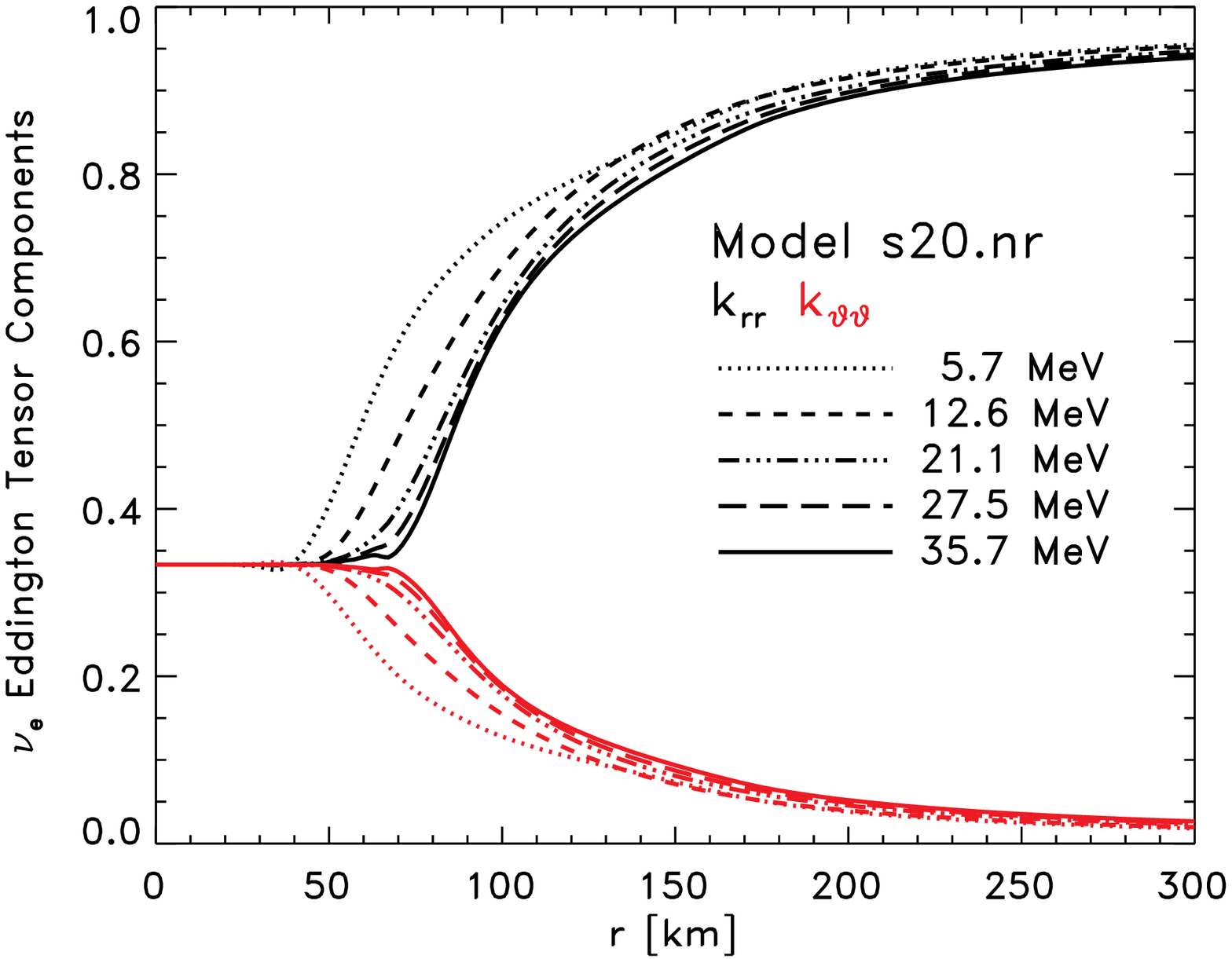}
\includegraphics[width=5.95cm]{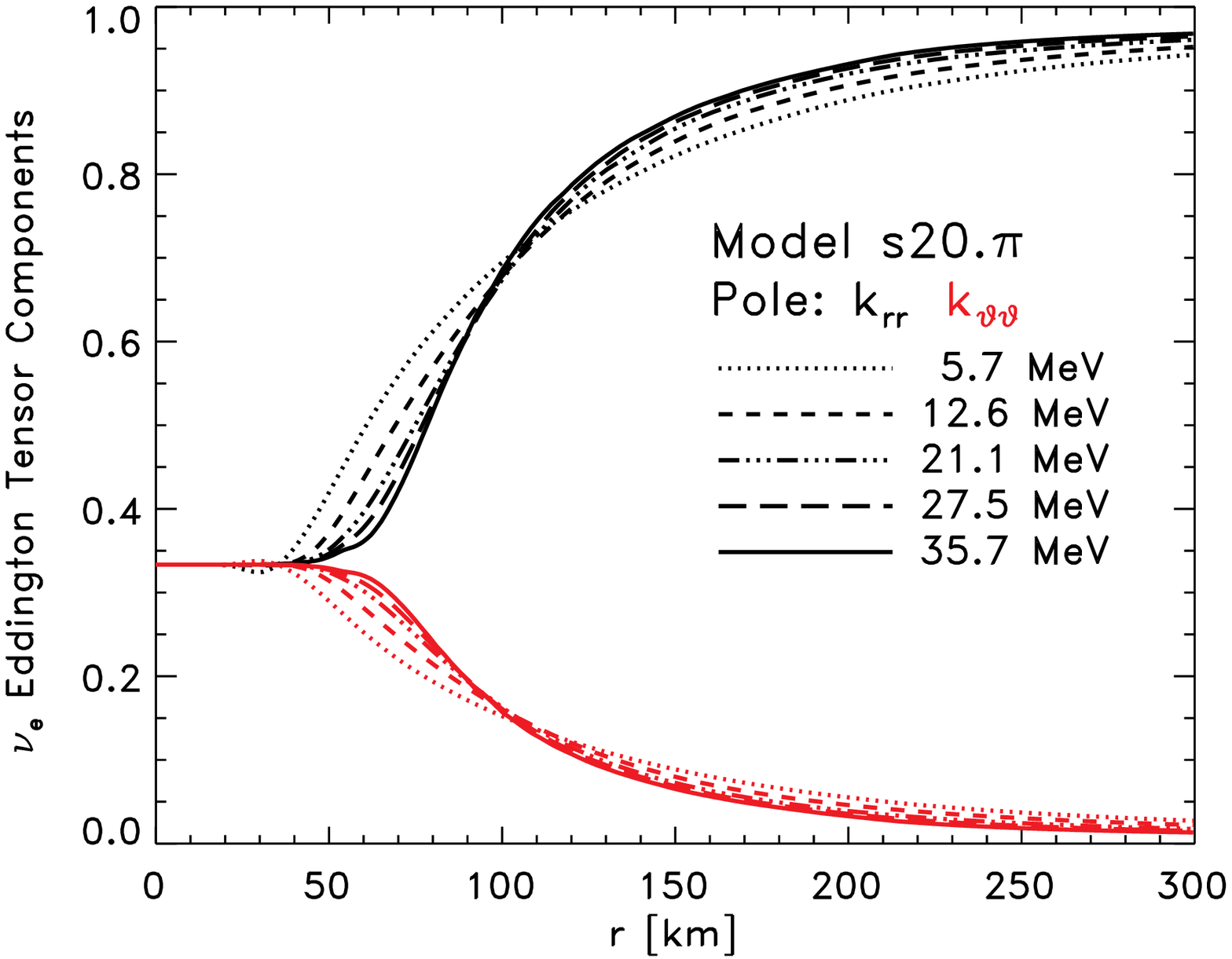}
\includegraphics[width=5.95cm]{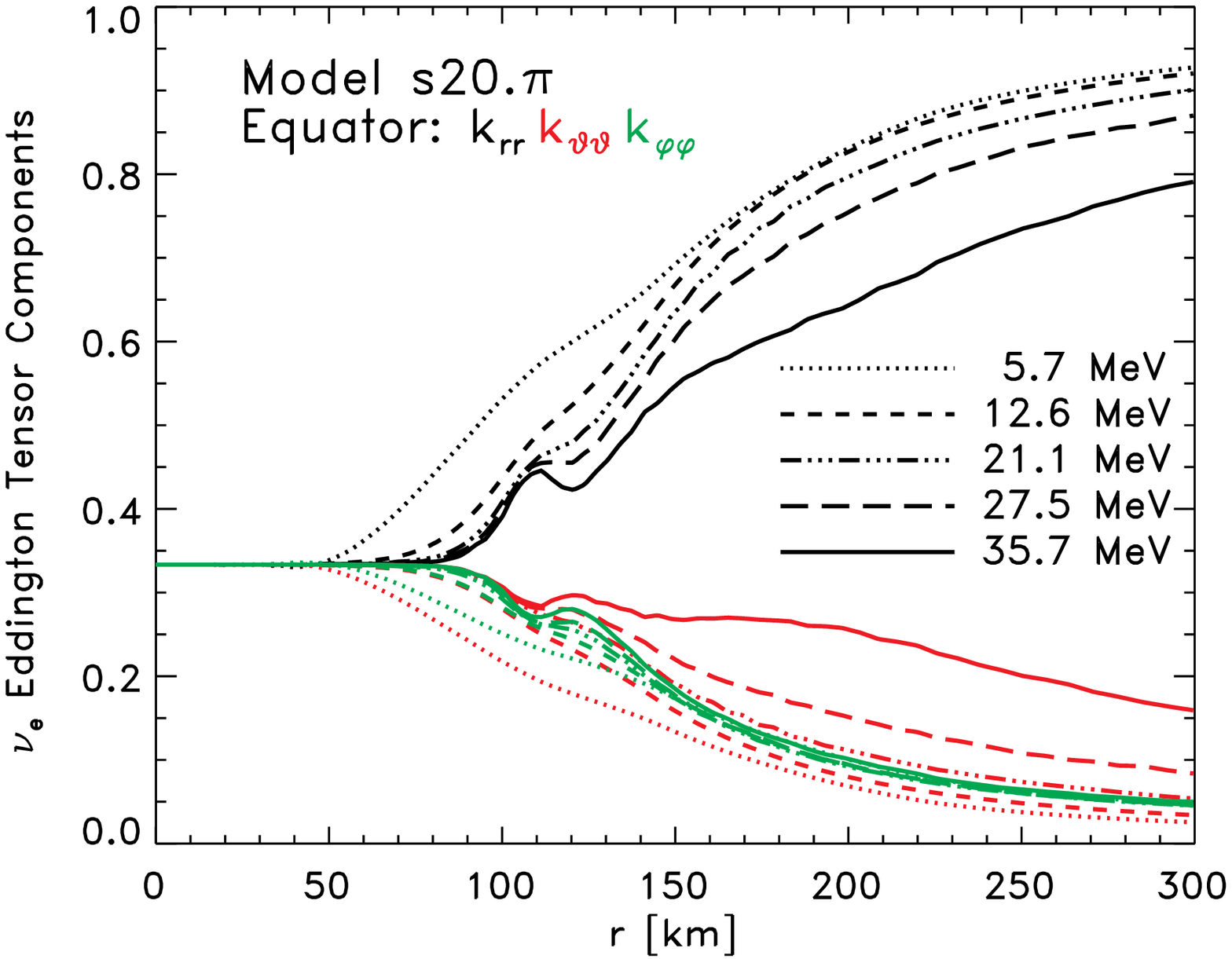}
\caption{Normalized Eddington tensor ${\textsf k}$ components in
spherical coordinates as a function of neutrino energy $\varepsilon_\nu$
and spherical radius $r$. {\bf Left}: Angular-averaged $k_{rr}$ and
$k_{\vartheta\vartheta}$ for electron neutrinos in model
s20.nr. $k_{\varphi\varphi}$ is not shown, but has essentially
identical behavior to $k_{\vartheta\vartheta}$.  The diagonal
components start out with $1/3$ at small radii, as expected for the
prevailing isotropic radiation fields. With increasing radius
(and decreasing density), the local radiation field becomes more
anisotropic and forward-peaked. This occurs at progressively larger
radii with increasing $\varepsilon_\nu$ and is reflected by the
increasing $k_{rr}$ and the decreasing $k_{\vartheta\vartheta}$ in the
plot. The off-diagonal component $k_{r\vartheta}$ is not shown, does
not exhibit clear systematics, and is generally a factor of 10--100
smaller than the diagonal components.  {\bf Center}: Same as left,
but showing profiles extracted from regions near the pole in the
rapidly-rotating model s20.$\pi$. Interior to $\sim$100~km, $k_{rr}$,
and $k_{\vartheta\vartheta}$ show the same systematics with
$\varepsilon_\nu$ as in the nonrotating model. However, at larger radii
they are reversed, $k_{rr}$ and $k_{\vartheta\vartheta}$ exhibiting
greater isotropy for lower $\varepsilon_\nu$. See text for discussion.
{\bf Right}: Equatorial profiles of $k_{rr}$, $k_{\vartheta\vartheta}$
and $k_{\varphi\varphi}$ for electron neutrinos in model s20.nr. Due to
rotational flattening of the PNS, the transition to free streaming
occurs over a much larger range of radii near the equator.
$k_{\vartheta\vartheta}$ shows a significantly 
larger variation as a function of energy than $k_{\varphi\varphi}$.
\label{fig:eddy}} 
\end{figure*}

\subsection{Eddington Factors}
\label{section:Eddy}

The radiation-pressure tensor ${\textsf K}_\nu$, also known as
the Eddington tensor, represents the second angular moment of the
specific intensity and is defined by eq.~(\ref{eq:eddy}).
In the following, we use its normalized variant 
${\textsf k}_\nu = {\textsf K}_\nu / J_\nu$.

In spherical symmetry, ${\textsf k}_\nu$ is diagonal and has a single
independent component, the Eddington factor $k_\nu$.  For isotropic
radiation, $k_\nu = 1/3$ and ${\textsf k}_\nu =
\mathrm{diag}(1/3,1/3,1/3)$, while in the streaming regime, $k_\nu =
1$ and ${\textsf k}_\nu = \mathrm{diag}(1,0,0)$.  In the transition
from isotropy to free streaming, $k_\nu$ generally varies from $1/3$
to $1$, but in special cases, e.g., enhanced radiation perpendicular
to the radial direction, may assume values below $1/3$.  Note that one
of the common assumptions of MGFLD is the Eddington closure, setting
$k_\nu = \frac{1}{3}$ everywhere.

In axisymmetry and ignoring velocity-dependent terms, the Eddington tensor
has four independent components whose individual meaning depends on
the coordinates chosen\footnote{Off-diagonal components of the
Eddington tensor can be related to radiation shear viscosity
\citep{mihalas:99}, which we do not consider here.}.  We assume and
transform to spherical coordinates for our discussion, since they make
the interpretation of the components most straightforward.

In Fig.~\ref{fig:eddy}, we present radial profiles of normalized
Eddington tensor components at selected electron-neutrino energies
$\varepsilon_\nu$ in models s20.nr and s20.$\pi$.  The nonrotating model
can be considered nearly spherically symmetric, and, hence, should and
does exhibit the expected Eddington-factor systematics. At small
radii and high densities, where neutrinos and matter are in
equilibrium, $k_{rr} = k_{\vartheta\vartheta} = k_{\varphi\varphi} =
\frac{1}{3}$ and with increasing radius, $k_{rr} \rightarrow 1$ and
$\{k_{\vartheta\vartheta},k_{\varphi\varphi}\} \rightarrow 0$.  As
expected from the basic decoupling hierarchy, the value of the
Eddington tensor components is a strong function of $\varepsilon_\nu$.
Lower-$\varepsilon_\nu$ neutrinos decouple at higher densities, and, hence,
have Eddington tensor components which depart from $\frac{1}{3}$ at
smaller radii than $\nu_e$s of higher energy.
This systematics applies, of course, to $\bar{\nu}_e$s 
and ``$\nu_\mu$''s as well.
The off-diagonal component
$k_{r\vartheta}$ is zero in the isotropic region, does not exhibit clear
systematics, and stays an order-of-magnitude smaller than the diagonal
components for all $\varepsilon_\nu$ and species.

The rotating model s20.$\pi$ has a postshock configuration that is far
from spherically symmetric (Fig.~\ref{fig:intro2D}).  We present in
Fig.~\ref{fig:eddy} separate plots for its Eddington tensor components
in regions near the pole and near the equator.  In the polar regions
and at small radii ($r\,\sless\,$100~km), the Eddington tensor
components show the same qualitative behavior as in model s20.nr. At
larger radii, however, the systematics are reversed and
lower-$\varepsilon_\nu$ electron neutrinos have more isotropic radiation
fields (smaller $k_{rr}$) than their higher-$\varepsilon_\nu$
counterparts. Analyzing their radiation fields and matter coupling in
detail, we find that this surprising feature is a consequence of
electron capture and the polar compactness (large density gradient due
to rotation) of the supernova core.  Electron capture near the shock
leads to isotropic neutrino emission that can locally isotropize the
radiation field in semi-transparent regions.  With decreasing density
and temperature, the mean energy of neutrinos emitted by capture
processes shifts to lower $\varepsilon_\nu$.  This leads to greater local
isotropization of lower-$\varepsilon_\nu$ neutrinos, which in turn is
reflected in the more slowly increasing $k_{rr}$ of these
neutrinos. This interpretation is confirmed by the fact that we do not
find any such feature in the Eddington tensor components of the
``$\nu_\mu$'' neutrinos that are not produced in 
capture processes. We also do not observe significant isotropization
in the $\bar{\nu}_e$ radiation fields, since the emission of
$\bar{\nu}_e$s by positron capture on neutrons is weaker due to
the lower positron abundance.

In regions of model s20.$\pi$ near the equator where the PNS is most
extended, the neutrino radiation fields stay isotropic to large radii
and decouple from matter only slowly with radius. Since the matter
densities in the equatorial plane stay roughly a factor of four larger
than in the polar regions, the cross-over feature in $\{k_{rr},
k_{\vartheta\vartheta}, k_{\varphi\varphi}\}$ does not appear and
these components follow the standard decoupling
hierarchy. Interestingly, and different from in the nonrotating model,
$k_{\vartheta\vartheta}$ and $k_{\varphi\varphi}$ show quantitatively
distinct variation with $\varepsilon_\nu$, the latter exhibiting
significantly less variation with $\varepsilon_\nu$ at any given radius.
The interpretation of this observation is not straightforward, but we
suggest that it can be attributed to the fact that in model s20.$\pi$ the
radiation field at any given point on the equator of the
rotationally-flattened core and for any $\varepsilon_\nu$ and neutrino
species varies locally less in the $\vartheta$ direction than in the
$\varphi$ direction. This, in combination with the fact that on the
equator the radiation field asymptotically peaks into the 
($\vartheta=0$,$\varphi=0$) direction, results on average in smaller
$k_{\varphi\varphi}$ with less spread in energy than exhibited by
$k_{\vartheta\vartheta}$.
The off-diagonal component $k_{r\vartheta}$ (not shown in
Fig.~\ref{fig:eddy}) vanishes in $k_{rr} = k_{\vartheta\vartheta} =
k_{\varphi\varphi} = \frac{1}{3}$ regions, but can become relatively
large at greater radii (up to $\sim$0.2 in magnitude; increasing with
$\varepsilon_\nu$ and radius) and flips sign at the equator. The
interpretation of $k_{r\vartheta}$ is not straightforward, since its
magnitude depends on the choice of coordinates. We do not
attempt to study it, nor its implications for neutrino shear
viscosity, in any detail.

\begin{figure}[t]
\centering
\includegraphics[width=8.5cm]{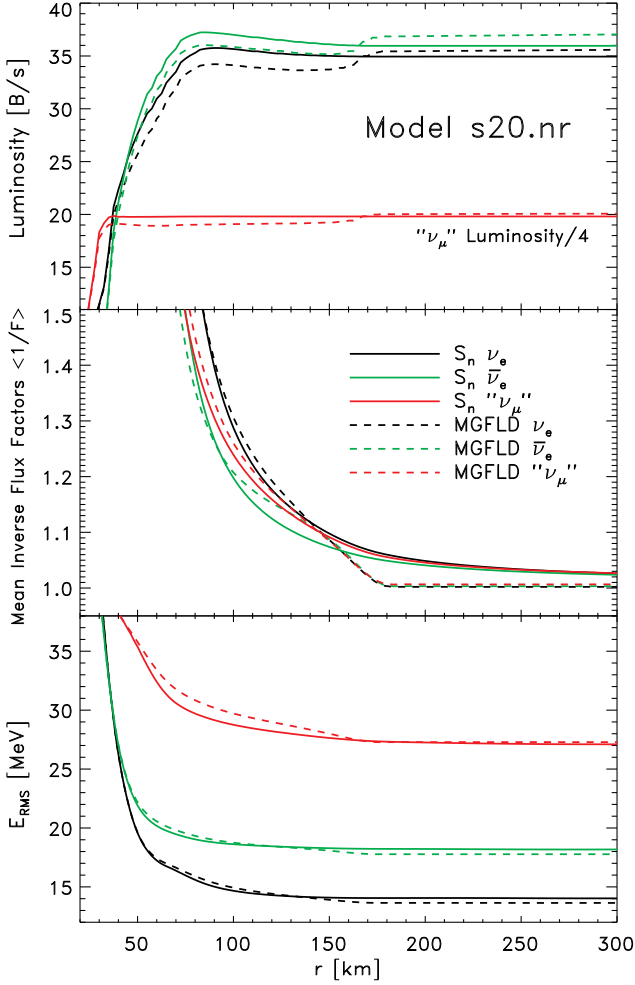}
\caption{ S$_n$--MGFLD comparison for the nonrotating model s20.nr at
160~ms after bounce. All S$_n$ results where obtained with a 16
$\vartheta$-angle calculation.  See text for details and discussion.
Top panel: Neutrino luminosity as a function of radius and broken down
into the three neutrino species considered. The ``$\nu_\mu$'' neutrinos
dominate in luminosity and their luminosity profiles are scaled by
a factor of 1/4 to preserve the overall scale of the plot. 
Center panel: Angle-averaged energy-mean inverse
neutrino flux factor profiles.  Bottom panel: RMS neutrino energy
profiles.
\label{fig:s20nr_line}}
\end{figure}

\subsection{Global Radiation Field Diagnostics: Luminosities,
Spectra, Flux Factors, and Neutrino Energy Deposition}
\label{section:stills}

So far we have studied aspects of neutrino transport inaccessible to
MGFLD. We now go on to discuss radiation field diagnostics that
facilitate a \sn--MGFLD comparison. For further reference
and comparison with previous studies \citep{janka:92,messer:98,
burrows:00}, we define the neutrino luminosity per species 
$L_{\nu_i}$ at spherical radius $r$, 
\begin{equation}
L_{\nu_i}(r) = \oint d\omega \int d\varepsilon_\nu\, F_r(r,\varepsilon_\nu,\nu_i) r^2\,, 
\end{equation} 
where $F_r$ is the spectral radial neutrino flux in species $\nu_i$ 
at energy $\varepsilon_\nu$. $d\omega$ is the
spatial solid-angle element, $d\omega = 2\pi \sin\theta d\theta$ in
axisymmetry. Furthermore, we define the
mean inverse flux factor $\langle 1 / {\textsf F}_{\nu_i}\rangle$,
\begin{equation}
\bigg\langle \frac{1}{{\textsf F}_{\nu_i}} \bigg\rangle = 
\frac{c \int d\varepsilon_\nu E(\varepsilon_\nu,\nu_i)}
{\int d\varepsilon_\nu F_r(\varepsilon_\nu, \nu_i)}\,,
\end{equation}
where $E(\varepsilon_\nu,\nu_i) = 4\pi c^{-1} J(\varepsilon_\nu,\nu_i)$
is the spectral neutrino energy density,
and the neutrino RMS energies are
\begin{equation}
E_{\mathrm{RMS},\nu_i} = \sqrt{{\frac{\int d\varepsilon_{\nu_i}
\varepsilon_{\nu_i}^2 J(\varepsilon_{\nu_i})}
{\int d\varepsilon_{\nu_i} J(\varepsilon_{\nu_i})}}}\,.
\end{equation}

The above three quantities are particularly useful diagnostics, since the
$\varepsilon_\nu$-averaged energy deposition rate by charged-current
absorption of $\nu_e$ and $\bar{\nu}_e$ on neutrons and protons scales
linearly with their product (\citealt{messer:98}).

\vspace*{.5cm}

\subsubsection{Model s20.nr}
\label{section:s20nr}

In Fig.~\ref{fig:s20nr_line}, we plot neutrino luminosities
$L_{\nu_i}$, angle-averaged mean inverse flux factors, and the
angle-averaged $E_\mathrm{RMS}$ for the postbounce snapshot at 160~ms
of the nonrotating model s20.nr. The asymptotic total luminosity at
this time is $\sim$150~B~s$^{-1}$ and is already dominated by the
thermally-produced ``$\nu_\mu$''s that cool the PNS, but contribute
little to the heating in the gain region, since they cannot take part
in charged-current absorption processes. In this quasi-spherically
symmetric model, we define a spherical gain radius $r_\mathrm{gain}$
as the radial position beyond which net neutrino energy deposition
occurs. At 160~ms after bounce, $r_\mathrm{gain} \simeq $~90~km and
the gain region extends almost out to the shock at $\sim$175~km.  The
MGFLD luminosities in Fig.~\ref{fig:s20nr_line} are systematically
lower by $\sim$5\% for $\nu_e$s, $\sim$3.5\% for $\bar{\nu}_e$s, and
$\sim$4\% for ``$\nu_\mu$''s, but qualitatively resemble the \sn\
luminosity profiles in the gain region. At around the shock position,
all MGFLD luminosities increase by $\sim$5\%. This is a due to the
combination of the artificially spread-out shock (over $\sim$4--5
zones), the rapid change of the inverse neutrino mean-free path in the
spread-out shock and the implementation of the flux limiter in
VULCAN/2D.  Since this MGFLD artefact occurs right at the shock, it
can have only little influence on the heating in the gain region, but
leads to somewhat overestimated asymptotic luminosities in the MGFLD
case.

The center panel of Fig.~\ref{fig:s20nr_line} shows the
$\varepsilon_\nu$-averaged inverse flux factors for the three neutrino
species in the MGFLD and \sn\ steady-state calculations of model
s20.nr.  For isotropic radiation $\langle 1 / {\textsf
F}_{\nu_i}\rangle$ tends to infinity, while it approaches one when the
radiation field becomes forward-peaked at low optical depth.
Focussing on the gain region between $r_\mathrm{gain}$ and the shock
position, we find that MGFLD yields mean inverse flux factors that are
up to $\sim$5\% larger for $\bar{\nu}_e$s (less for the other
species) in the inner gain region. At radii $\sgreat\,$150~km, the
MGFLD $\langle 1 / {\textsf F}_{\nu_i}\rangle$ quickly drops to 1
(free streaming), becoming up to 8\% lower than the \sn\ values in the
outer gain region. We note that ``$\nu_\mu$'' interact only via
neutral-current weak interactions, hence, decouple from matter at
higher densities and temperatures. Next in the decoupling hierarchy
are electron anti-neutrinos followed by electron neutrinos.  Both \sn\
and MGFLD realize this hierarchy at radii below $\sim$150~km, beyond
which MGFLD rapidly transitions to free streaming irrespective of
neutrino species.

\begin{figure}[t]
\centering
\includegraphics[width=8.5cm]{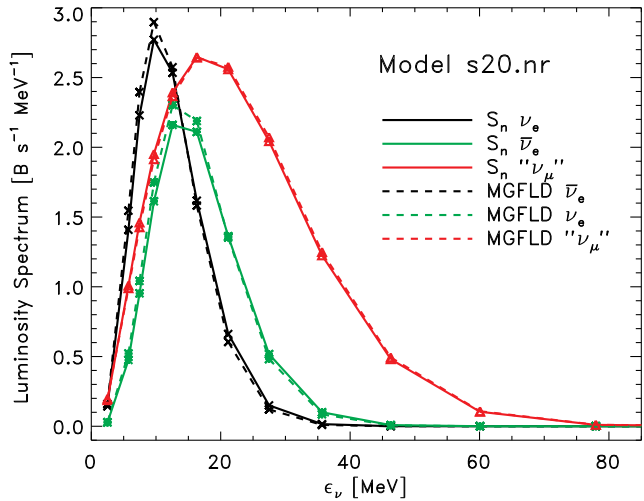}
\caption{ Neutrino luminosity spectra extracted at a radius of 500~km for
$\nu_e$, $\bar{\nu}_e$, and ``$\nu_\mu$'' neutrinos at 160~ms after bounce in
model s20.nr. Solid lines correspond to S$_n$ results, while dashed
lines are obtained using MGFLD. The spectra have the canonical shape
and the quantitative behavior found in nonrotating intermediate-time
postbounce supernova calculations~(e.g., \citealt{thompson:03}) with
the ``$\nu_\mu$''-neutrinos peaking at the highest energies, since they
decouple from the fluid at the smallest radii. MGFLD and S$_n$ spectra
agree closely in shape, but  MGFLD is overestimating slightly the
total asymptotic luminosity (cf. Fig~\ref{fig:s20nr_line}).
\label{fig:s20nr_spec160}}
\end{figure}

\begin{figure}[t]
\centering
\includegraphics[width=8.5cm]{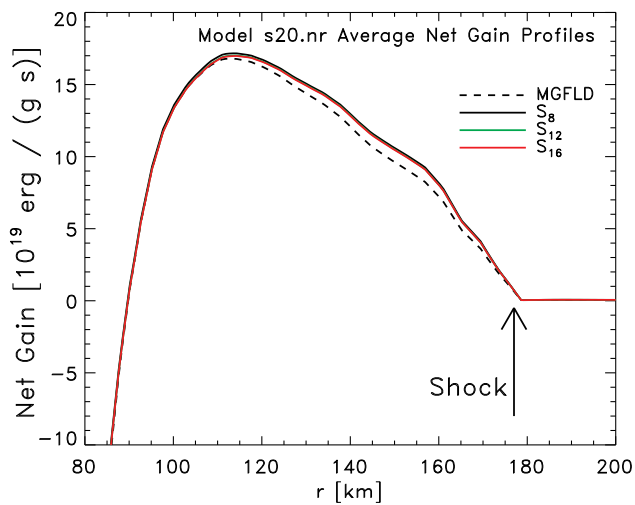}
\caption{Angle-averaged specific neutrino net gain profile in the 
s20.nr model at 160~ms after core bounce. Shown are the MGFLD results, as
well as results from steady-state \sn\ calculations with 8, 12, and 
16 $\vartheta$-angles, corresponding to a total number of angular zones of 40,
92, and 162. The gain region extends from $\sim$90~km to the shock position
at $\sim$175~km. The three different \sn\ resolutions yield net gain profiles
that agree very well (relative differences below 1\% even for S$_8$). The MGFLD
calculation underestimates the total net gain in the outer gain region by 
at most 10\% locally and $\sless\,$5\% on average.
\label{fig:s20nr_gainprof}}
\end{figure}

\begin{figure}[h!]
\centering
\includegraphics[width=8.8cm]{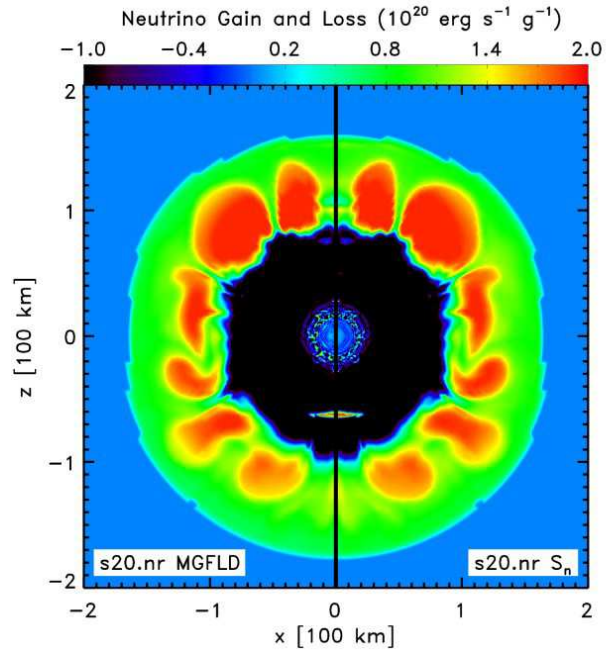}
\caption{ 2D colormap of the specific (per gram) net gain
distribution in model s20.nr at 160~ms after core bounce. The left half
of the plot depicts the MGFLD result, \sn\ is shown on the right.
The differences between S$_n$ and MGFLD are marginal at this time in this
model and are practically indiscernible by eye. As a consequence of
convection in the gain region and the onset of the SASI, even this
nonrotating model exhibits significant angular and radial variations in the
neutrino energy deposition not captured by the average profiles
in Fig.~\ref{fig:s20nr_gainprof}.
\label{fig:s20nr_net_gain160}}
\end{figure}

\begin{figure*}[t!]
\includegraphics[width=6.0cm]{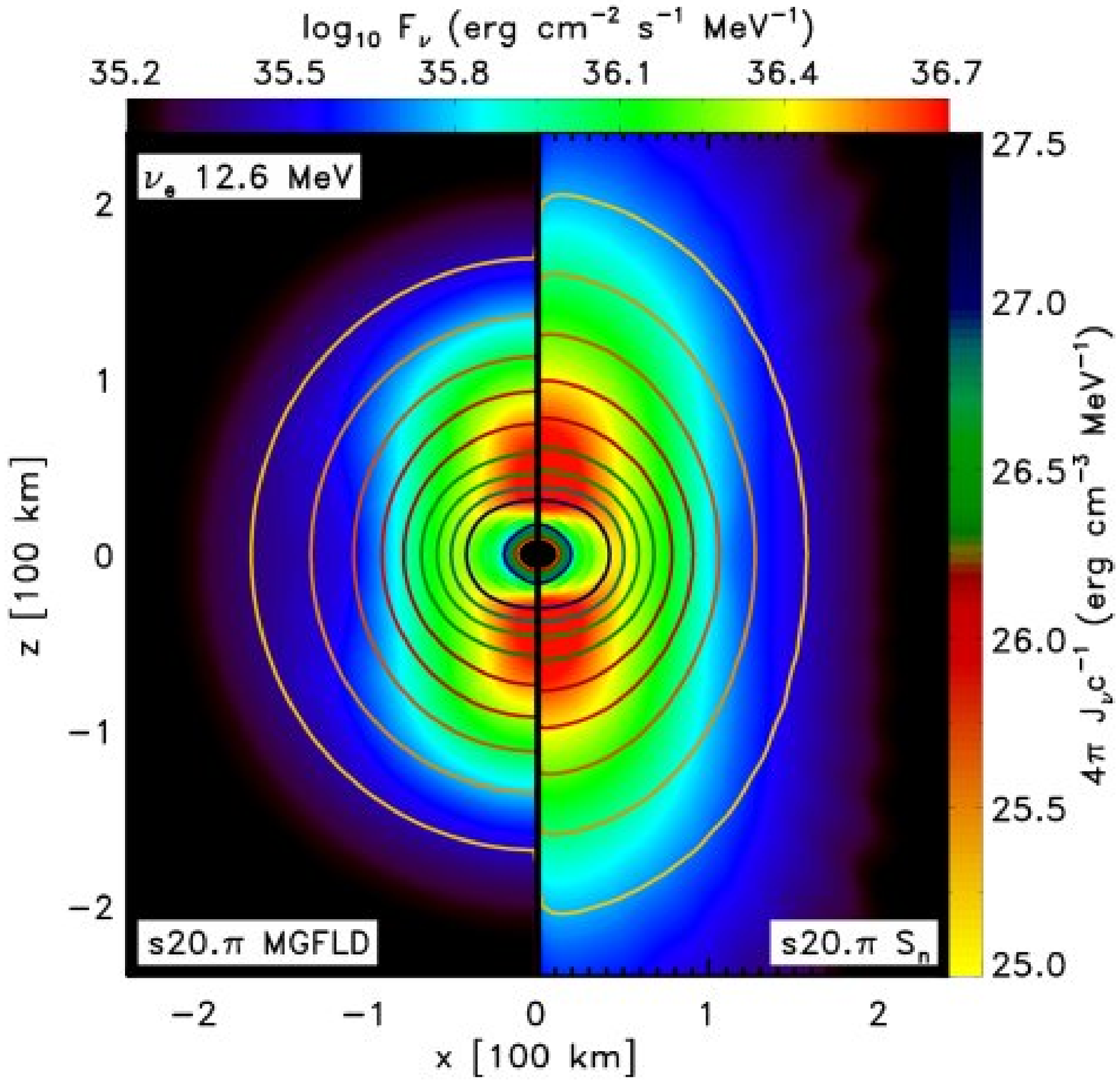}
\includegraphics[width=6.0cm]{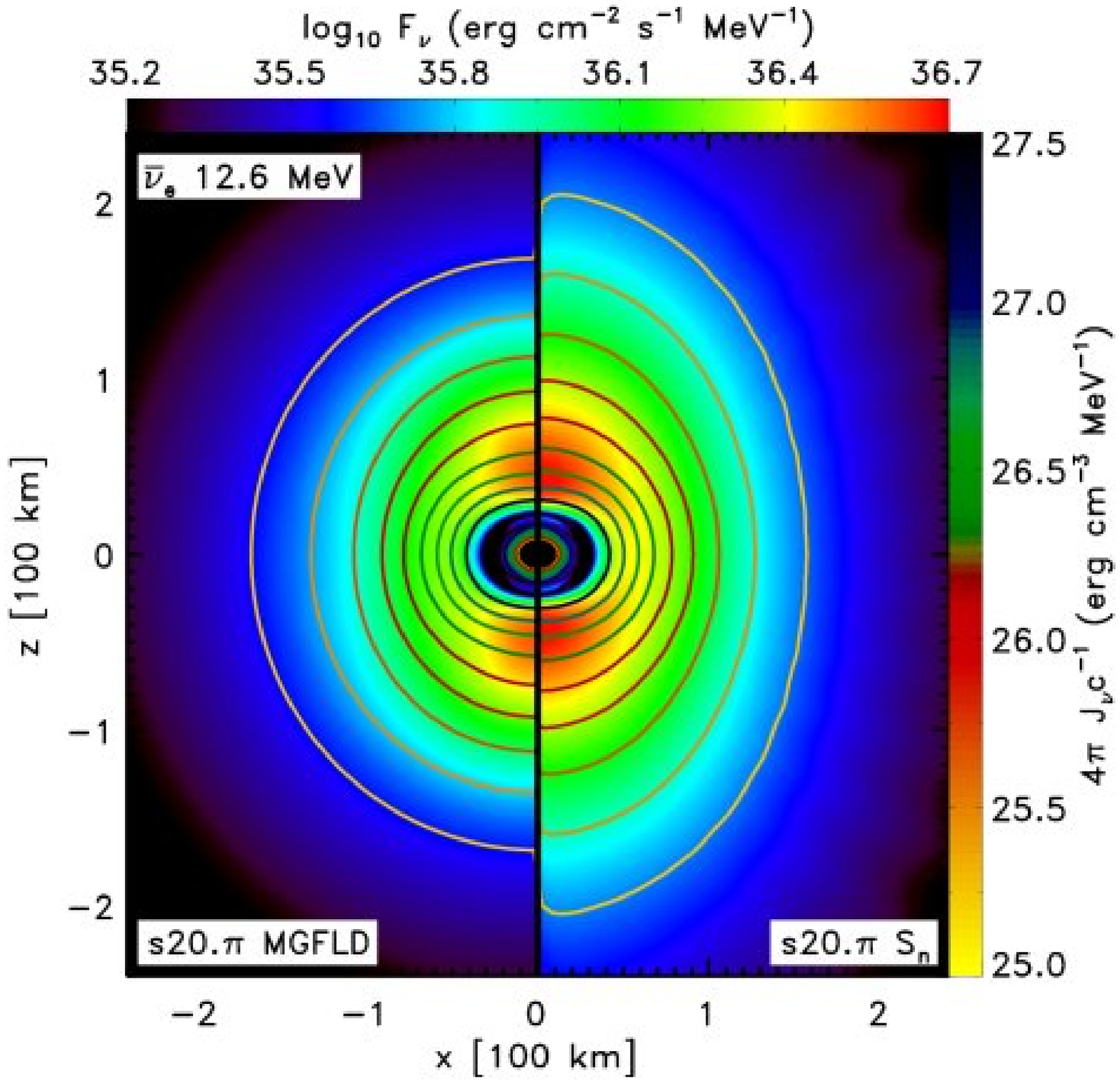}
\includegraphics[width=6.0cm]{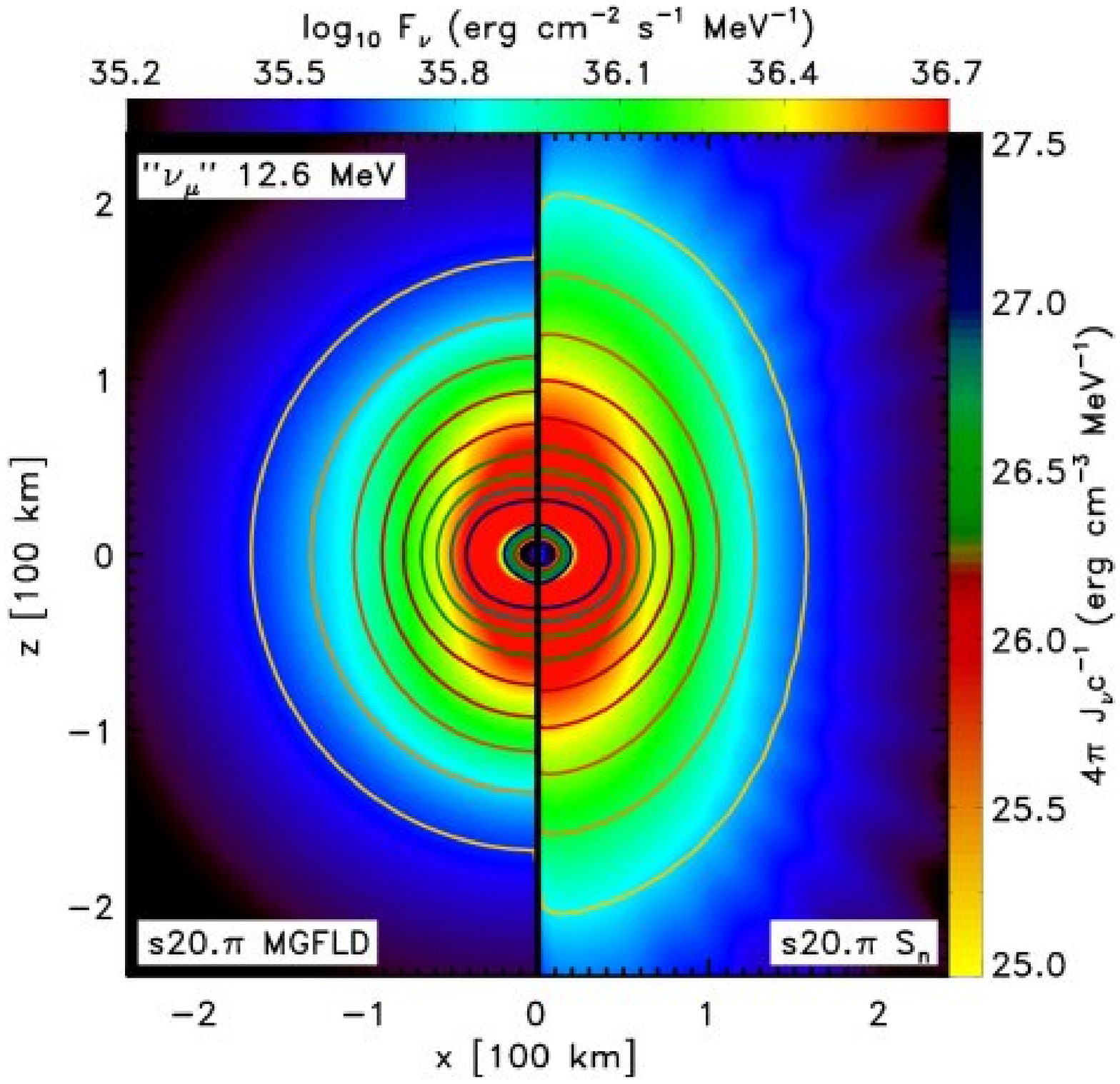}
\caption{Colormaps of the radial spectral flux at $\varepsilon_\nu
=$~12.6~MeV of $\nu_e$ (left), $\bar{\nu}_e$ (center), and
``$\nu_\mu$'' (right) neutrinos in the rapidly-rotating model
s20.$\pi$ at 160~ms after bounce.  Isoenergy density contours ($4\pi
c^{-1} J_\nu$, vertical color legend) are superposed.  The left half
of each panel displays the MGFLD result -- \sn\ is shown in the right
half. The radiation fields are oblate in the PNS core and deform to a
prolate shape further out. Note that \sn\ predicts a prolateness of
the radiation field to much greater radii than MGFLD does. The latter
leads to nearly spherically symmetric radiation fields at radii
greater than $\sgreat\,$150--200~km independent of neutrino species.
\label{fig:s20pi_fluxes}}
\end{figure*}

\begin{figure}[t]
\centering
\includegraphics[width=8.5cm]{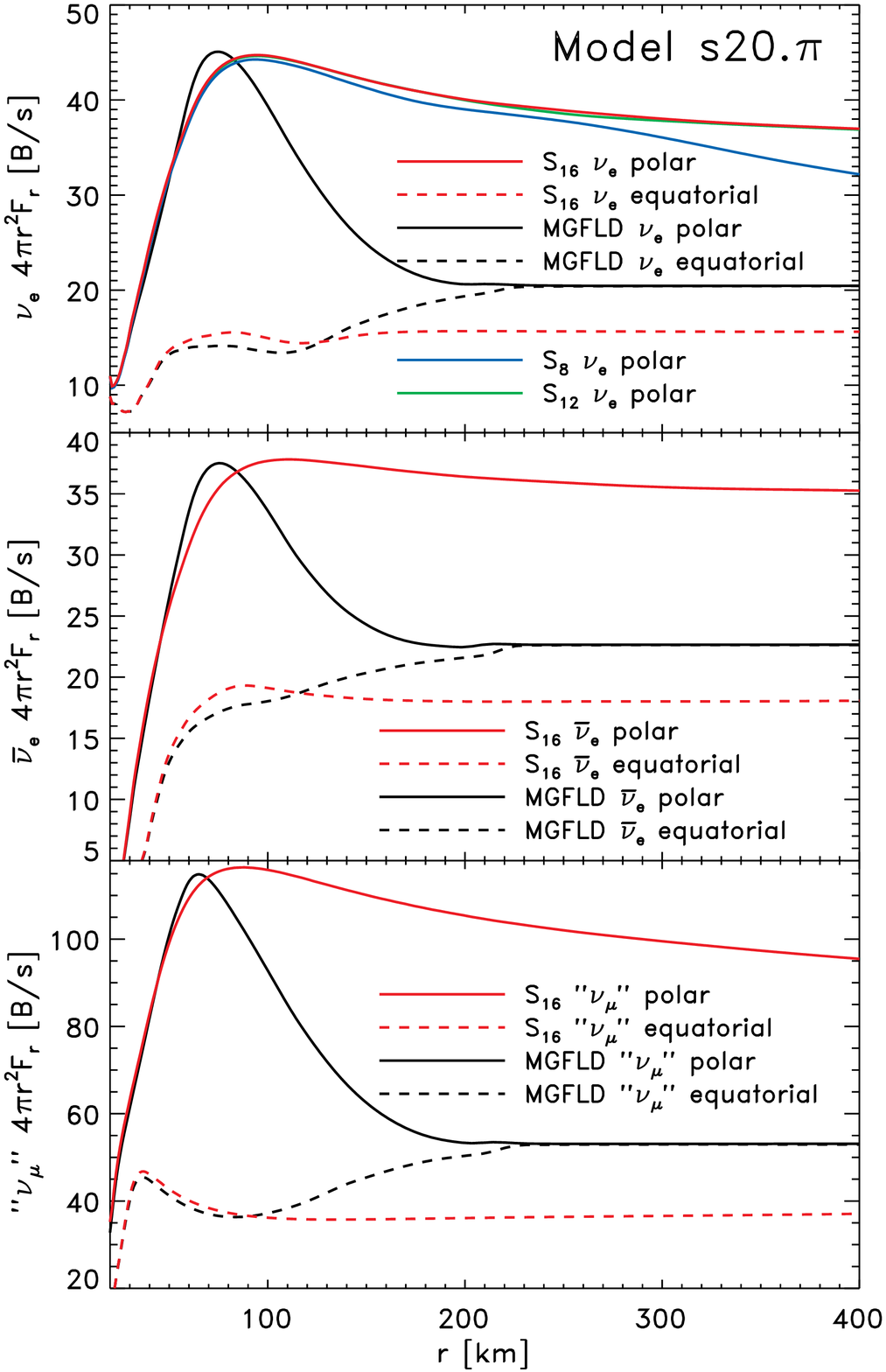}
\caption{ Radial neutrino ``luminosity'' profiles ($4\pi r^2 F_r$) as
seen by observers near the pole (solid lines) and near the equator (dashed
lines) in model s20.$\pi$ at 160~ms after bounce. Red
graphs correspond to \sn\ results, black graphs depict MGFLD results.
Top, center, and bottom panels show $L_\nu$ for $\nu_e$, $\bar{\nu}_e$,
and ``$\nu_\mu$,'' respectively. All \sn\ results were obtained with
$n = 16$, but for comparison we also plot in the top panel polar
profiles that were obtained with S$_{8}$ and S$_{12}$ and find that
both S$_{16}$ and S$_{12}$ are very well converged, while S$_{8}$
has troubles at radii greater than $\sim$200~km. However, it agrees very well
at smaller radii with the higher-resolution \sn\ calculations.
\label{fig:s20pi_lum160}}
\end{figure}

The behavior we observe with radius of the luminosity and mean-inverse
flux factor agrees with the general findings of \cite{messer:98}.  In
particular, we agree with their assessment that the artificially
accelerated transition to free streaming in MGFLD occurs not at the
neutrinospheres (which are generally below the gain region), but at
relatively large radii within which most of the neutrino source is
enclosed.

In the bottom panel of Fig.~\ref{fig:s20nr_line} 
we present profiles of the RMS neutrino energy  
for all species in MGFLD and \sn\ snapshots of model s20.nr. The
corresponding luminosity spectra (extracted at 500~km) are shown 
in Fig.~\ref{fig:s20nr_spec160}.
Both MGFLD and \sn\ capture the energy systematics that 
is set essentially by the matter temperature in the decoupling
region. Neutrino species that decouple at smaller radii (higher
densities and temperatures) have higher RMS energies and
harder spectra than neutrinos decoupling at larger radii.
Quantitative differences in RMS energies and in the spectra
between MGFLD and \sn\ are small, the slightly higher MGFLD
spectral luminosities being mostly a result 
of the artificially enhanced MGFLD luminosities
near and beyond the shock.

We now conclude our discussion of the 160~ms-postbounce snapshot of
model s20.nr by considering the instantaneous neutrino energy
deposition rates.  Figure~\ref{fig:s20nr_gainprof} depicts
angle-averaged radial profiles of the specific neutrino
heating/cooling rates in units of erg~(g s)$^{-1}$. The region of net
gain extends from $\sim$90~km to the shock radius and the chief
contribution to the heating comes from charged-current
$\bar{\nu}_e$-capture processes on protons, exceeding the
corresponding $\nu_e$-capture on neutrons by a factor of two and more
in the narrow radial interval from 145 to 175~km.  MGFLD
underestimates the specific net gain in the angle-averaged
radial profile by at most 10\% locally and by $\sim$5\% on average at
radii greater than $\sim$110~km.  The integral total net gain
predicted by S$_{16}$ is 2.13~B~s$^{-1}$. This is only 3\% larger than
the MGFLD value of 2.07~B~s$^{-1}$. We note in passing that S$_8$
overestimates the integrated gain rate by at most $\sim$1.6\% while S$_{12}$
agrees with S$_{16}$ to better than $\sim$0.3\%.

Figure~\ref{fig:s20nr_net_gain160} depicts the 2D distribution of
neutrino heating and cooling in the snapshot of model s20.nr
considered here. Regions of net gain range from green to red, cooling
regions are blue to black. The colormap demonstrates the somewhat
misleading character of angle-averaged profiles. While we find that
there is little spatial angular variation in the neutrino radiation
field, the neutrino--matter coupling depends strongly on 
angular position, and energy deposition is generally greatest in
regions of high entropy (cf. Fig~\ref{fig:intro2D}).

\subsubsection{Model s20.$\pi$}
\label{section:s20pi}

As we discussed in the context of the Eddington tensor in
\S\ref{section:Eddy} and as may be guessed from the significant
rotational deformation of the core in model s20.$\pi$
(Fig.~\ref{fig:intro2D}), the radiation field in this model exhibits a
strong rotationally-induced asymmetry between pole and equator.  In
Fig.~\ref{fig:s20pi_fluxes}, we present 2D colormaps of the radial
spectral flux component (in erg s$^{-1}$ cm$^{-2}$ MeV$^{-1}$) and
isoenergy-density contours ($4\pi J_\nu / c$ in erg cm$^{-3}$
MeV$^{-1}$) at a representative $\varepsilon_\nu$ of 12.6~MeV and for all
species. Numbers for both \sn\ and MGFLD are compared side by side.
The global radiation-field anisotropy systematics are qualitatively
similar to what was found in the previous MGFLD rotating core-collapse
study of \cite{walder:05}. At small radii, the radiation field (energy
density) follows the density distribution and is oblate, but in the
snapshot at 160~ms after bounce shown in Fig.~\ref{fig:s20pi_fluxes}
has a pole--equator ratio of only 1:2. This ratio increases as the PNS
cools and contracts. The polar compactness of the PNS core leads to a
decoupling of matter and neutrinos at smaller radii in regions near
the pole, resulting there in greater spectral fluxes at higher
neutrino energies and in a prolate distribution of neutrino fluxes
and isoenergy-density contours.

\begin{figure}[t]
\centering
\includegraphics[width=8.5cm]{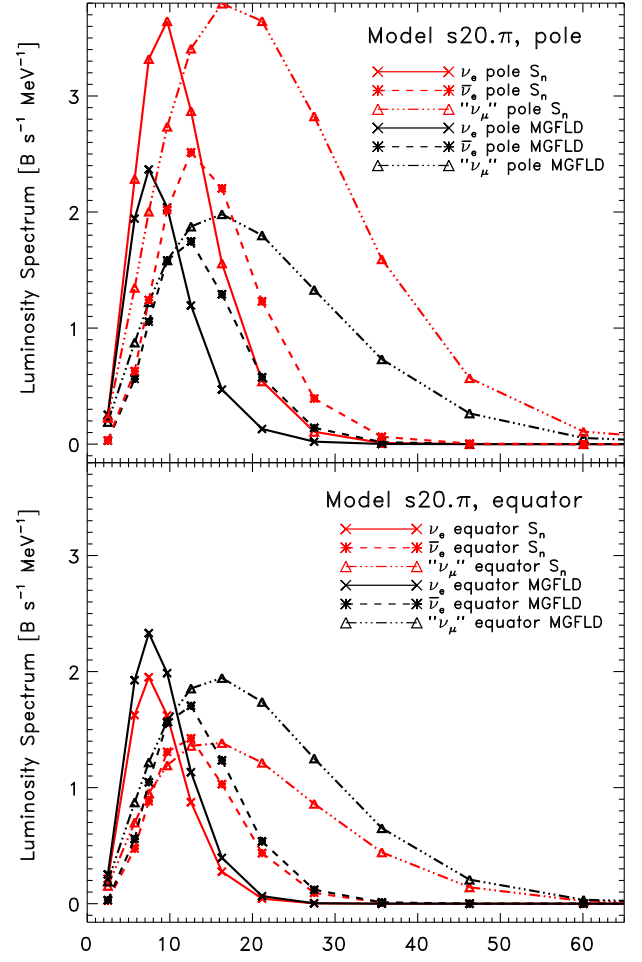}
\caption{{\bf Top}: Neutrino ``luminosity'' spectra ($4\pi r^2
F_r(\varepsilon_\nu)$) in \sn\ (red) and MGFLD (black) variants of model
s20.$\pi$ as seen by observers near the pole.
 $\nu_e$ spectra have solid lines, $\bar{\nu}_e$ spectra are
shown in dashed lines, and ``$\nu_\mu$''s have dashed-dotted
spectra. The spectra are taken from a S$_{16}$ calculation at a
radius of 300~km at 160~ms after core bounce.
{\bf Bottom:}  ``Luminosity'' spectra seen by equatorial observers.
\label{fig:s20pi_spec160}}
\end{figure}

The most striking difference between the \sn\ and MGFLD radiation
fields presented in Fig.~\ref{fig:s20pi_fluxes} is the former's much
greater prolateness at large radii for all species (and all energies,
though we show only $\varepsilon_\nu =$~12.6~MeV).  With the MGFLD
prescription, the prolateness of the flux is muted and does not extend
to large radii. Though the radiation fields are smoothed out at radii
$\sgreat\,$150~km by MGFLD, the \sn\ fluxes and energy densities
remain prolate through the entire postshock region and beyond. At
radii outside $\sim$200~km, the typical striping pattern of
\sn\ (\citealt{castor:04}) becomes visible, though not 
yet dominant.

In Fig.~\ref{fig:s20pi_lum160}, we plot line profiles of the polar and
equatorial ``luminosities'' ($4\pi r^2 F_r$) of each neutrino
species. Profiles obtained with \sn\ and MGFLD are shown. The
asymptotic ``luminosities'' obtained with S$_n$ have pole-to-equator
ratios of 2.2 ($\nu_e$), 1.8 ($\bar{\nu_e}$), and 2.4 (``$\nu_\mu$'').
MGFLD smoothes out these large asymmetries, yielding higher equatorial
and significantly lower polar ``luminosities'' at radii greater than
$\sim$100~km. This is consistent with the more qualitative findings
based on Fig.~\ref{fig:s20pi_fluxes}.  We note that the MGFLD variant
of VULCAN/2D still conserves total flux and energy. For the \sn\
calculation, we find total asymptotic luminosities of
21.1~B~s$^{-1}$ for $\nu_e$ neutrinos (MGFLD:
20.4~B~s$^{-1}$),
22.7~B~s$^{-1}$ for $\bar{\nu}_e$ neutrinos (MGFLD:
22.6~B~s$^{-1}$), and
53.0~B~s$^{-1}$ for ``$\nu_\mu$'' neutrinos (MGFLD:
52.3~B~s$^{-1}$). Hence, \sn\ and MGFLD total
luminosities per species agree very well (and differ at most by
$\sim$3.5\% in the $\nu_e$ case), while their flux distributions
disagree significantly.

Figure~\ref{fig:s20pi_spec160}, depicting polar and equatorial
``luminosity'' spectra ($4\pi r^2 F_r(\varepsilon_\nu)$), reveals that in the
\sn\ calculation (polar: black graphs, equatorial: red graphs) the
neutrino radiation emerging from the PNS and postshock environments
through the polar region not only has greater fluence, but also a
significantly different and -- in the $\nu_e$ case -- a significantly
harder spectrum.  $\nu_e$ neutrinos decouple at the largest
radii. Their ``luminosity'' spectrum observed by a polar observer peaks at
$\varepsilon_\nu \sim$9.5~MeV, while for an observer near the equator it
peaks at $\sim$7.5~MeV. Both $\bar{\nu}_e$ and ``$\nu_\mu$'' neutrinos
(which decouple further in) exhibit a smaller variation in peak
energy from pole to equator. The MGFLD calculation, on the other hand,
shows much smaller variations in neutrino energy and flux between
pole and equator (green and blue graphs, respectively).
 We note in passing that the emerging
neutrino spectra of model s20.$\pi$ are systematically softer by up to
$\sim$10\% in each species than those of the nonrotating model s20.nr
presented in Fig.~\ref{fig:s20nr_spec160}. This is a direct
consequence of the rotationally-induced lower overall compactness of
the PNS in model s20.$\pi$.

The RMS neutrino energies in model s20.$\pi$ show the same overall
qualitative behavior and decoupling hierarchy discussed in the context
of model s20.nr. Hence, we do not show them here, but rather state
quantitative results. They do, of course, trace the strong
pole--equator asymmetry that we observe in the radiation field.  The
RMS energies in the 160~ms \sn\ snapshot are 11.5~MeV (pole) and
10.5~MeV (equator) for $\nu_e$, 16.7~MeV (pole) and 15.2~MeV (equator)
for $\bar{\nu}_e$, and 25.8~MeV (pole) and 24.8~MeV (equator) for
``$\nu_\mu$'' neutrinos. The MGFLD values converge at pole and equator
to 10.5~MeV ($\nu_e$), 15.2~MeV ($\bar{\nu}_e$), and 25.0~MeV
(``$\nu_\mu$'').

\begin{figure}[t]
\centering
\includegraphics[width=8.5cm]{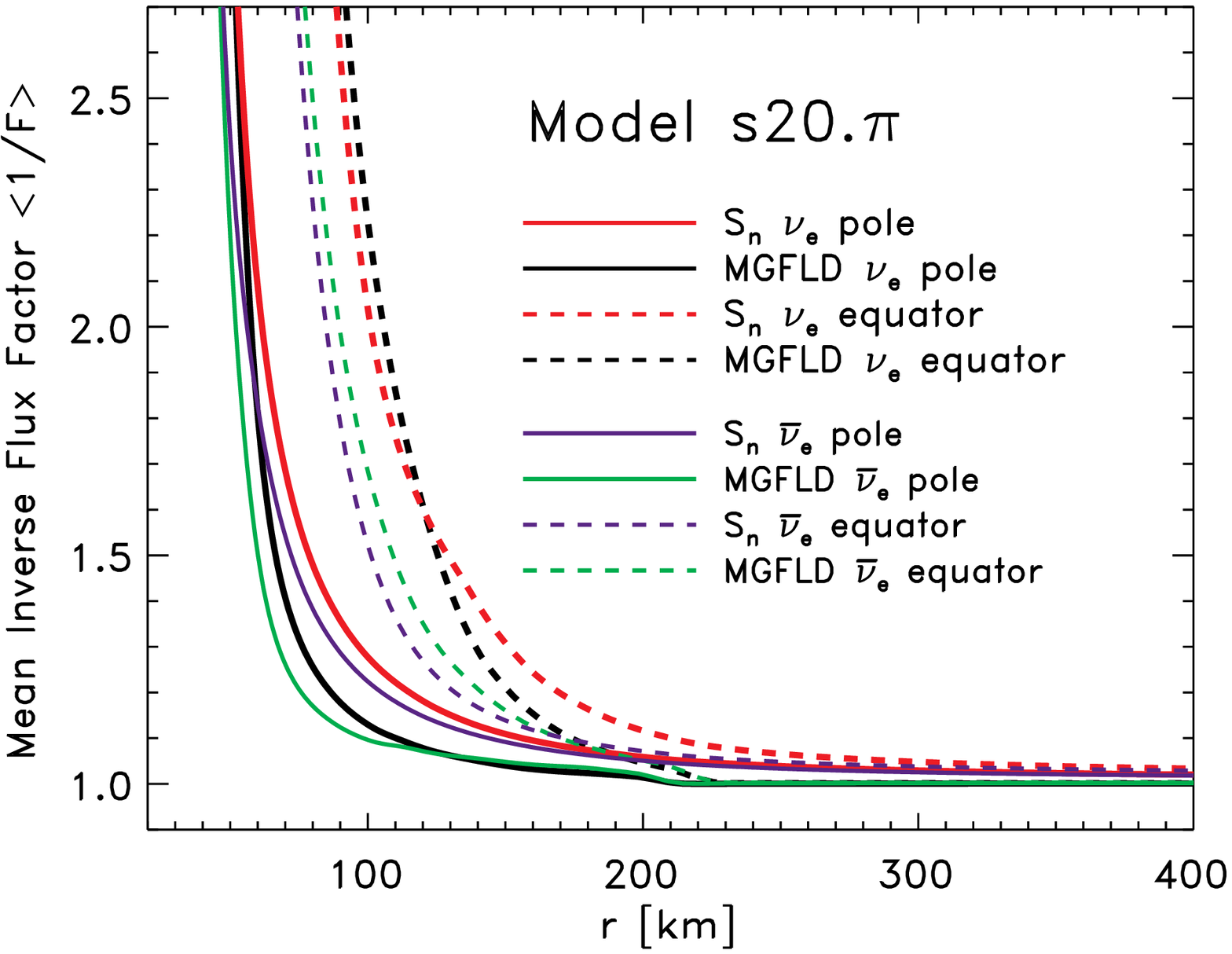}
\caption{ 
Mean inverse flux factors in model s20.$\pi$
at 160~ms after bounce in polar regions (solid lines) and equatorial
regions (dashed lines). Shown are profiles for $\nu_e$ neutrinos
obtained with \sn\ (red) and MGFLD (black), as well as profiles
for $\bar{\nu}_e$ neutrinos (\sn\ blue, MGFLD green). \sn\ 
and MGFLD graphs agree well inside $\sim$50~km at the pole
and inside $\sim$80~km in equatorial regions. For $\nu_e$ neutrinos,
\sn\ yields systematically larger mean inverse flux factors in
polar and equatorial regions. For $\bar{\nu}_e$, however, \sn\ predicts
larger mean inverse flux factors in polar regions, yet transitions
slightly faster than MGFLD to free streaming in equatorial
regions.\label{fig:s20pi_FF160}}
\end{figure}

\begin{figure}[t]
\centering
\includegraphics[width=8.5cm]{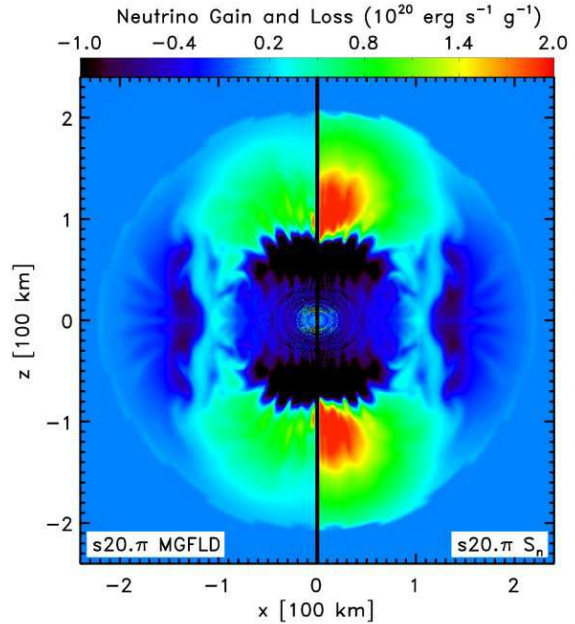}
\caption{ Colormaps of energy- and species-integrated specific
neutrino energy deposition and loss rates in the rotating model
s20.$\pi$ at 160~ms after core bounce (in units of
erg~s$^{-1}$~g$^{-1}$).  The left section of the plot depicts the
MGFLD result and the right shows the result of the \sn\
calculation. Note the distinctively enlarged polar gain regions and
greater specific gain of the S$_n$ result compared to the MGFLD
calculation. This is in part a consequence of the larger polar
neutrino fluxes and overall greater flux asymmetry in the S$_n$ model
(see Fig.~\ref{fig:s20pi_fluxes}).  A feature prevalent in both S$_n$
and MGFLD versions of this rapidly rotating core is an extended loss
region between the shock and the small gain region at low latitudes
(cf. Fig.~\ref{fig:s20pi_netgain2}). The material in the loss region
is still proton rich ($Y_e $\sgreat0.4) and efficiently captures
electrons as it advects in, radiating away a significant flux of
neutrinos (see, e.g., the increase in the equatorial ``luminosity''
between 120 and 150~km in the \sn\ variant of this model, visible in
the top panel of Fig.~\ref{fig:s20pi_lum160}). Note that both MGFLD
and \sn\ exhibit a very small artefact (lower gain/loss) at the
symmetry axis associated with imperfect numerics/regularization.
\label{fig:s20pi_netgain}}
\end{figure}

\begin{figure}[t]
\centering
\includegraphics[width=8.5cm]{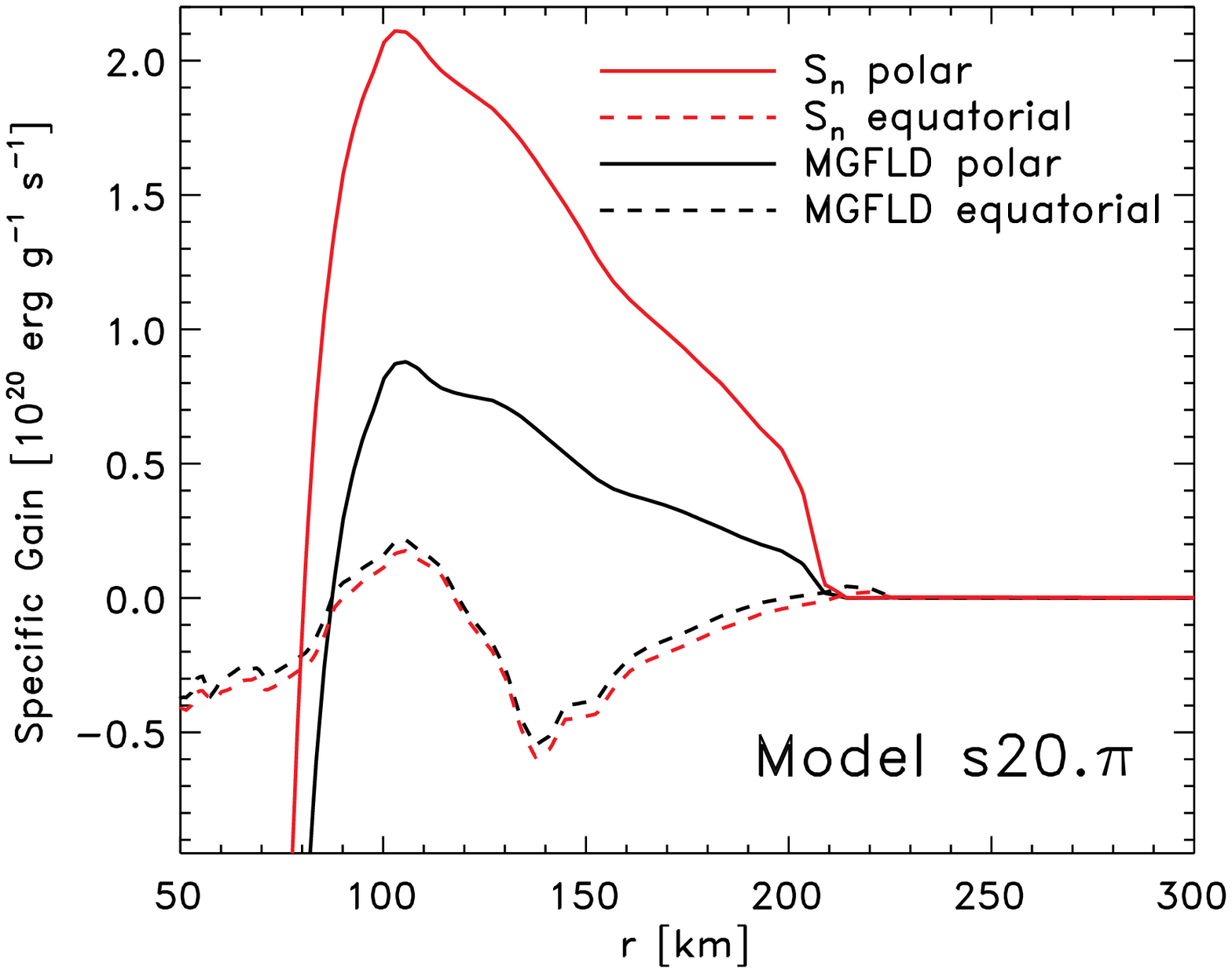}
\includegraphics[width=8.5cm]{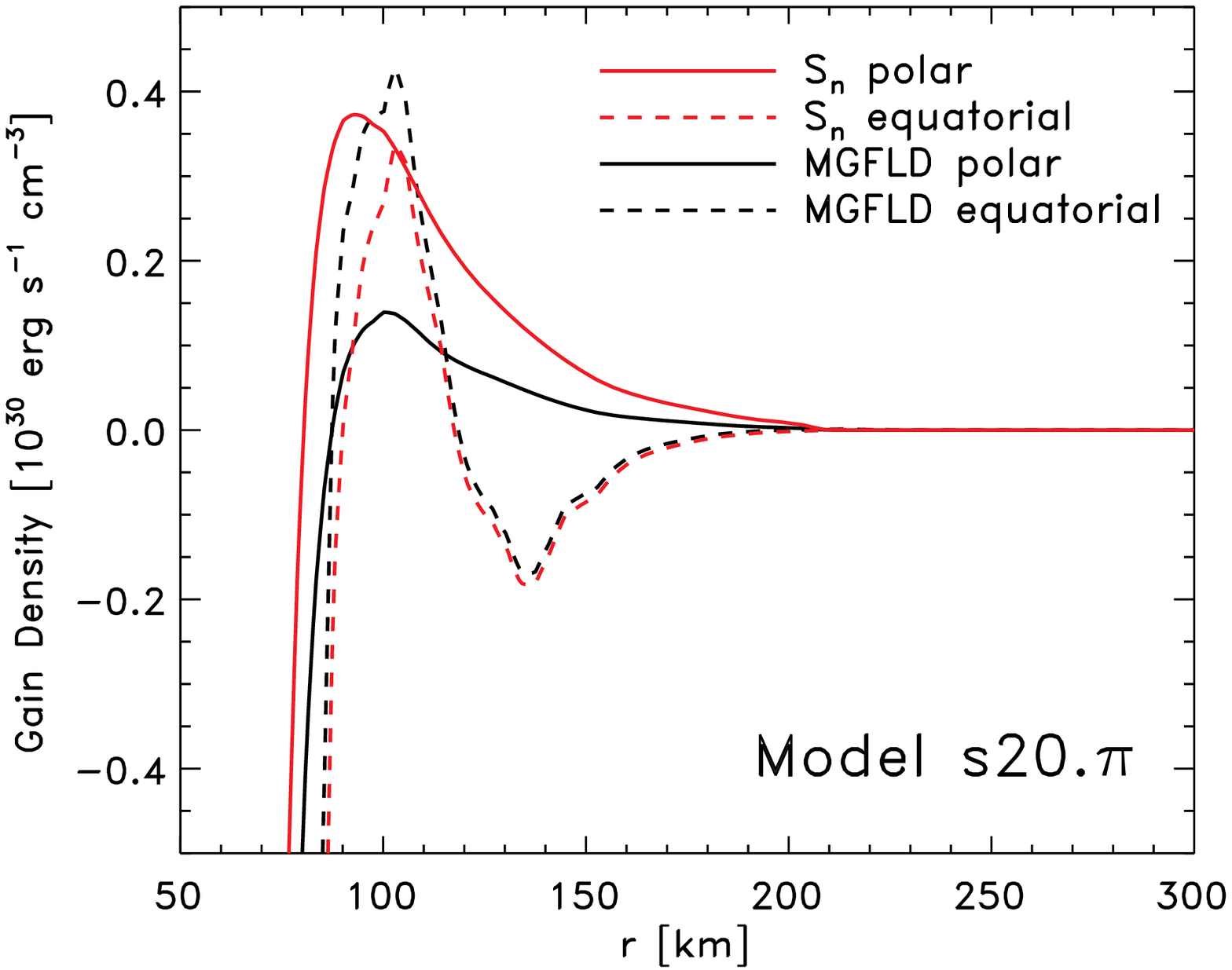}
\caption{{\bf Top}: Averaged specific radial neutrino gain and loss
profiles in model s20.$\pi$ at 160~ms after core bounce. Shown are
results from the S$_n$ (red) and MGFLD (black) calculations. Both
polar and equatorial radial profiles are obtained by averaging over
20$^\circ$ wedges. As is already clear from
Fig.~\ref{fig:s20pi_netgain}, S$_n$ yields significantly greater polar
specific neutrino energy gain than MGFLD. The S$_n$ gain region
extends further in by $\sim$10~km and the gain is more than a factor
of two larger in the interval from $\sim$90 to 200~km.  Given the
larger flux asymmetry in the S$_n$ calculation
(Fig.~\ref{fig:s20pi_fluxes}), less neutrino flux is going through
regions of low latitude, resulting in the \emph{lower} specific gain at low
latitudes predicted by S$_n$.  {\bf Bottom}: Neutrino gain density
(density-weighted specific gain). Due to rapid rotation higher
densities obtain out to larger radii at low latitudes. This results in
a partial reversal of the picture presented by the upper panel;
weighted by density, the neutrino gain (now per unit volume) in the
equatorial wedge becomes comparable to that near the poles.
Furthermore, equatorial regions, since they subtend the largest solid
angles, contribute most to the volume integral. The integral numbers
for the net gain in the polar wedge (counting both poles) for S$_n$
(MGFLD) are 0.17~B~s$^{-1}$
(0.047~B~s$^{-1}$) and in the equatorial wedge
are 0.35~B~s$^{-1}$
(0.47~B~s$^{-1}$). The total integrated net gain is
1.603~B~s$^{-1}$ and
1.637~B~s$^{-1}$ for S$_n$ and MGFLD,
respectively. These numbers are surprisingly close given the large
qualitative and quantitative \emph{local} differences in the neutrino
gain distribution.
\label{fig:s20pi_netgain2}}
\end{figure}

In Fig.~\ref{fig:s20pi_FF160}, we plot polar and equatorial mean
inverse flux factor profiles for $\nu_e$ and $\bar{\nu}_e$ neutrinos
in our steady-state snapshot for model s20.$\pi$. Results from MGFLD
and S$_{16}$ runs are shown. A free-streaming radiation field has an
inverse flux factor of one. Due to the steeper density gradient in
polar regions, neutrinos decouple from matter at smaller radii than at
the equator. While MGFLD must handle the decoupling and increased
forward-peaking of the radiation field via the flux limiter, \sn\
can track it self-consistently.  For $\nu_e$ neutrinos and along the
poles, \sn\ predicts significantly greater mean inverse flux factors
with shallower slopes than MGFLD, indicating a more gradual transition
to free streaming than predicted by the flux limiter.  In the radial
interval of $\sim$60--100~km, the relative difference is
$\sim$12--19\%, decreasing to $\sim$6--12\% out to 200~km.  In
equatorial regions, the $\nu_e$ radiation field is somewhat more
forward-peaked in the \sn\ calculation at radii below $\sim$120~km,
beyond which MGFLD transitions quickly to free streaming while \sn\
approaches it more gradually, exhibiting $\sim$6--8\% larger mean
inverse flux factors in the outer postshock region. For $\bar{\nu}_e$
neutrinos, the behavior of the mean inverse flux factors in polar
regions essentially mirrors that observed for the $\nu_e$s.  In
equatorial regions, the \sn\ mean inverse flux factor of the
$\bar{\nu}_e$s stays below that using MGFLD out to 165~km, beyond which
the MGFLD $\bar{\nu}_e$ radiation field rapidly transitions
to free streaming. At 180~km, the MGFLD $\bar{\nu_e}$ mean inverse 
flux factor is $\sim$1\% smaller than that predicted by \sn. At 220~km,
this difference has grown to $\sim$5\%.

Having established the overall neutrino radiation-field
characteristics in the 160-ms postbounce snapshot of model s20.$\pi$,
we now turn our focus to the neutrino cooling and heating rates in
this model. We have found little difference in the net neutrino
heating between \sn\ and MGFLD variants in the 160-ms postbounce
snapshot of the nonrotating model s20.nr. However, based on the
differences between \sn\ and MGFLD in neutrino fluxes, RMS energies, and
flux factors we have highlighted in this section, we may expect
to find significant differences in the neutrino heating rates for
model s20.$\pi$.

Figure~\ref{fig:s20pi_netgain} depicts 2D colormaps of the neutrino
energy gain and loss rate per unit mass (accounting for all energies
and species), computed for the 160-ms postbounce snapshot of model
s20.$\pi$ using \sn\ (left panel) and MGFLD (right panel). At low
latitudes near the equator, \sn\ and MGFLD agree very well to the eye.
In regions near the pole, both MGFLD and \sn\ show a pronounced region
of net loss at $z$-coordinates between $\sim$40 and $\sim$80~km,
beyond which a region of net gain (colors light blue and green to red)
prevails out to the shock position at $\sim$230~km. While the
gain region has roughly the same physical extent in MGFLD and \sn, the
latter yields significantly higher energy deposition rates.  This is
particularly the case in the lower gain region at polar angles below
$\sim$20$^\circ$ and at radii between $\sim$80 and 150~km, where the
\sn\ gain rate is larger by a factor of two and more. The top panel in
Fig.~\ref{fig:s20pi_netgain2} provides a more quantitative comparison
of \sn\ and MGFLD gain/loss rates, since it contrasts average specific
gain/loss profiles obtained from polar and equatorial 20$^\circ$
wedges.  In the polar region, the \sn\ gain region begins at a radius
of $\sim$80~km (MGFLD: $\sim$88~km) and the \sn\ specific gain rate
magnitude exceeds the MGFLD numbers by a factor of 2.6 at 100~km,
increasing to 3.2 at 200~km.  Near the equator, net energy deposition
occurs only in a small radial interval of $\sim$90--120~km and the
MGFLD specific gain rate is larger by 80\% at 95~km, 41\% at 100~km,
and 26\% at 110~km. The net energy loss between $\sim$120--210~km
(captured by both \sn\ and MGFLD) results from strong electron capture
that dominates energy deposition by neutrino absorption.

\begin{figure*}[t]
\centering
\includegraphics[width=8.5cm]{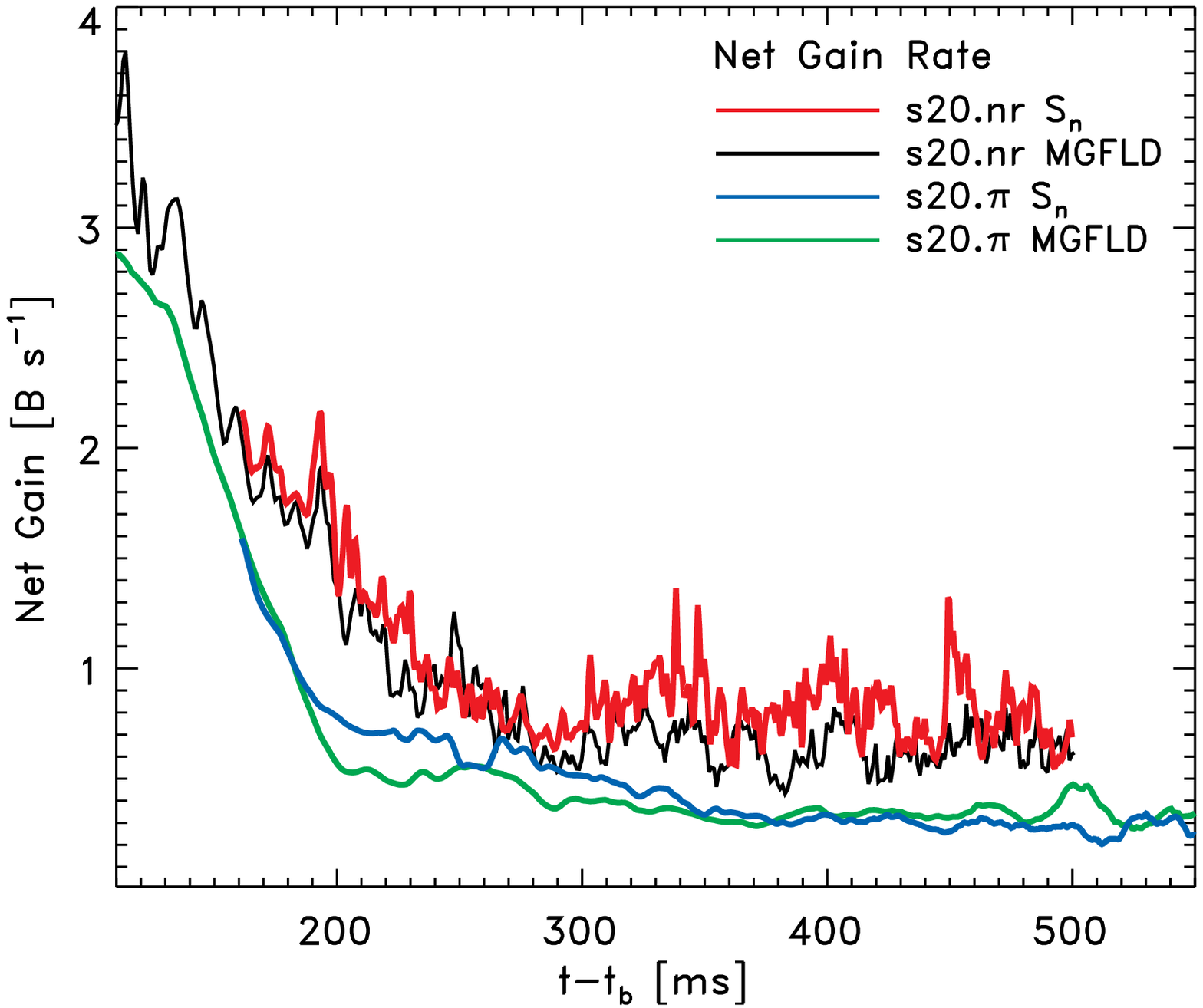}
\includegraphics[width=8.5cm]{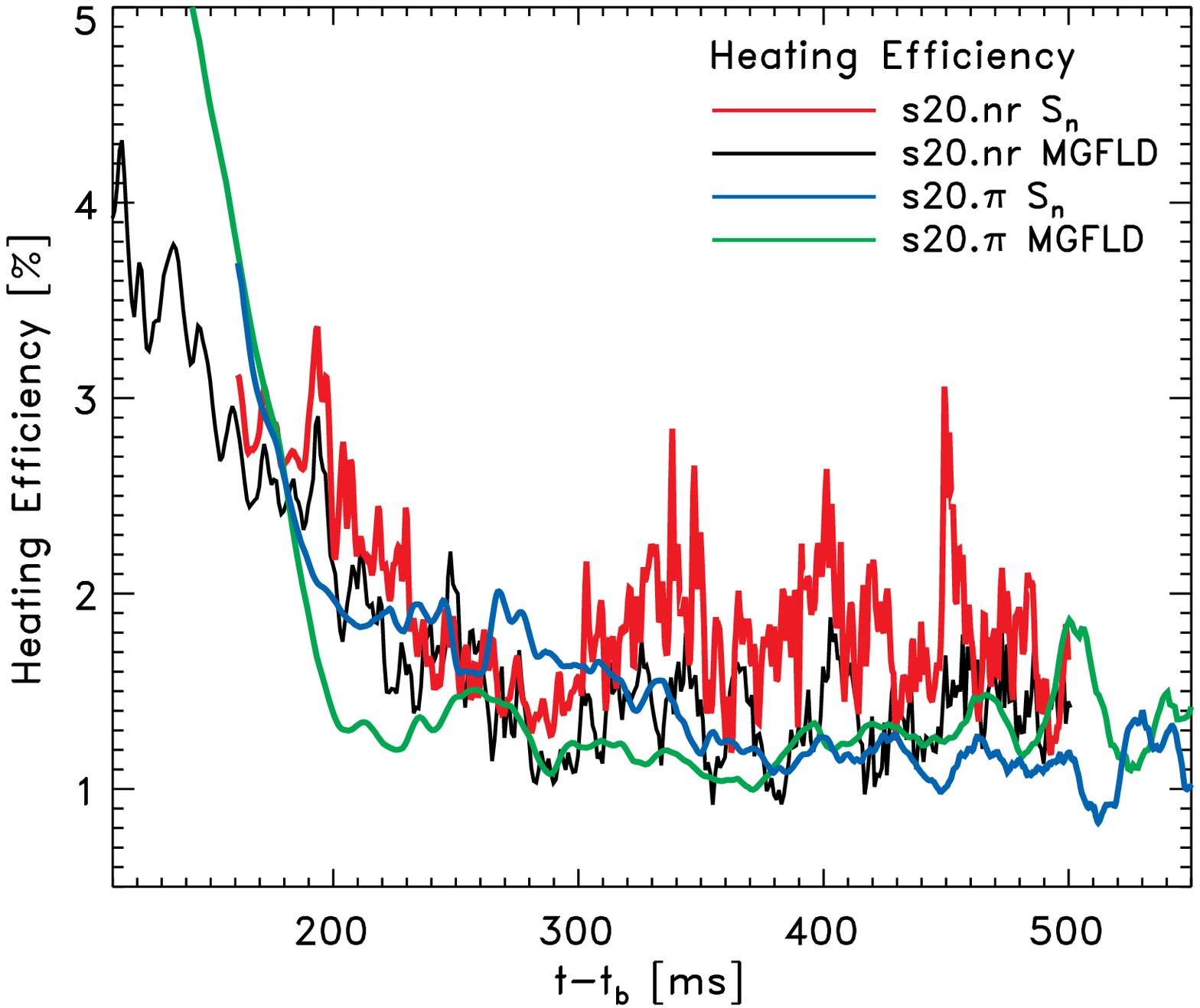}
\caption{{\bf Left:} Evolution of the total neutrino net gain rate as
a function of postbounce time in the \sn\ and MGFLD variants of models
s20.nr and s20.$\pi$. At postbounce times before $\sim$300~ms in model
s20.nr, \sn\ yields a net gain rate that is larger by (on average)
$\sim$10--15\% than that predicted by MGFLD. As the SASI becomes more
pronounced at postbounce times $\sgreat$~300~ms, the \sn\ net gain
begins to more significantly exceed that of MGFLD, averaging out at
$\sim$20-30\% larger values than the MGFLD net gain rate.
In model s20.$\pi$, \sn\ and MGFLD net gain rates stay very
close in the first $\sim$30~ms of evolution, yet depart when
the \sn\ variant approaches its new dynamical equilibrium
(see Fig.~\ref{fig:shockrad}) and provides for a larger gain region
(mass and volume). This leads to a net gain rate that is larger
by $\sim$20-25\% (on average in the postbounce interval
from 200--350~ms). At later times, the MGFLD calculation, approaching
the \sn\ variant's postshock extent (Fig.~\ref{fig:shockrad})
produces larger net gain rates due to its larger equatorial neutrino
fluxes at similar hydrodynamic configuration.
{\bf Right:} Heating efficiency evolution in the two models with their
\sn\ and MGFLD variants. We define the heating efficiency as the 
ratio of total neutrino net gain rate and the sum of electron and
anti-electron neutrino luminosities. 
\label{fig:s20_gain}}
\end{figure*}

The observed local differences in neutrino energy deposition between
\sn\ and MGFLD are due primarily to the vastly different degree to
which the two schemes capture the global pole-equator asymmetry of the
radiation field in the rapidly-rotating postbounce supernova core of
model s20.$\pi$. \sn\ yields much larger fluxes in the polar direction
than MGFLD, but predicts lower neutrino fluxes in equatorial
regions~(cf.\ Fig.~\ref{fig:s20pi_lum160}). Differences in the radial
mean-inverse flux factors and RMS energies are much smaller, and,
hence, are of only secondary importance.  The \sn\ steady-state
snapshot yields an integrated gain rate of 1.603~B~s$^{-1}$ while
MGFLD predicts 1.637~B~s$^{-1}$ for the s20.$\pi$ snapshot under
consideration.  This corresponds to $\sim$2.1\% \emph{more} energy
deposition per unit time in the MGFLD calculation.  Given the above
discussion, the reader may be surprised by these numbers.  The
explanation consists of two factors.  Owing to rotation, the amount of
mass per unit volume (i.e., the rest-mass density) is higher at any
given equatorial radius than at the same radius in the polar
direction. Plotting the neutrino gain/loss rate per unit volume
instead of per unit gram, the bottom panel of
Fig.~\ref{fig:s20pi_netgain2} clearly shows the rotation-induced
enhancement of the energy deposition (per unit volume) near the
equator and the larger gain rate per unit volume predicted by MGFLD at
small to intermediate radii.  The second factor is the simple fact
that the volume of the equatorial gain regions is much larger than
that of the polar gain regions.

As we shall discuss in the following section, the large local
differences in neutrino heating between the \sn\ and MGFLD snapshots
have a dynamical consequence for the rapidly rotating model and lead
to a significant polar expansion of the shock in the \sn\ postbounce
evolution calculation.

\section{Results: Evolution Calculations}
\label{section:evolution}

In order to study differences between \sn\ and MGFLD in a
time-dependent postbounce setting, we follow our relaxed 160-ms
\sn\ models in fully coupled radiation-hydrodynamics
fashion for $\sim$340~ms (model s20.nr) and 390~ms
(model s20.$\pi$) of postbounce time. In parallel with the \sn\ runs, we
continue their MGFLD counterparts for the same time span.

\subsection{Model s20.nr}

\begin{figure*}[t]
\centering
\includegraphics[width=8.5cm]{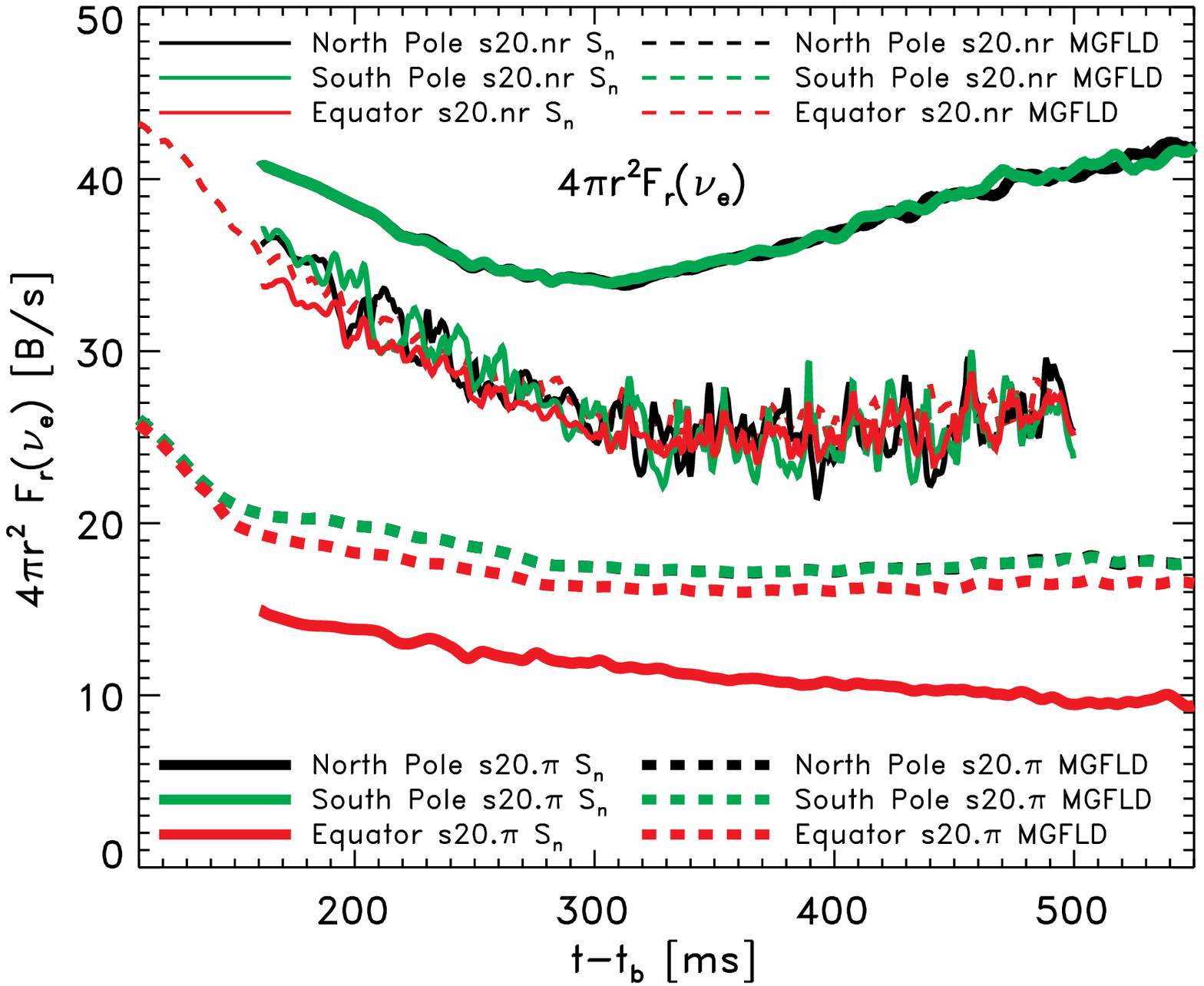}
\includegraphics[width=8.3cm]{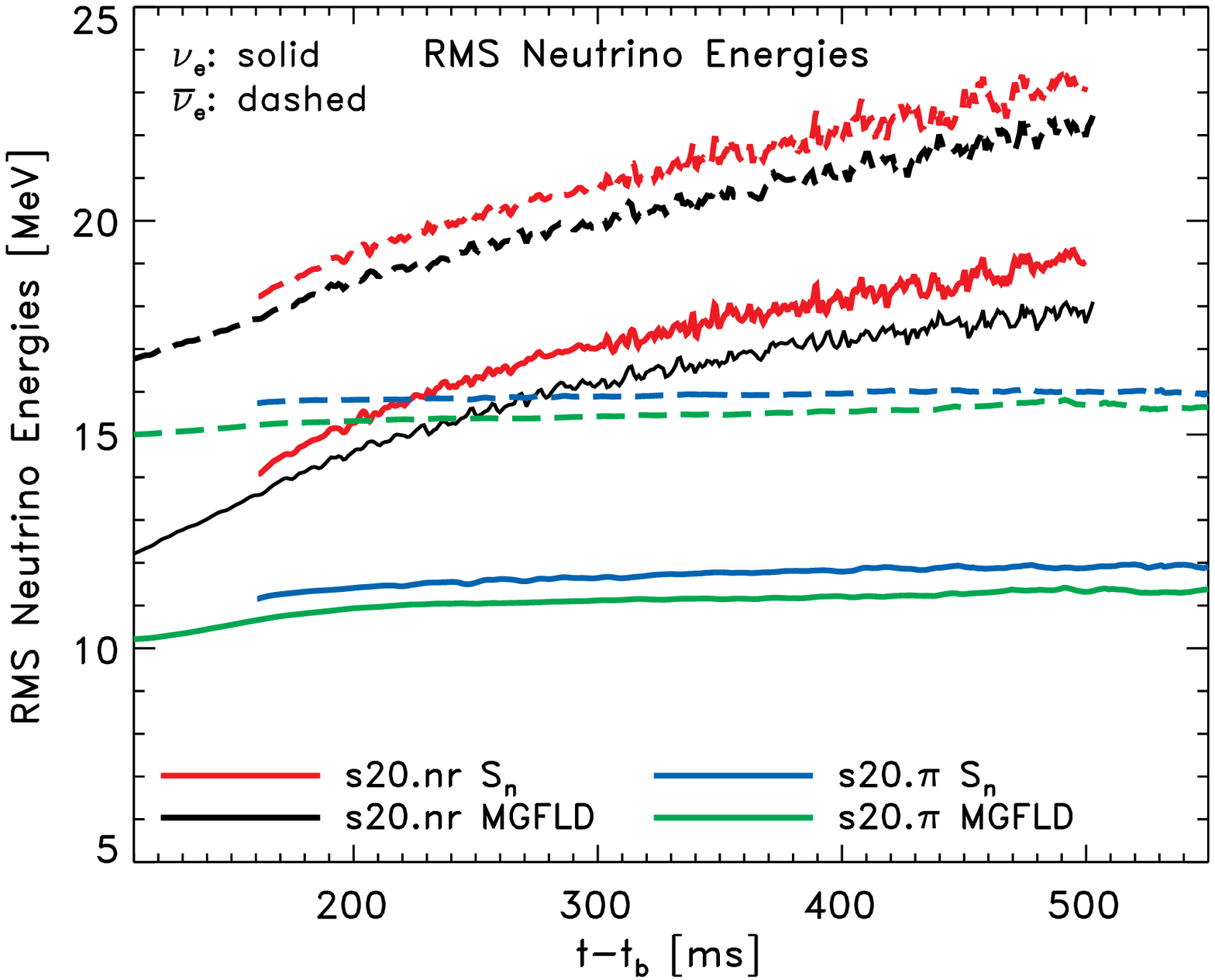}
\caption{{\bf Left:} $\nu_e$ ``luminosities'' ($4\pi r^2 F_r$) as a
function of postbounce time as seen by observers located at a
spherical radius of 250~km along the north pole (black lines), south
pole (green lines), and in the equatorial plane (red lines) in \sn\
(solid lines) and MGFLD (dashed lines) variants of model s20.nr (thin
lines) and s20.$\pi$ (thick lines). Note that the south pole, north
pole, and equator MGFLD ``luminosities'' in model s20.nr (thin dashed
lines) are very similar. Their lines are indistinguishable.  The same
holds for the south and north pole MGFLD ``luminosities'' in model
s20.$\pi$ (thick black and green dashed lines).
{\bf Right:} Angle-averaged RMS energies of $\nu_e$ (solid lines) and
$\bar{\nu}_e$ (dashed lines) neutrinos as a function of postbounce
time in the \sn\ and MGFLD simulations the two 
models. \sn\ predicts systematically higher RMS neutrino
energies in both models.
\label{fig:s20_lum_rms}}
\end{figure*}

\begin{figure*}
\centering
\includegraphics[width=8.5cm]{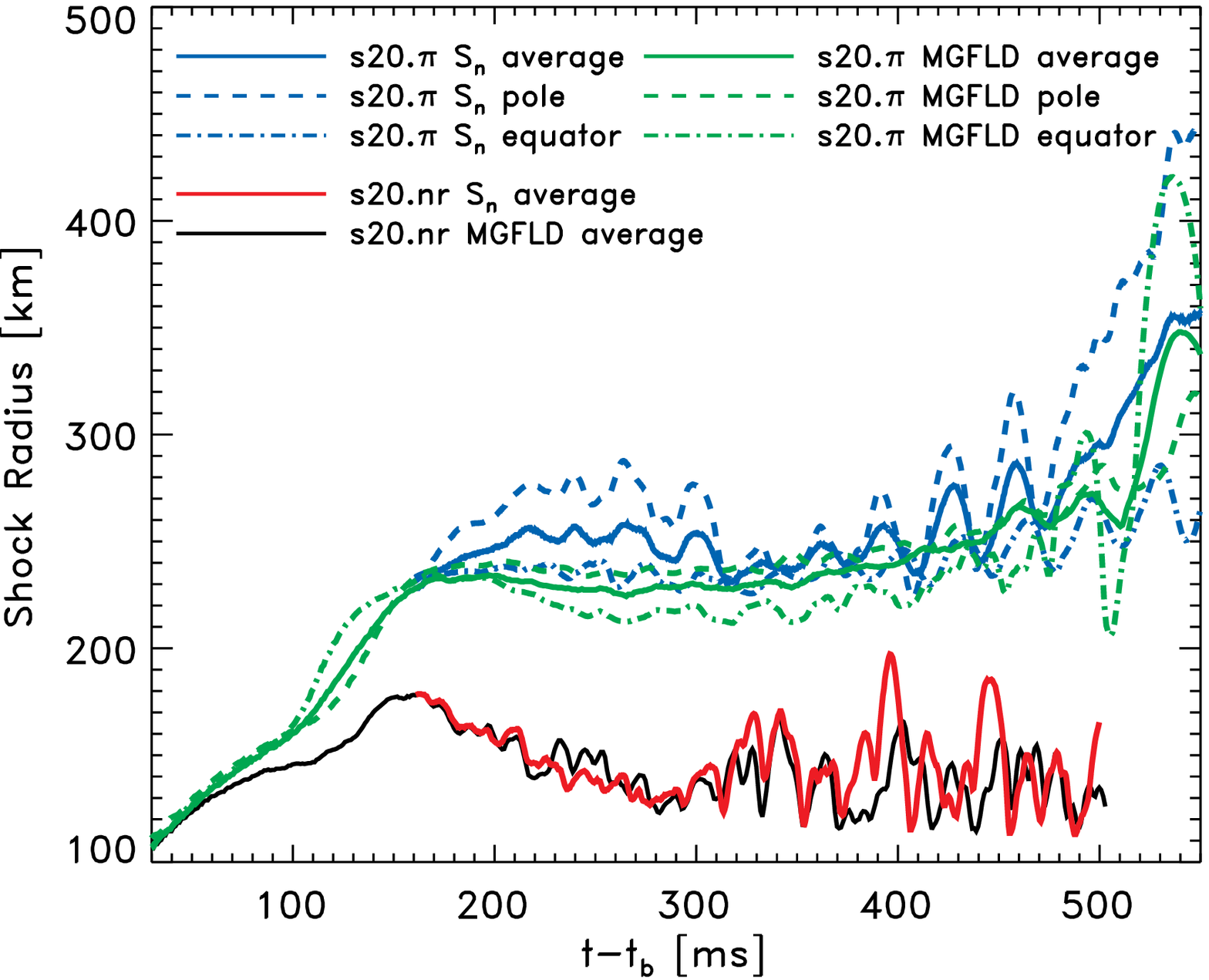}
\includegraphics[width=8.5cm]{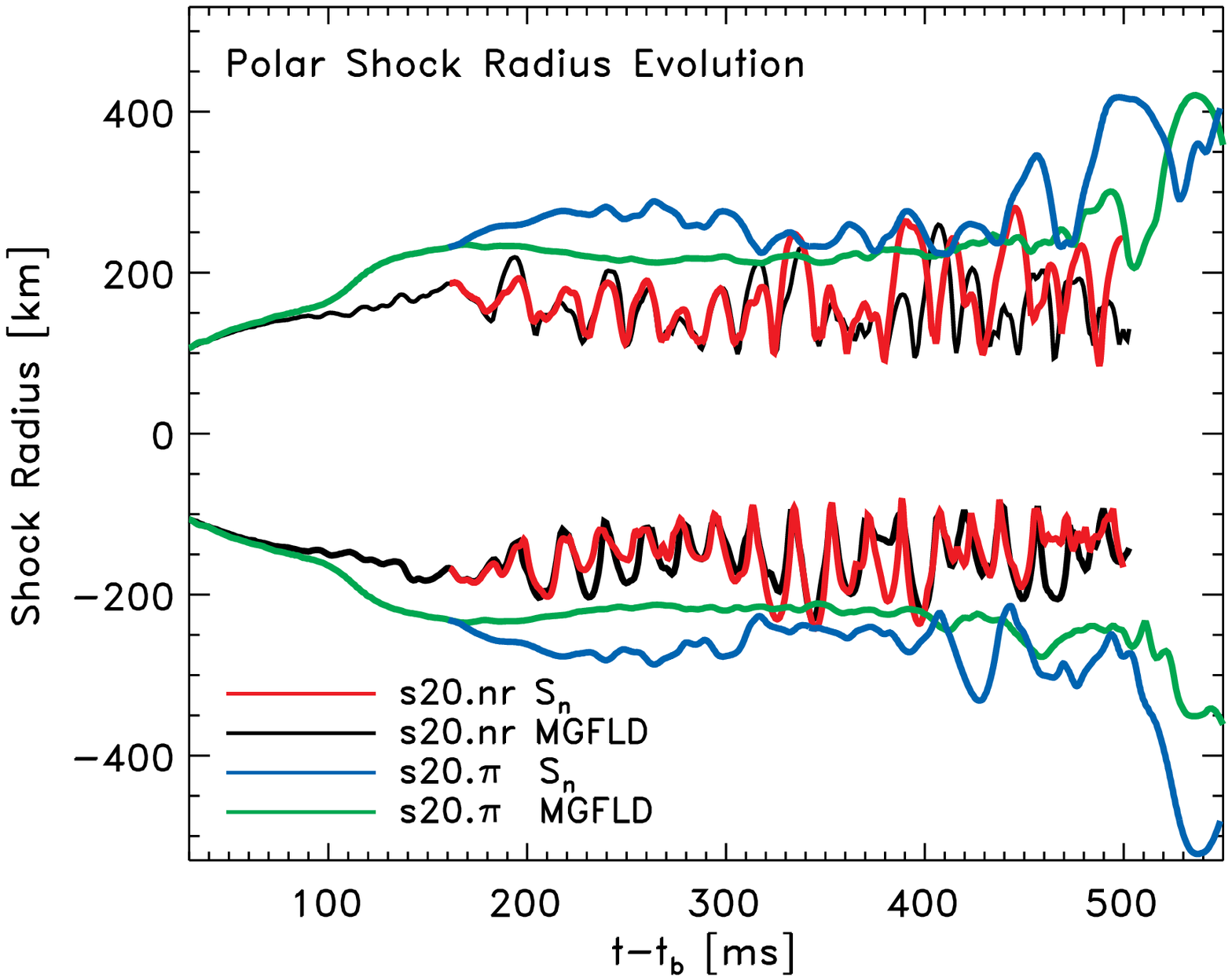}
\caption{{\bf Left}: Average shock radii as a function of postbounce
time in \sn\ (red) and MGFLD (black) variants of the nonrotating model
s20.nr.  Also shown are the overall average shock radius, the average
of south-pole and north-pole shock radii, and the equatorial shock
radius for the rapidly spinning model s20.$\pi$, again for \sn\ (blue)
and MGFLD (green). In model s20.nr, MGFLD and \sn\ show little
quantitative deviation from each other. In the s20.$\pi$ evolution,
however, a significant increase in the various shock radii is
noticable right at the beginning of the time-dependent \sn\
calculation. At later times MGFLD catches up and the average shock
radii approach each other.  The \sn\ variant exhibits larger
variations, indicating stronger SASI-like shock excursions. 
{\bf Right}: Evolution of the north-pole (positive) and south-pole
(negative) shock radii for the \sn\ and MGFLD variants of the two
models. Since the lowest-order and dominant mode of the 2D SASI is the
$\ell$=1 polar sloshing mode, the polar shock radii are good
indicators of its strength and periodicity.  Note the initial
suppression, but late-time development of SASI-like polar shock
excursions in the rotating model. 
\label{fig:shockrad}}
\end{figure*}

\begin{figure*}[t]
\centering
\includegraphics[width=5.8cm]{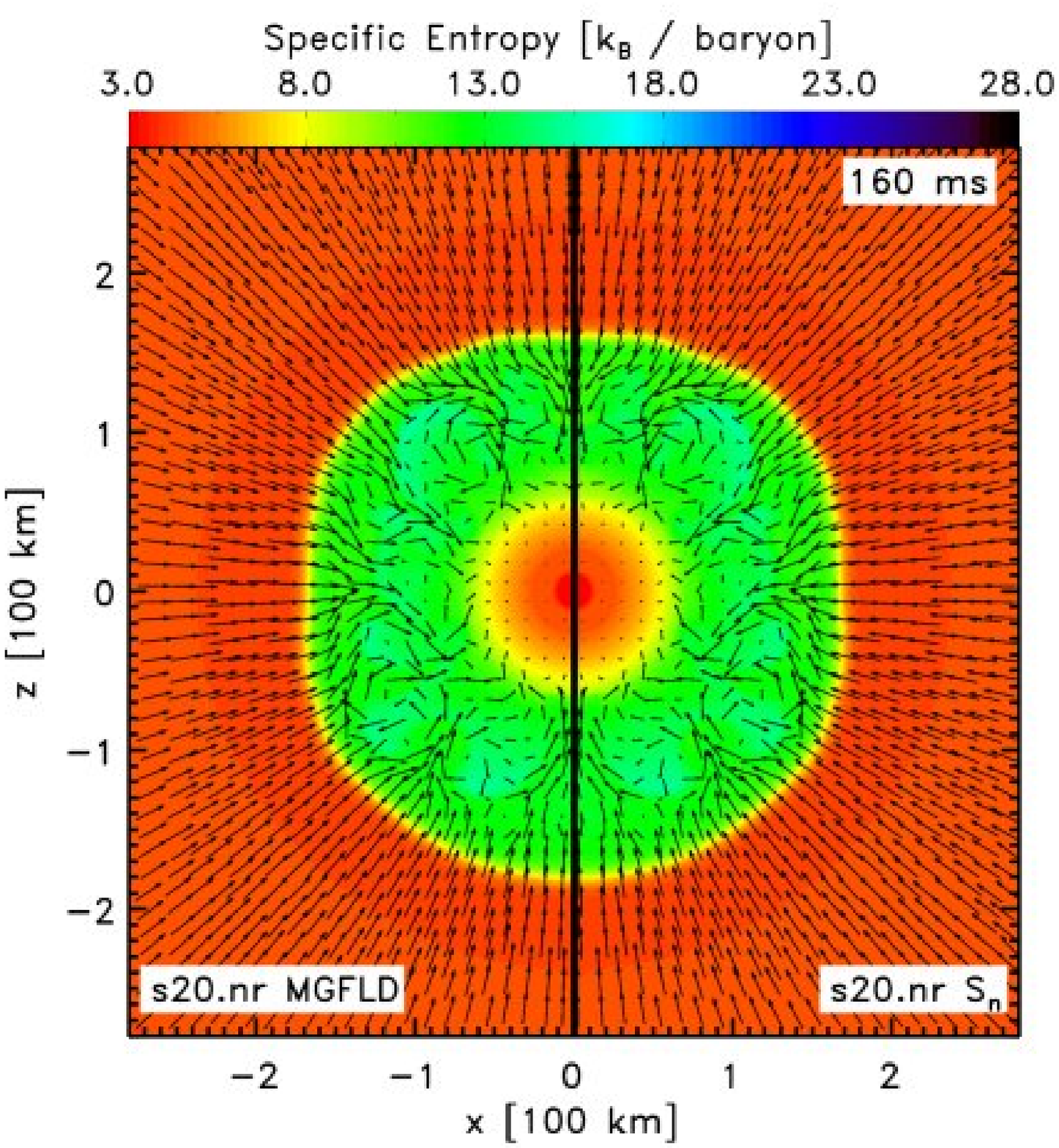}
\includegraphics[width=5.8cm]{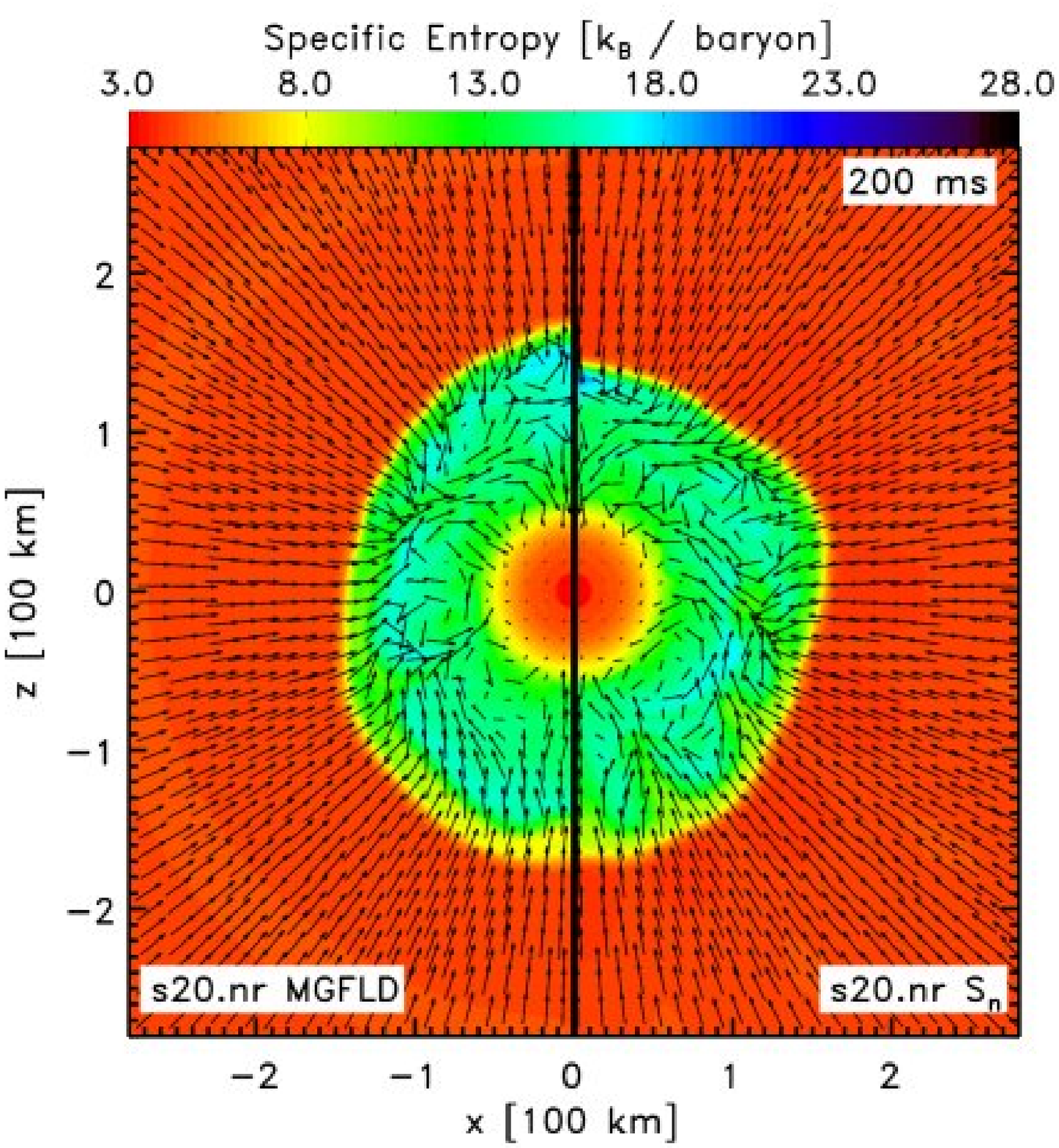}
\includegraphics[width=5.8cm]{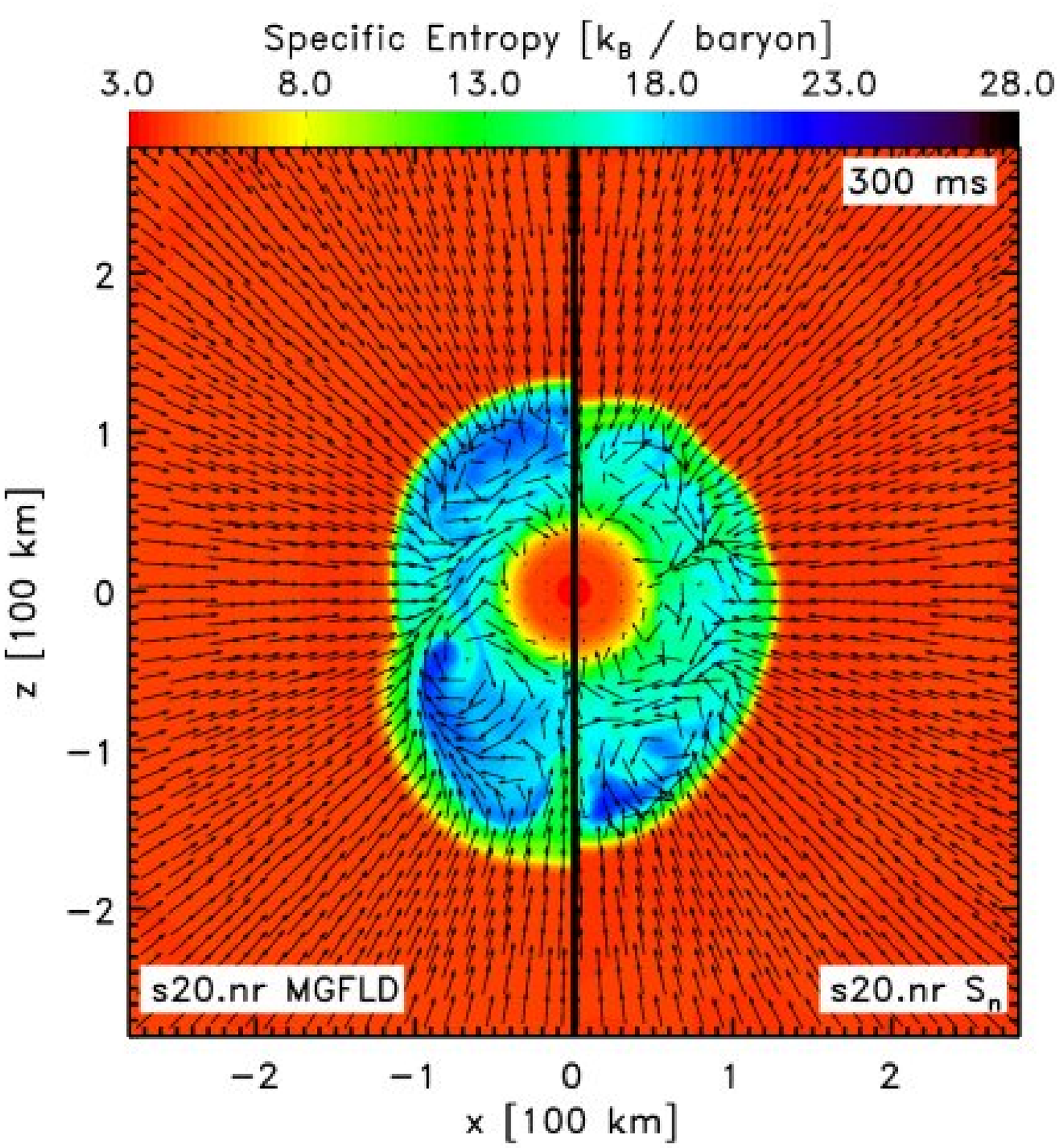}

\includegraphics[width=5.8cm]{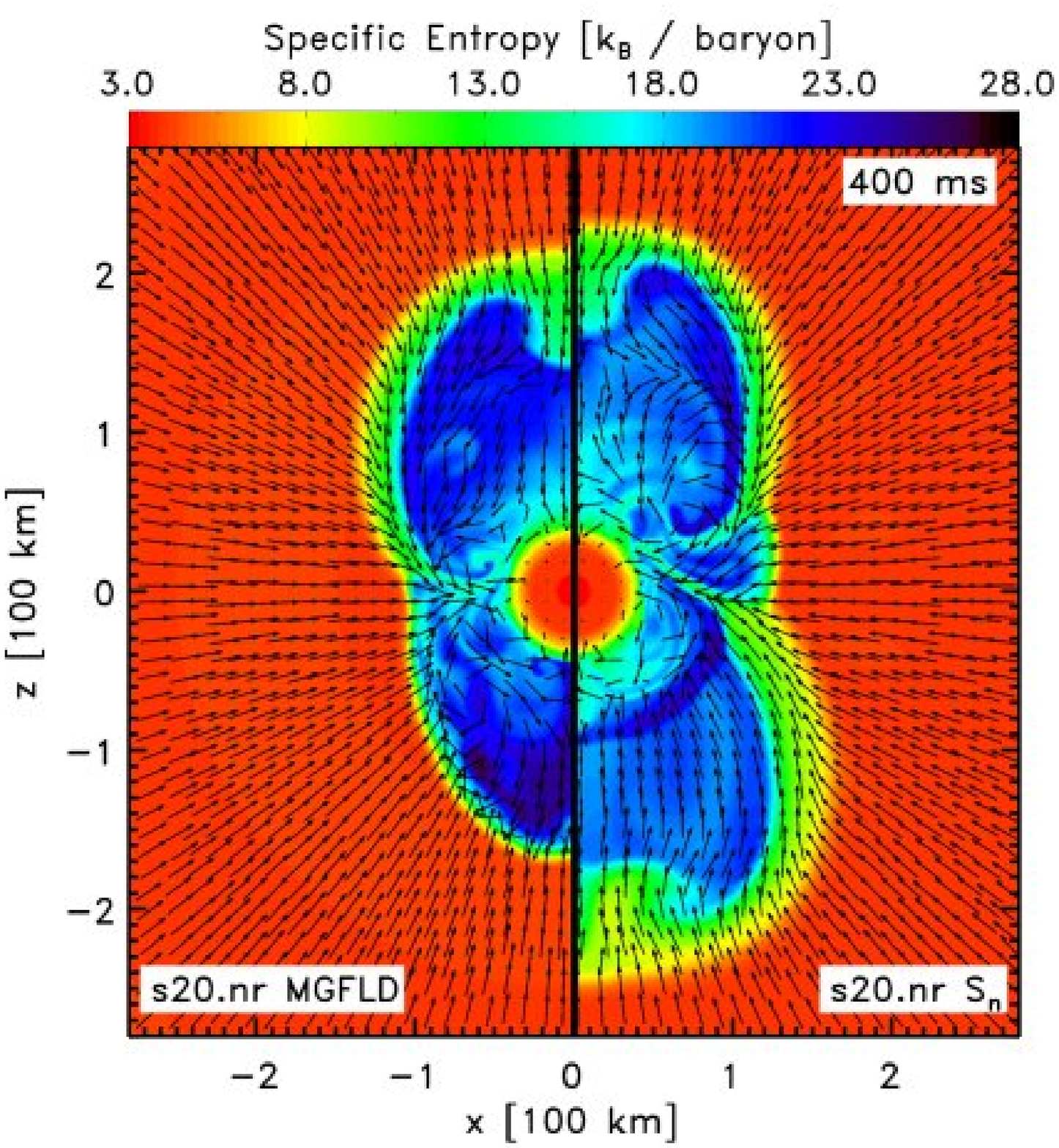}
\includegraphics[width=5.8cm]{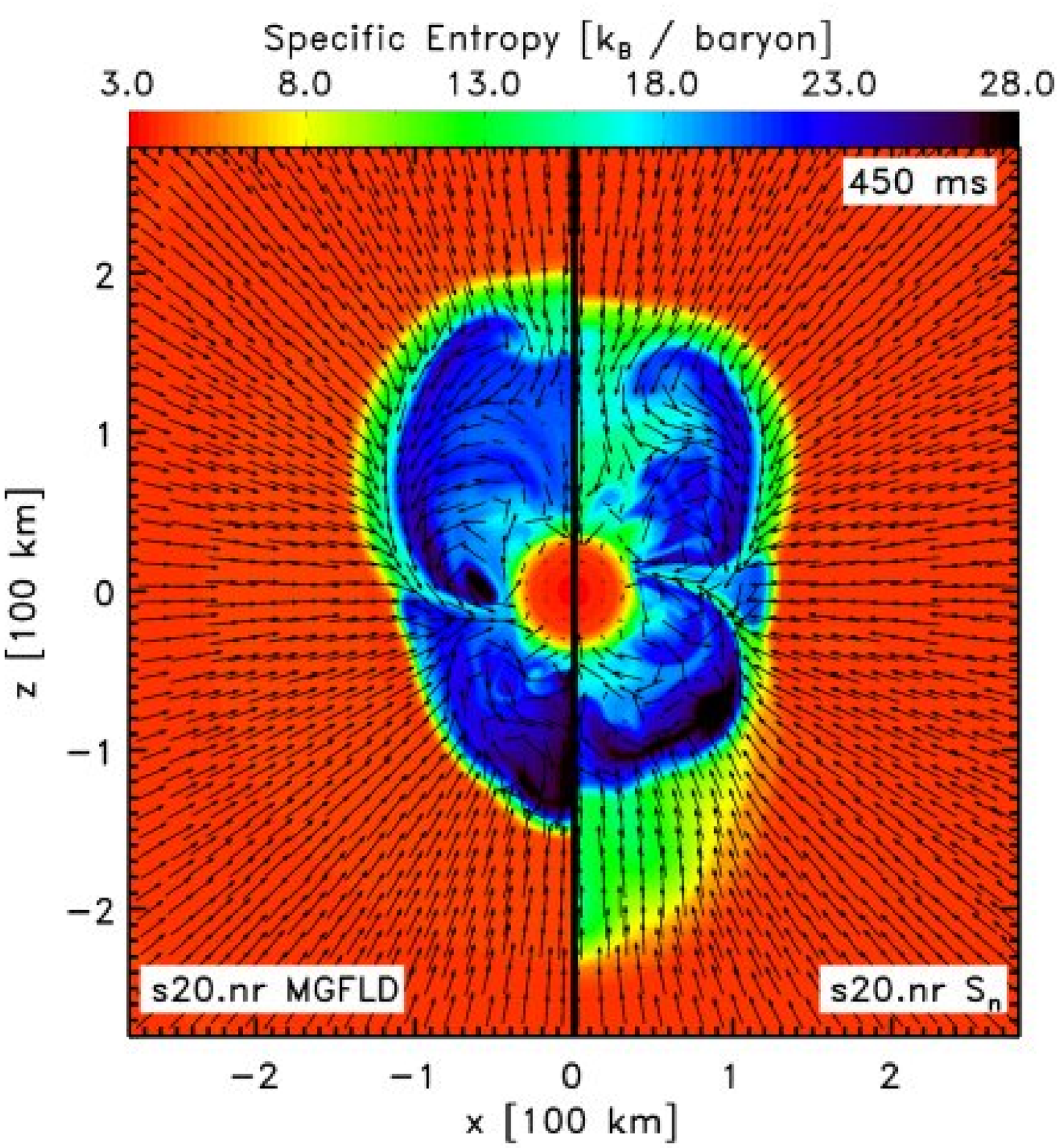}
\includegraphics[width=5.8cm]{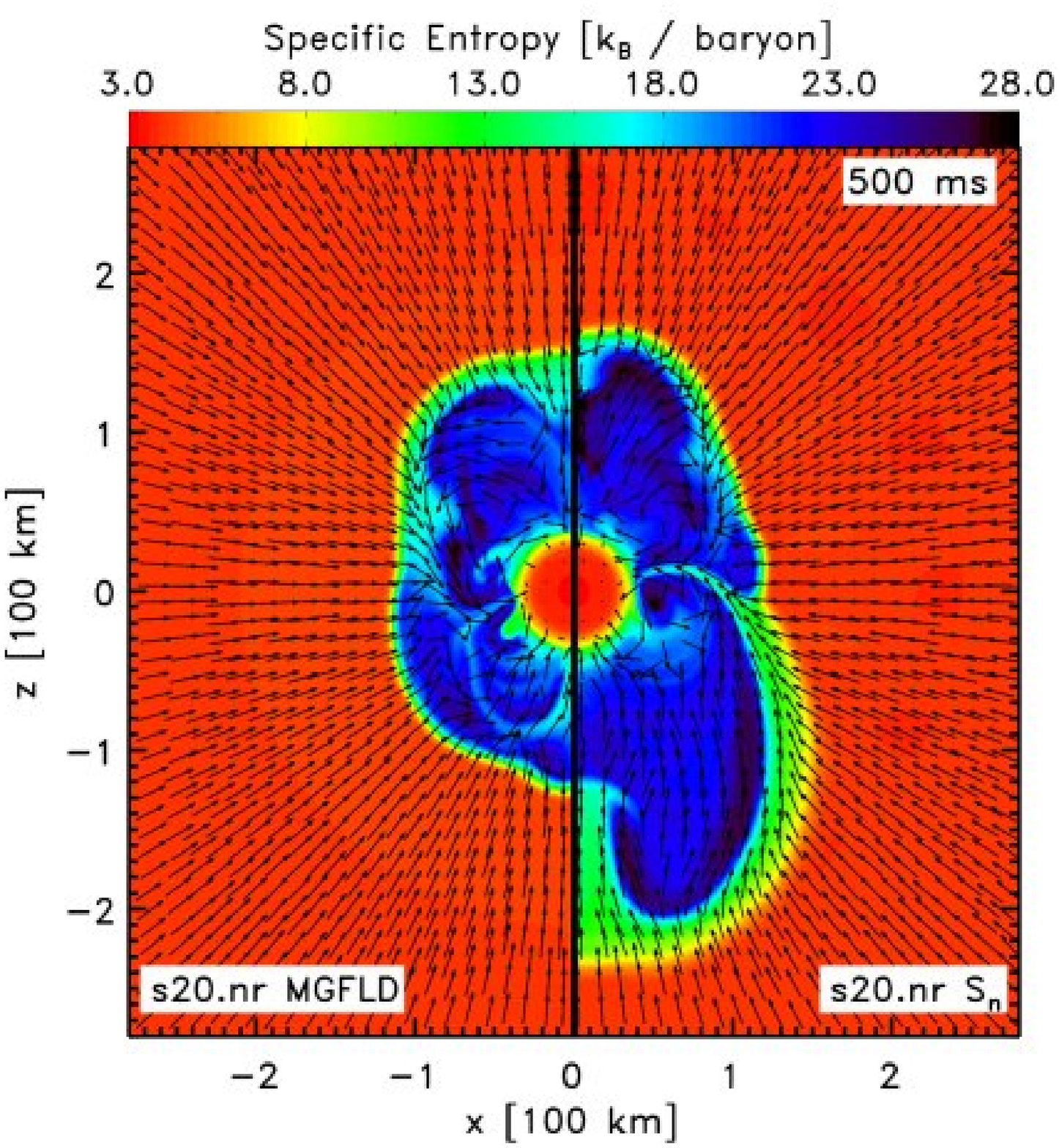}
\caption{2D entropy colormaps portraying the postbounce evolution of
model s20.nr between 160~ms (top-left panel) and 500~ms (bottom-right
panel) after core bounce. Fluid-velocity vectors are superposed to
provide an impression of the flow. Each panel's left-hand side
corresponds to the MGFLD calculation and each panel's right-hand side
shows the \sn\ result. The time of each panel is given relative to the
time of core bounce. The sequence of panels portrays the canonical
development of the SASI in the nonrotating axisymmetric context. \sn\
and MGFLD evolution agree very well in the early SASI phases, but
deviate in detail at later times, while still
exhibiting the same overall SASI dynamics. \label{fig:s20nr_evol2D}}
\end{figure*}

Since we begin the MGFLD and \sn\ calculations from an identical
hydrodynamic configuration at 160~ms after bounce, any qualitative
or quantitative differences in their evolutions must ultimately be due
to differences in the neutrino heating and cooling between \sn\ and
MGFLD.

In the left panel of Fig.~\ref{fig:s20_gain}, we display the time
evolution of the integral neutrino energy deposition (net gain) in the
gain region of model s20.nr. The net gain systematically declines at
early postbounce times, due (a) to the declining neutrino luminosity
and (b) to the rapid settling of accreting material into the net loss
region near the PNS core (cf. Fig.\ 7 of \citealt{marek:07}).  
At later times, SASI-modulated convection increases
the dwell time of accreting outer core material in the gain layer and
the slope of the net gain evolution flattens. Both \sn\ and MGFLD
track these systematics without qualitative difference. The \sn\
calculation predicts on average $\sim$5--10\% higher net gain in the
postbounce interval from $\sim$160~ms to $\sim$220~ms.  Between
$\sim$220~ms and $\sim$280~ms, MGFLD and \sn\ net gain rates agree to
within a few percent. Towards the end of this interval, the net gain
of the \sn\ calculation grows and settles at values that are on
average 20--30\% higher than those of the MGFLD run. This trend is
confirmed by the right panel of Fig.~\ref{fig:s20_gain}, which portrays
the heating efficiency, defined as the ratio of net gain to the
sum of $\nu_e$ and $\bar{\nu}_e$ luminosities.

The left panel of Fig.~\ref{fig:s20_lum_rms} depicts the temporal
evolution of the $\nu_e$ ``luminosities'' ($4\pi r^2 F_r$) as seen by
observers situated at 250~km along the north pole and south pole as
well as in the equatorial plane of models s20.nr and s20.$\pi$. Here
we focus on model s20.nr and note for the \sn\ variant that north pole
(thin solid black lines) and south pole (thin solid green lines)
``luminosities'' agree (on average) in magnitude, but exhibit
oscillations about their temporal average that are roughly out of
phase by half a cycle. The MGFLD calculation (thin dashed lines), on
the other hand, does exhibit some short-period ``luminosity''
variations, yet shows no appreciable difference between poles and
equator.

The time at which \sn\ begins to yield systematically larger neutrino
heating rates (Fig.~\ref{fig:s20_gain}) coincides with the growth of
the SASI-related shock excursions to large amplitudes 
(Fig.~\ref{fig:shockrad}). This suggests that the increased
heating is related at least in part to the \sn variant's ability to
better capture radiation field asymmetries (see also the discussion
in \S\ref{section:s20pi}), induced at late times by the rapidly varying
shock and postshock hydrodynamics in this model. Other factors that
contribute to the increased heating in the \sn\ calculation are the
higher RMS neutrino energies (by $\sim$5\%; shown in the right panel
of Fig.~\ref{fig:s20_lum_rms}) and the more gradual transition of the
\sn\ neutrino radiation field to free streaming in the postshock
region (see \S\ref{section:s20nr}).

Figure~\ref{fig:s20nr_evol2D} contrasts \sn\ and MGFLD simulations of
model s20.nr by means of colormaps depicting the specific entropy
distributions in the two variants. To visualize the hydrodynamic flow,
we superpose fluid velocity vectors.  Each panel of this figure
corresponds to a specific postbounce time and each panel's
left-hand-side depicts the state of the MGFLD calculation, while the
right-hand-side depicts the corresponding \sn\ calculation. The figure
covers a postbounce interval from 160~ms (top left) to 500~ms (bottom
right). At the beginning of the runs, the SASI-driven deviation from
sphericity of the stalled shock is mild, but grows with time, showing
$\ell=1$ excursions now generally recognized as characteristic of the
SASI\footnote{At least in detailed 2D models. \cite{iwakami:08}
carried out an exploratory 3D numerical study with nonrotating 
progenitors that suggests that in
the 3D case the $\ell$=1 dominance still obtains, yet reaches smaller
relative amplitudes, since not only higher $\ell$ modes, but also $m$
modes, may now contain power. However, \cite{yamasaki:08}, who
performed a perturbative study without symmetry constraints, argued
that in the 3D case with rotation, a dominant $m=1$ ($m=2$) mode is
likely to emerge in the case of slow (rapid) rotation.}
\citep{scheck:08, marek:07,bruenn:06,burrows:07a}.

As expected from the discussion of the s20.nr 160-ms postbounce
steady-state snapshot in \S\ref{section:snapshots}, \sn\ and MGFLD
variants of this model do not differ significantly in the early SASI
phase. However, at later SASI stages, in particular at postbounce
times $\sgreat$~300--350~ms, the simulations diverge, showing different
local qualitative and quantitative behavior within the overall SASI
theme. This is also reflected in Fig.~\ref{fig:shockrad}, which
depicts the evolution of the average shock radius, as well as the
shock radii along north pole and south pole.  The shock positions in
the \sn\ and MGFLD simulations remain close and the SASI stays
practically in phase (right panel of Fig.~\ref{fig:shockrad}) until
$\sim$~350~ms after bounce. Only then do they begin to show
significant departures from each other. The SASI in the \sn\
calculation appears more pronounced at later times, exhibiting larger
local (in time) shock excursions.  Yet, quite surprisingly, given the
significant increase in neutrino energy deposition, the \sn\
calculation does not exhibit any increase in the average shock radius,
nor does it appear to be any closer to explosion than its
MGFLD counterpart.

\subsection{Model s20.$\pi$}

\begin{figure*}[t]
\centering
\includegraphics[width=5.8cm]{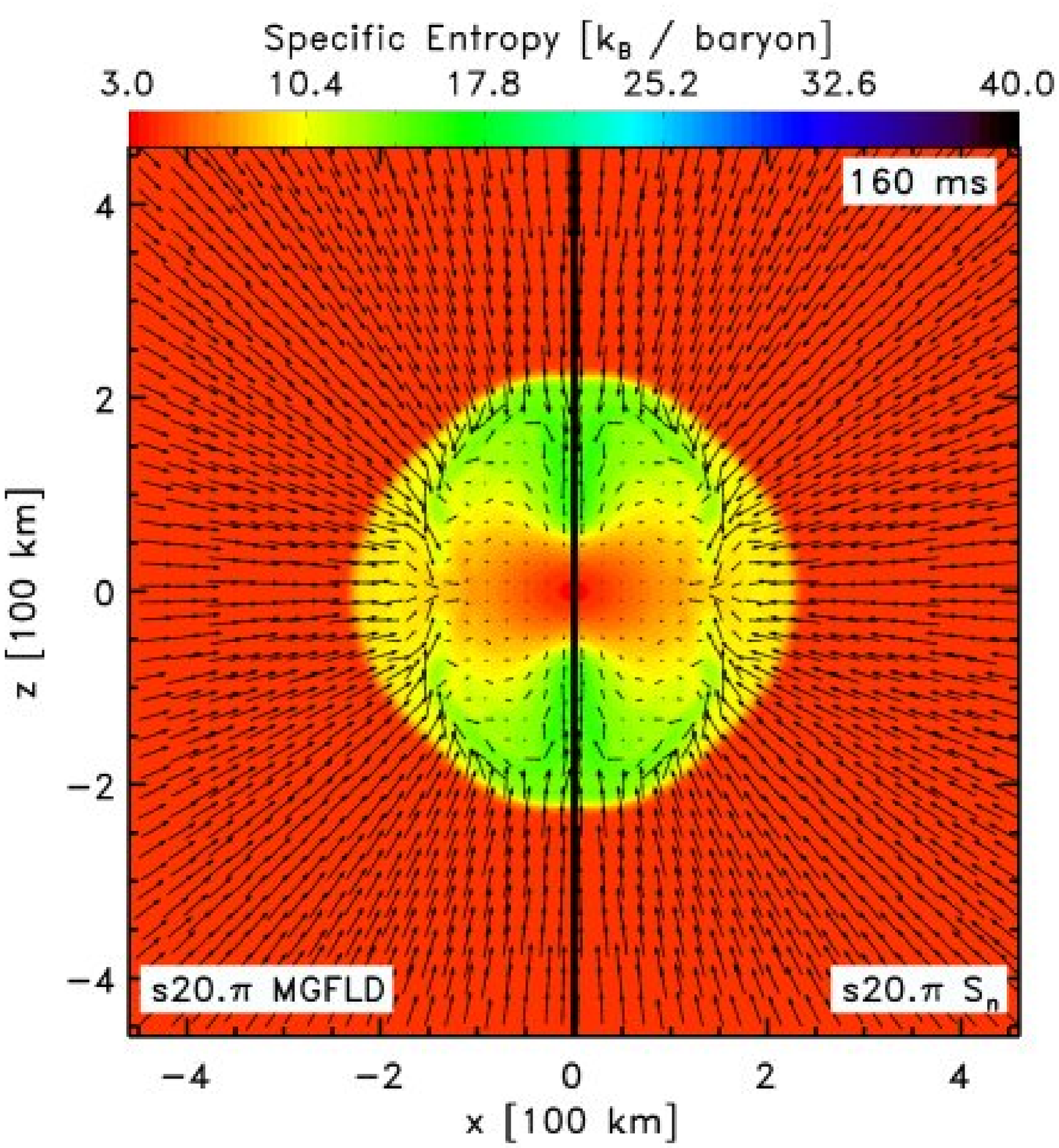}
\includegraphics[width=5.8cm]{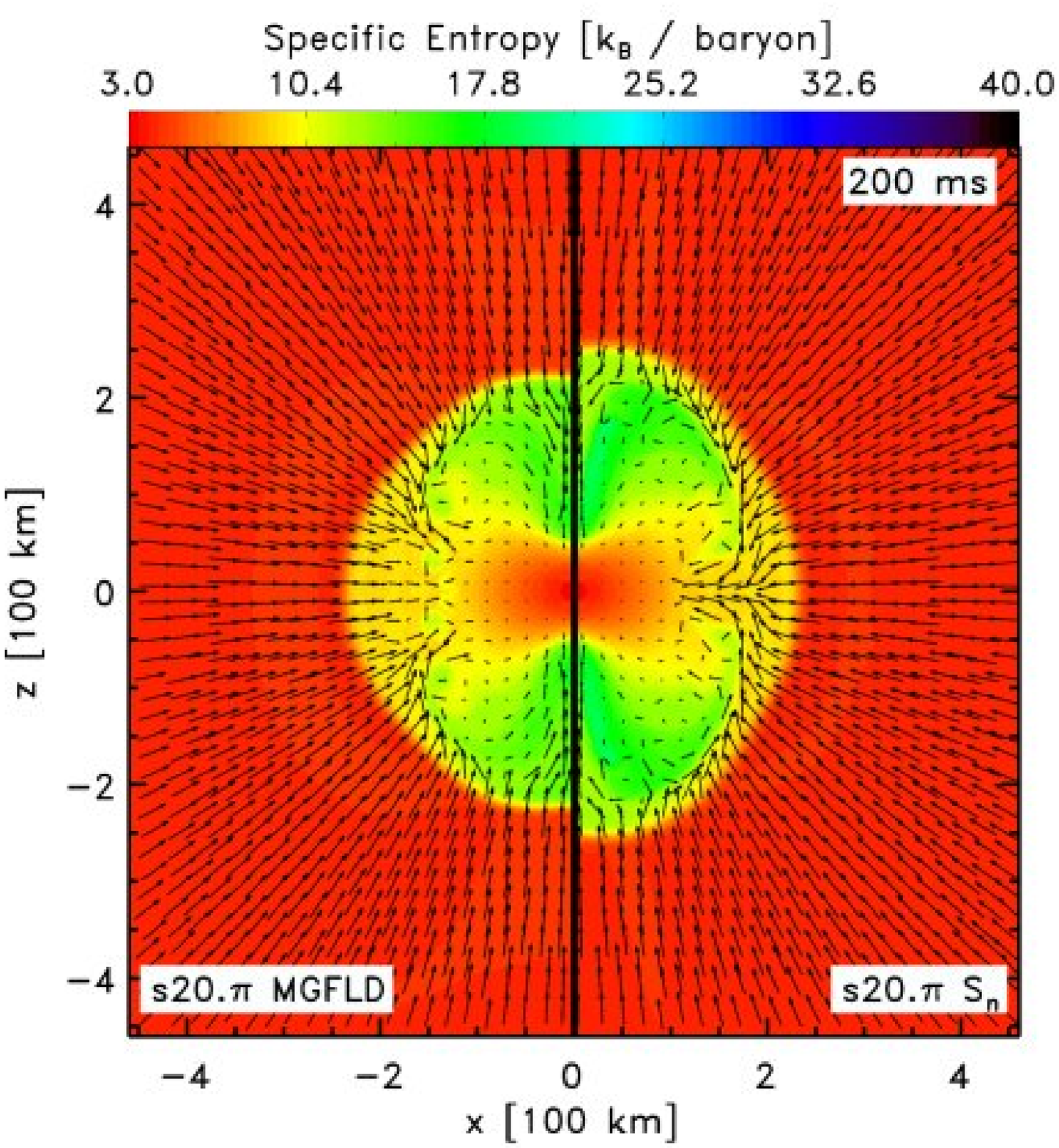}
\includegraphics[width=5.8cm]{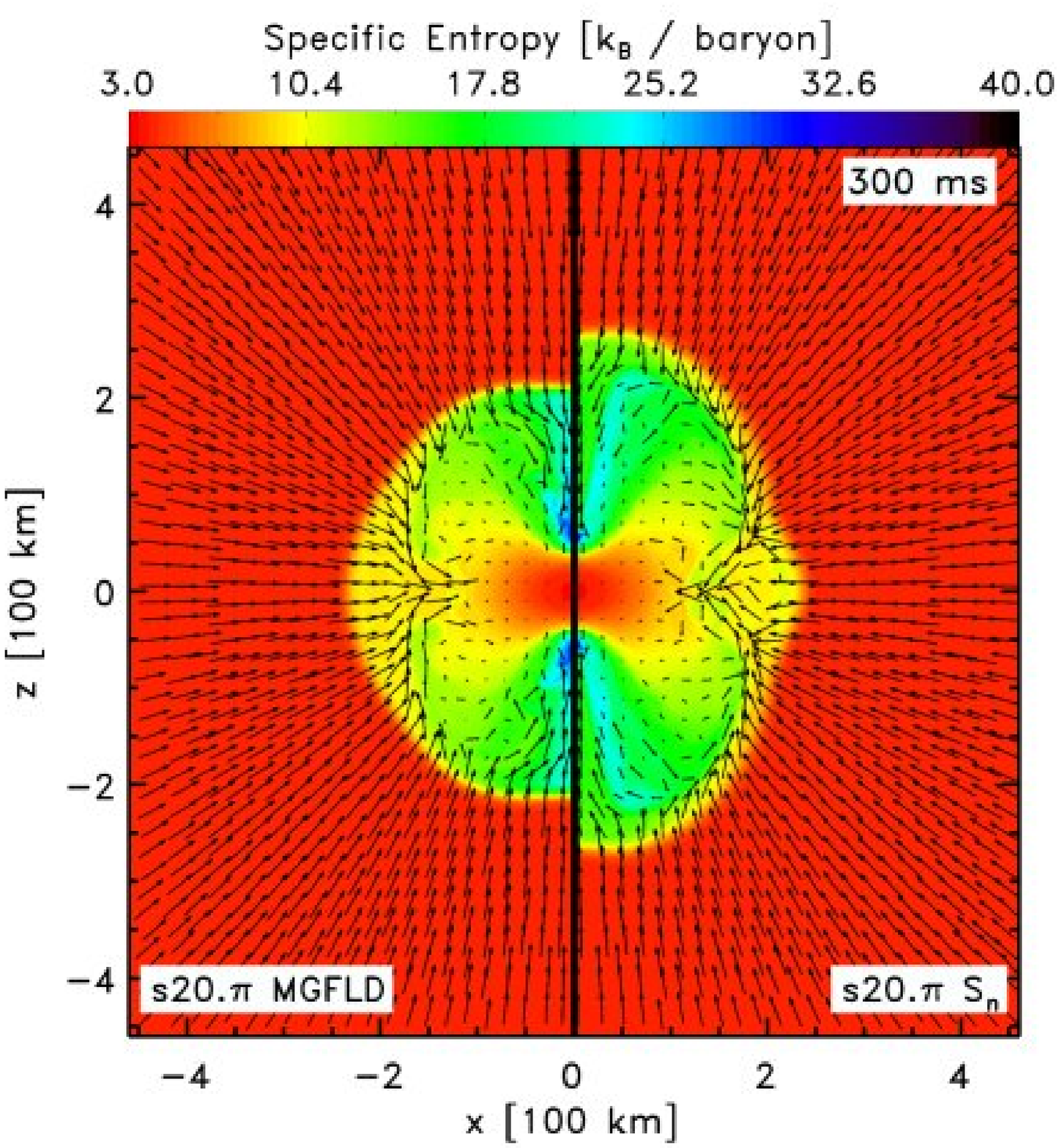}

\includegraphics[width=5.8cm]{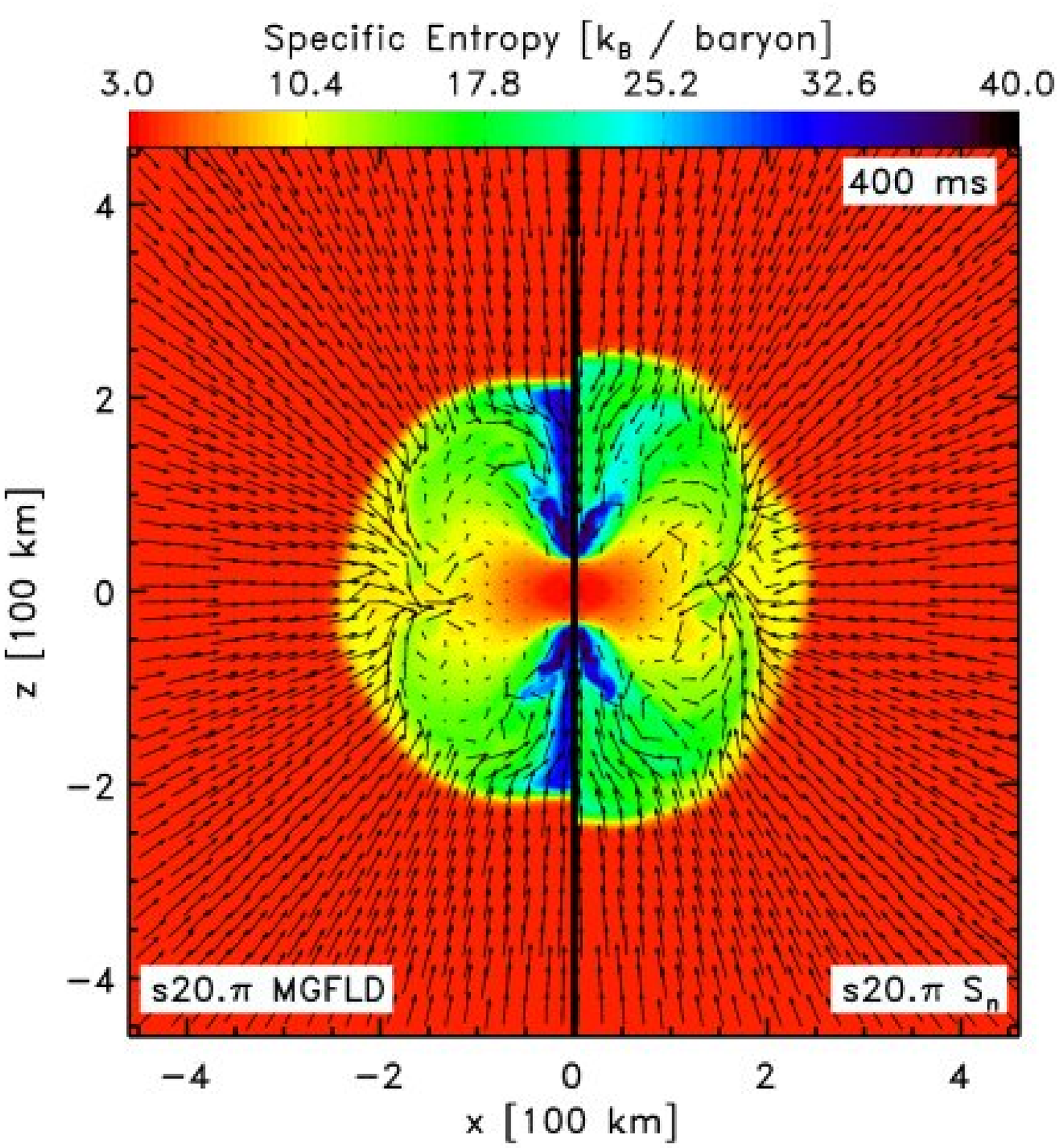}
\includegraphics[width=5.8cm]{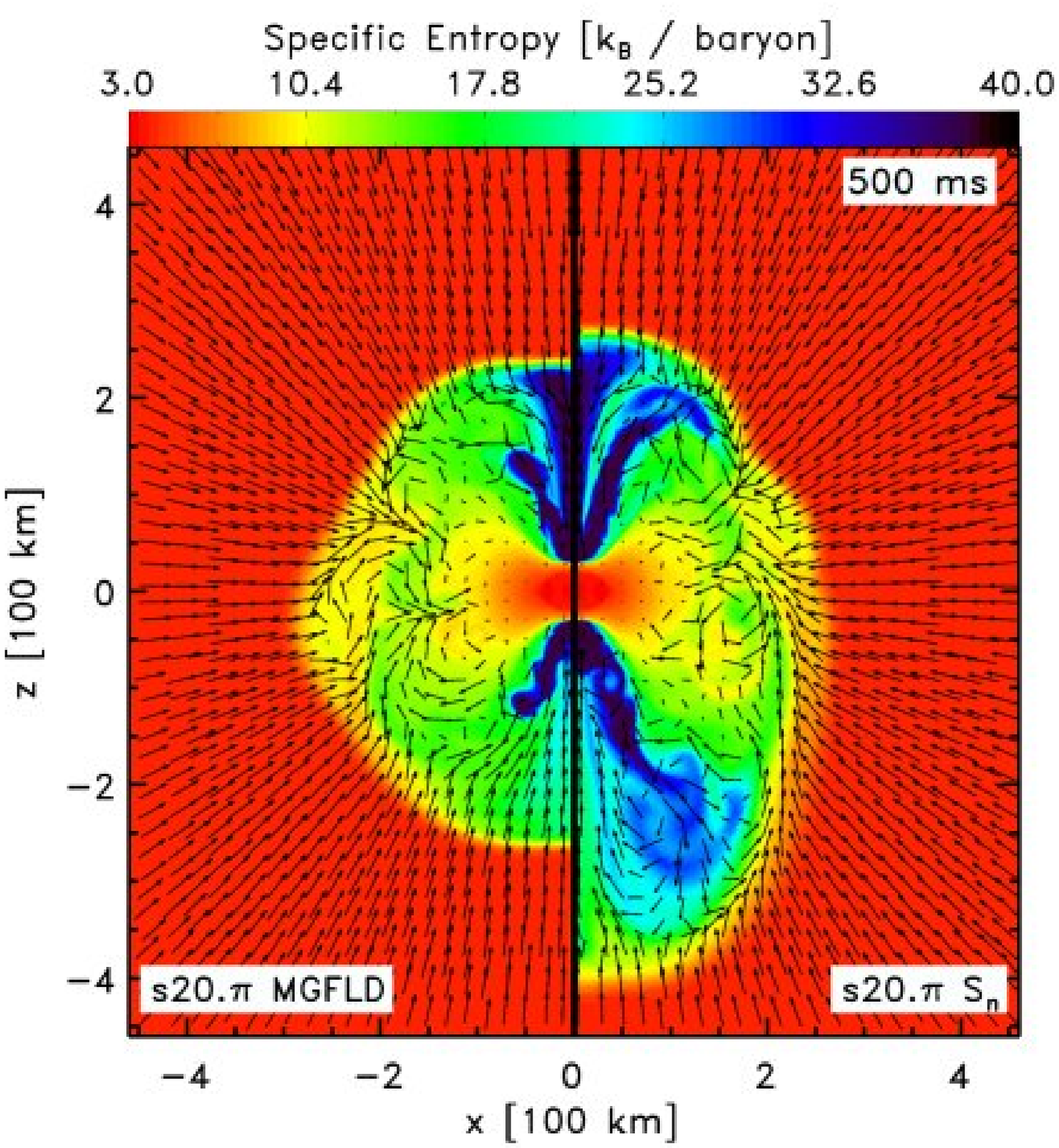}
\includegraphics[width=5.8cm]{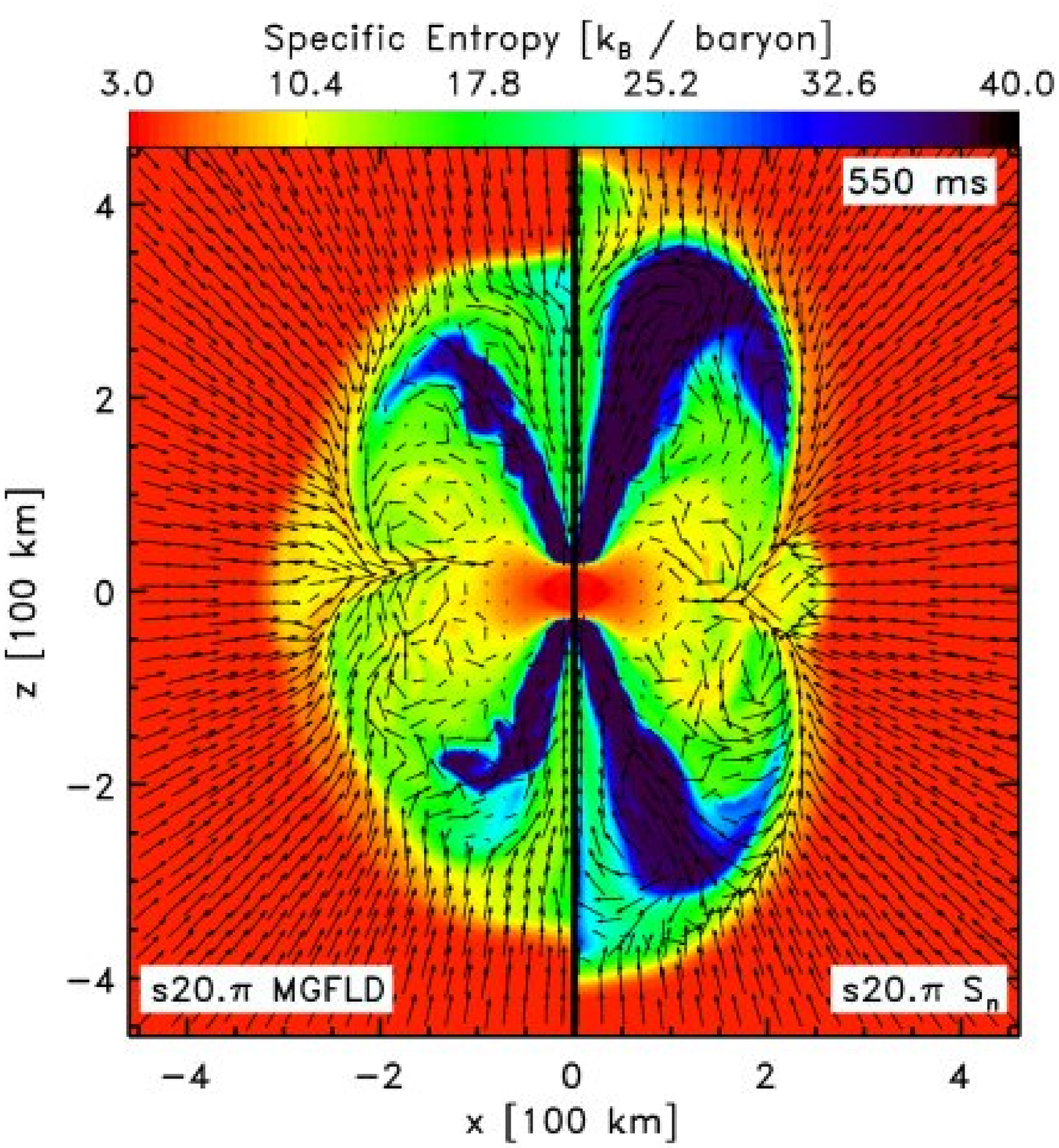}
\caption{2D entropy colormaps portraying the postbounce evolution of
the rapidly-rotating model s20.$\pi$ between 160~ms (top-left panel)
and 550~ms (bottom-right panel) after core bounce. Fluid-velocity
vectors are superposed to relay an impression of the flow and convey
the partial suppression of convective overturn in regions of
positive specific angular momentum gradient. As in
Fig.~\ref{fig:s20nr_evol2D}, we plot the MGFLD result on the left-hand
side and the \sn\ result on the right-hand side of each panel. Easily
discernible is the immediate increase in the polar shock radius
in the \sn\ calculation.  This is a direct consequence of the
increased polar neutrino heating in this variant
(Figs.~\ref{fig:s20pi_netgain} and \ref{fig:s20pi_netgain2}). At
intermediate times, \sn\ and MGFLD shock positions grow closer, but 
later on in the postbounce evolution, the \sn\ variant begins to
develop larger top-bottom SASI-like asymmetry and polar shock excursions at
earlier time than its MGFLD counterpart.\label{fig:s20pi_evol2D}}
\end{figure*}

The diagnosis of the radiation-hydrodynamic evolution of the rapidly
spinning model s20.$\pi$ is less straightforward than for the
nonrotating model s20.nr. As discussed in \S\ref{section:s20pi}, rotation
creates a global pole-equator asymmetry in the hydrodynamics of this
model. MGFLD and \sn\ track the effect of globally asymmetric
matter distributions on the neutrino radiation field to different
degrees. In the steady-state snapshot at 160~ms, \sn\ predicts
stronger neutrino heating in polar regions, yet weaker heating 
in the higher-density, larger-volume equatorial regions.

The polar, equatorial, and angle-averaged shock positions portrayed by
Fig.~\ref{fig:shockrad} show that the hydrodynamics responds
immediately to the increased polar heating in the \sn\ calculation by
a pronounced expansion of the shock along the poles. This
expansion lasts for $\sim$40~ms, after which the shock has
expanded by $\sim$20\% from $\sim$230~km to $\sim$275~km on both
poles. It stagnates at this radius and subsequently contracts
again when feedback of the hydrodynamics to the neutrino microphysics
leads to increased cooling (cf.\ the increased polar neutrino emission
shown in Fig.~\ref{fig:s20_lum_rms}). The increased
postshock volume also results in a larger gain region and increased
(compared to MGFLD) total neutrino energy deposition and heating
efficiency. However, this increased heating is not able to sustain the
large postshock volume. The shock slowly recontracts in the
postbounce interval from $\sim$250~ms to $\sim$380~ms and eventually
settles at radii similar to those obtained by the MGFLD shock.

In Fig.~21, we present a sequence of 2D entropy colormaps with
superposed velocity vectors, portraying the postbounce evolution of
model s20.$\pi$ from 160~ms on. The rapid rotation in this model not
only partially stabilizes convection, but also weakens and delays the
growth of the characteristic $\ell=1$ SASI\footnote{But see the work
of \cite{yamasaki:08}, who find via perturbative analysis that in 3D,
rotation enhances the development of azimuthal $m=1$ and $m=2$
SASI-related spiral structure.}. Since larger shock radii are
associated with an increased growth rate of the SASI (e.g.,
\citealt{foglizzo:07,scheck:08}), the \sn\ variant begins to develop
periodic shock excursions along the symmetry axis at much earlier
times than the MGFLD simulation (Fig.~\ref{fig:shockrad}).  However,
at times later than $\sim$400~ms, the MGFLD model picks up the
large-amplitude SASI as well and both calculations exhibit large-scale
radial shock excursions beyond $\sim$400~km along the pole
(Fig.~\ref{fig:shockrad}).  The average shock radius increases in both
calculations in this late postbounce phase.  We observe neither such
large shock excursions nor a systematic late-time increase of the
average shock radius in the nonrotating model. The observed behavior
is most likely due to the rapid rotation and the resulting rarefaction
of the polar regions that reduces, in particular at late times, the
ram pressure of accretion and allows for the more pronounced SASI.

In the left panel of Fig.~\ref{fig:s20_lum_rms}, we contrast the
$\nu_e$ ``luminosities'' ($4\pi r^2 F_r$) seen by observers located at
a radius of 250~km above the north pole, the south pole, and in the
equatorial plane of model s20.$\pi$. The MGFLD variant predicts a
pole-equator flux asymmetry of $\sless$10\% that is roughly
constant with time. The \sn\ calculation yields a very different
picture. Polar and equatorial ``luminosities'' at 250~km (i.e., near the
shock) are vastly different (cf. \S\ref{section:s20pi}).  Over time,
the equatorial ``luminosity'' decreases while the ``luminosity'' along the
poles is enhanced.  At $\sim$200~ms, polar and equatorial ``luminosities''
differ by a factor of $\sim$3. By $\sim$500~ms, this factor has grown
to 4.  In addition, the \sn\ simulation shows SASI-induced variations
in north and south-pole ``luminosities'' that grow to $\sim$3--5\% at late
times and are not tracked in the MGFLD variant. These variations are
akin those reported for the nonrotating model s20.nr, yet have longer
periods, since the large shock excursions in model s20.$\pi$ occur on
longer timescales.

As in the nonrotating model, we also find in model s20.$\pi$ that \sn\
yields systematically higher RMS neutrino energies for all species
and at all times. However, as shown in the right panel of
Fig.~\ref{fig:s20_lum_rms}, the angle-averaged RMS energies do not
exhibit a significant increase in the time interval covered by our
simulations. This, again, is due to rapid rotation which slows down the
PNS's contraction. Not shown in Fig.~\ref{fig:s20_lum_rms}, but
present in the \sn\ variant throughout its postbounce evolution, are 
$\sim$10--20\% (roughly constant in time and independent of
species) higher RMS energies for neutrinos emitted from polar regions
compared to those emitted from the PNS equator. This is consistent
with our analysis of the neutrino spectra and RMS neutrino energies
for the 160~ms postbounce steady-state snapshot presented in
\S\ref{section:s20pi}.

We end our postbounce simulations of model s20.$\pi$ with \sn\ and
MGFLD at 550~ms after bounce. Though within roughly the same
qualitative picture, the two approaches to neutrino transport yield
appreciable differences in the postbounce radiation-hydrodynamics
evolutions.  Importantly, and in contrast to our findings for the
nonrotating model, \sn\ in model s20.$\pi$ does not lead to
systematically higher integral neutrino energy deposition, and at late
postbounce times, shows a volume-integrated heating rate that is even
$\sim$30\% lower (on average) than in its MGFLD counterpart.

\section{Summary and Discussion}
\label{section:summary}

Using the code VULCAN/2D \citep{livne:04,burrows:07a,livne:07}, we
perform long-term full-2D multi-angle, multi-group neutrino
radiation-hydrodynamic calculations in the core-collapse supernova
context. Based on postbounce hydrodynamic configurations from MGFLD
simulations, we first compute 2D angle-dependent (\sn) steady-state
solutions for models without precollapse rotation and with rapid
rotation ($\Omega_0 = \pi$~rad~s$^{-1}$). From these snapshots, we
numerically follow the radiation-hydrodynamics evolution with \sn\
neutrino transport, tracking the nonrotating model to 500~ms and the
rotating model to 550~ms after bounce.

Done for the first time in 2D, we investigate in
detail the angle-dependent specific intensities and neutrino radiation
fields. We compute angular moments of the specific
intensity, including the Eddington tensor, and introduce Hammer-type
map projections to visualize the angle dependence of the specific
intensity. These we employ to demonstrate the decoupling systematics
of the neutrinos and the gradual transition to free-streaming of the
radiation fields with decreasing optical depth.

We compare our \sn\ simulations with MGFLD counterparts.  We find for
both models and at all times that the \sn\ specific intensity
distributions transition less rapidly from isotropy to free-streaming
in the semi-transparent outer postshock regions.  \sn\ yields mean
inverse flux factors and RMS neutrino energies in these regions that
are $\sim$10\% larger than those obtained with MGFLD.  In the context
of the neutrino mechanism of core-collapse supernova explosions,
differences in the net neutrino energy deposition rates between MGFLD
and multi-angle \sn\ transport are of greatest interest.  In the
quasi-spherical early postbounce phase of the nonrotating model, we
find that \sn\ predicts a 5--10\% greater neutrino energy deposition
rate than MGFLD.  At later times, when the SASI has reached large
amplitudes and globally deforms the postshock region, we find that
\sn\ yields consistently larger (up to 30\% on average) 
energy depositions and leads to significantly larger temporary
shock excursions around average shock radii that do not depart
much from those in the MGFLD calculation.

Convection on small and intermediate scales and SASI on large
scales, are the key agents of the breaking of spherical symmetry in
nonrotating (or slowly rotating) core-collapse supernovae. While we
observe no large qualitative differences in the growth and dynamical
evolutions of convection and SASI between nonrotating \sn\ and MGFLD
models, we find that the imprint of the asymmetric hydrodynamics on
the neutrino radiation fields is captured with greater detail by the
multi-angle transport scheme. For the late-time, heavily
SASI-distorted postbounce core, \sn\ predicts asymptotic neutrino
fluxes that have variations with time and angle of 5--10\% in
magnitude. MGFLD is able to capture the temporal variations of the
neutrino luminosity, but smoothes out the angular flux variations at
large radii/low optical depths.

Rapid rotation leads to large deviations from spherical symmetry and a
rotationally-deformed PNS emits, by von Zeipel's law of gravity
darkening, a greater neutrino flux along its rotational axis than
through its equatorial regions
\citep{janka:89a,janka:89b,kotake:03,walder:05,buras:06b,
dessart:06b}.  We find that both 2D MGFLD and \sn\ yield similar
radiation fields and pole-equator flux ratios at radii smaller than
$\sim$100~km. At larger radii, the MGFLD radiation fields sphericize
and show little pole-equator asymmetry in their asymptotic
variables. \sn, on the other hand, captures large pole-equator flux
ratios of up to 4:1 at late times and predicts polar neutrino spectra
that are harder in peak energy (RMS energy) than on the equator by up
to 30\% (10--15\%) for $\nu_e$ neutrinos, and somewhat less for the
other species. All this results in a neutrino energy deposition rate
per unit mass in polar regions that is locally up to $\sim$2.5--3
times higher when multi-angle transport is used. This increased polar
neutrino heating has a dynamical effect on the postbounce evolution,
leading to rapid shock expansion in the polar regions and an earlier
onset of the (initially) rotationally-weakened SASI. However, at late
times, the SASI in the MGFLD calculation catches up and yields shock
excursions of a similar magnitude.

In summary, our results show that 2D multi-angle neutrino transport
manifests interesting differences with 2D MGFLD when addressing local
and global radiation field asymmetries associated with rapid rotation
and the non-linear SASI at late postbounce times. In addition,
multi-angle transport results in enhanced neutrino energy
deposition. The latter is most significant in the polar regions of
rapidly rotating postbounce configurations and affects dynamically the
postbounce evolution, including the growth of the SASI. However, in
the large postbounce interval covered by our simulations, the local
and global differences between multi-angle transport and MGFLD
calculations do not appear large enough to alter the overall
simulation outcome. Importantly, the multi-angle models do not appear
to be closer to explosion than their MGFLD counterparts.

Although we neglect velocity-dependent transport terms and coupling of
neutrino energy bins, we do not expect our conclusions to be altered
by their inclusion, since they are not likely to affect significantly
the differences between multi-angle transport and MGFLD. Further
significant limitations of our present study are the neglect of
general relativistic and MHD effects, the restriction to only one
finite-temperature nuclear EOS, the limited resolution in
momentum-space imposed by the computational cost of multi-angle
calculations, and the use of two spatial dimensions, plus
rotation. In the future, we will investigate
the dependence of our results (e.g., heating rates, radiation-field
asymmetries etc.) on the choice of flux limiter and will consider
different progenitor models.

The core-collapse supernova problem is one of many feedbacks.  Larger
heating rates and heating efficiencies than found in our models appear
to be necessary to break the feedback cycle between neutrino radiation
fields and hydrodynamics, revive the stalled shock, and unbind the
supernova envelope -- if the neutrino mechanism is to obtain in the
way presently envisioned. Future work will have to go beyond the
limitation of axisymmetry and must address in detail the entire
ensemble of possible factors relevant in the supernova problem,
including, but not limited to, 3D dynamics, multi-angle neutrino
transport with velocity dependence and inelastic $\nu_e$-$e^-$
scattering, progenitor structure, rotational configuration,
magnetohydrodynamics, convection, the SASI, PNS g-modes, general
relativity, the nuclear EOS, and neutrino-matter interactions.

\section*{Acknowledgements}

We acknowledge helpful discussions with and input from Jeremiah
Murphy, Ivan Hubeny, Casey Meakin, Jim Lattimer, Alan Calder, Stan
Woosley, Ed Seidel, Harry Dimmelmeier, H.-Thomas Janka, 
Kei Kotake, Thierry Foglizzo, Ewald
M\"uller, Bernhard M\"uller, Martin Obergaulinger,
Benjamin~D.~Oppenheimer, Thomas Marquart, and Erik Schnetter.  This
work was partially supported by the Scientific Discovery through
Advanced Computing (SciDAC) program of the US Department of Energy
under grant numbers DE-FC02-01ER41184 and DE-FC02-06ER41452.
C.D.O. acknowledges support through a Joint Institute for Nuclear
Astrophysics postdoctoral fellowship, sub-award no.~61-5292UA of NFS
award no.~86-6004791. E.L. acknowledges support by the Israel Science
Foundation (grant 805/04).  The computations were performed at the
local Arizona Beowulf cluster, on the Columbia SGI Altix machine at
the Ames center of the NASA High End Computing Program, at the
National Center for Supercomputing Applications (NCSA) under Teragrid
computer time grant TG-MCA02N014, at the Center for Computation and
Technology at Louisiana State University, and at the National Energy
Research Scientific Computing Center (NERSC), which is supported by
the Office of Science of the US Department of Energy under contract
DE-AC03-76SF00098.

\end{document}